\documentclass[twocolumn,secnumarabic,amssymb, nobibnotes, aps, prd, nofootinbib]{revtex4-2}

\usepackage[T1]{fontenc}
\usepackage[pdftex]{graphicx,color}
\usepackage{hyperref}
\usepackage{subfigure}
\usepackage{enumerate}
\usepackage[all]{hypcap} 
\usepackage{tabularx,multirow,booktabs,blindtext}

\hypersetup{pdfstartview={XYZ null null 0.99}, bookmarksnumbered=true, bookmarksopen=true, colorlinks=true, linkcolor=black, citecolor=black, urlcolor=black, filecolor=black, anchorcolor=black, runcolor=black}

\begin{document}

\title{Promoting the transition to quantum thinking: development of a secondary school course for addressing knowledge revision, organization, and epistemological challenges}

\author{Giacomo Zuccarini}%
\email{giacomo.zuccarini@unipv.it}
\affiliation{Department of Physics, University of Pavia, Via Bassi 6, Pavia 27100, Italy}
\author{Marisa Michelini}%
\affiliation{Department of Mathematics, Computer Science and Physics, University of Udine, via delle scienze 206,
33100 Udine, Italy}
\date{\today}

\begin{abstract}
We describe the development of a course of quantum mechanics for secondary school designed to address the challenges related to the revision of classical knowledge, to the building of a well-organized knowledge structure on the discipline, and to the development of a plausible picture of the quantum world. The course is based on a systemic approach to conceptual change, which relies on its analysis in the transition from classical to quantum mechanics, and coordinates cognitive and epistemological aspects. We show how our approach drives the derivation of design principles, how these principles guide the development of the instructional sequence and of its strategies, how their implementation requires the blending of different research perspectives and learning systems. The first challenge is addressed through a path of revision of classical concepts and constructs which leverages prior knowledge according to the dynamics of each notion in theory change. The second by adopting a framework that promotes the construction of a unifying picture of quantum measurement across contexts. The third by designing the course around a modelling process that engages students in epistemic practices of the theoretical physicist, such as generating and/or running thought experiments, and mathematical modelling in a purely theoretical setting. All is aimed to help students accept the quantum description of the world as a plausible product of their own inquiry. This process is assisted by the discussion of the facets of the foundational debate that are triggered by each of the suggested interpretive choices, with the goal to promote an awareness of its cultural significance, of the limits the chosen stance, of the open issues. Data on the cycles of refinement illustrate how a set of activities have been made effective in addressing the challenges at a local level, and the process by which the learning outcomes informed their development.

\end{abstract}

\maketitle

\section{Introduction} \label{Sec:Intro}
Research on the teaching and the learning of quantum mechanics (QM) holds a special position in physics education and science education at large, since it is at the crossroads of general research threads and key topics in the field.

First of all, both students learning introductory science topics and students learning QM face a substantial challenge in achieving an effective knowledge revision. In theory change from classical mechanics (CM) to QM, basic classical terms, such as `measurement' and `state', undergo a shift in meaning. Students struggle to interpret the properties of their quantum counterparts, as reported by research conducted at different educational levels. Investigations on upper division students elicited several issues with the new features of ideal quantum measurement \cite{Zhu2012}; at a sophomore level, the interpretation of its probabilistic character, and as a result, of quantum uncertainty has been recognized as a major challenge to students \cite{Ayene2011};  in the context of photon polarization, research revealed difficulties to interpret the concept of quantum state, identified by secondary school students as a physical quantity \cite{Pospiech2021}. The nonintuitive nature of the new versions of the concepts and the impossibility to directly visualize quantum systems represent an educational bottleneck that can be overcome with the support of mathematical sense-making. For instance, the representation of the state of a system in a complex Hilbert space or configuration space is amenable, at least in part, to different forms of visualization. However, also familiar constructs such as `vector' and `vector superposition' change both in properties and representational role \cite{Pospiech2021}. Not surprisingly, students struggle to develop a consistent physical interpretation of the quantum version of these constructs: even at the beginning of graduate instruction, they have difficulties to identify the referent of vector superposition in QM, as they tend to associate it with mixed states, which can be described classically as lack of knowledge about the state of the system \cite{Passante2015}.

Another challenge faced both by introductory science students and physics majors enrolled in a QM course is the difficulty to overcome knowledge fragmentation, i.e., the organization of knowledge in small, disconnected pieces of contextual nature, which can be productively applied locally but which lack global coherence. This concerns, in the two cases, student interpretation of natural phenomena \cite{Vosniadou2008, diSessa2014}, and their knowledge of the quantum model \cite{Johnston1998}. Research conducted at the end of upper-division QM courses and at the beginning of graduate instruction suggests that student reasoning is strongly context-dependent \cite{Marshman2015}, and therefore that the development of a globally consistent knowledge structure may be only halfway even after prolonged periods of instruction. So far as we know, no investigation of this issue is available on secondary school students and non-physics-or-engineering majors who received traditional instruction on QM. However, the more limited scope of teaching/learning sequences designed for such student populations enhances the risk of promoting the construction of disconnected models valid only in the context of an individual phenomenon or experiment \cite{Malgieri2017}.

A challenge specifically related to the learning of QM is due to the controversial character of its scientific epistemology: the nature of the systems described by the mathematical formalism, the completeness or not of the information we can get on them, and the explanation of observations in the lab depend on the chosen interpretive stance. The traditional presentation of the theory comes with a seemingly counterintuitive picture of the world, which requires students to revise or renounce very basic tenets about nature such as the well-defined position of physical objects \cite[e.g.,][]{Griffiths2018}. Research indicates that an elementary understanding of the core concepts of the theory does not necessarily imply the acceptance of QM as a personally convincing description of physical reality \cite{Ravaioli2017}. In order to be accepted, the quantum model needs to be perceived by students as plausible and reliable \cite{Posner1982, Ravaioli2017}.

Overall, a major goal of physics education research on QM is helping students overcome the manyfold  challenges involved in learning QM. A recent review of educational research on QM at secondary and lower undergraduate level \cite{Lewerissa2017} divides the teaching approaches into those that focus on conceptual/mathematical understanding, on epistemic activities such as model-building, and on the interpretation of the theory. Favoring the development of an integrated conceptual understanding has been a basic aim of Malgieri \textit{et al.}, who implemented Feynman's sum-over-paths approach by using GeoGebra simulations, so as to allow secondary school students to analyze different experimental setups with the same conceptual tools \cite{Malgieri2017}. The course of Wittmann and Morgan for nonscience majors places special emphasis on personal epistemology (not to be confounded with scientific epistemology) as a means to help students work with nonintuitive content and to strengthen their understanding of scientific modelling \cite{Wittmann2020}. Baily and Finkelstein made of the controversial nature of the interpretation of QM and of the discussion of students' beliefs about it a topic onto itself, by designing a modern physics course for engineering majors aimed to help develop more consistent views of quantum phenomena, more sophisticated views of uncertainty, and greater interest in QM \cite{Baily2015}.

Given the growing consensus to shift the focus from difficulties to student resources, i.e., pieces of prior knowledge that can be productively used in the learning process \cite{Coppola2013, Goodhew2019}, researchers are starting to ask how to put conceptual, symbolic and epistemological resources of students in the service of learning QM \cite{Dreyfus2017, Dini2017}. However, as regards instructional materials on QM, there is a need to identify the links between specific sets of available knowledge elements and possible educational strategies, and to empirically test their effectiveness.

In order to find indications on how to address the aforementioned challenges and needs in the learning of QM, we chose to work in the framework of conceptual change (CC). Since the release of \textit{The Structure of Scientific Revolutions} by T. Kuhn \cite{Kuhn1962}, the theory change from CM to QM has been seen as an exemplary case of CC in the history of science \cite{Thagard1992}. Educational research shows that this is a central element also behind the challenges students face in learning QM \cite{Tsaparlis2009, Singh2015, Lewerissa2017}. With the rise of educational models of CC \cite{Posner1982, Potvin2020}, a number of researchers started to consider the problem of teaching QM as the design of strategies to effectively promote a conceptual change in individual learners \cite{Thagard1992, Kalkanis2003, Tsaparlis2013, Malgieri2017}.

In general, a CC approach to science teaching can be described as ``seeking to foster understanding and the adoption of scientific ideas as new systems of interpretation'' \cite{Amin2014} of natural phenomena.
In the context of introductory science topics, challenges related to knowledge revision \cite{Carey1999, Chi2013, Vosniadou2008}, fragmentation \cite{Vosniadou2008, diSessa2014}, epistemology \cite{Amin2014}, as well as the use of resources \cite{diSessa2014} have been widely investigated in this framework. Indeed, educational models of CC have been primarily developed to account for the transition from na\"{\i}ve to scientific knowledge, a process associated with the modification of conceptual structures formed in the context of lay culture. The transition from CM to QM, instead, requires changes in knowledge structures about a scientific theory, and developed as a result of instruction. While long-established models of CC can represent a valuable aid also in addressing this transition, there is a need to develop approaches to the teaching of QM that take into account the differences involved in learning a successive theory.

In addition, studies on CC are undergoing a systemic turn, with a growing number of researchers coming to understand this process in terms of multiple interacting elements, which include, e.g., cognitive aspects, epistemology, social interaction \cite{Amin2014}. Teaching-learning sequences on QM have prevalently focused on one or the other of these factors. Educational proposals based on a systemic approach, coordinating multiple factors at different levels of analysis, are currently lacking.

In this paper, we describe the design and refinement of a QM course for secondary school based on a more systemic approach, which relies on an analysis of CC in the learning of QM as a successive theory, and integrates cognitive and epistemological aspects. Our goal is to address all the aforementioned challenges, taking advantage of available knowledge elements and intuition in the process. The analysis of the cognitive aspects is aimed to provide an in-depth characterization of the first two challenges, suggesting how to leverage prior knowledge for achieving an effective revision, and how to identify conceptual tools suitable to make predictions on quantum processes across contexts. The epistemological analysis suggests to engage students in a theory-building activity, with the aim to propose strategies for promoting the development of a plausible picture of the quantum world, and is assisted in this process by the parallel development of an awareness of the foundational debate.

\section{Theoretical framework: the identification of the design principles} \label{Sec:2}

\subsection{Transition to quantum mechanics as a successive theory: cognitive aspects}
Our analysis was one source of inspiration for the development of a model of the transition from the understanding of a theory to the understanding of its successor presented by Zuccarini and Malgieri \cite{Zuccarini2022}. The model includes an exploration of the impact of theory change on various forms of challenges (cognitive, epistemological, affective) and the identification of strategies for promoting the understanding of the new content.

As regards the cognitive aspects, in the design of our course we focused only on the transition from CM to QM, and specifically on two cognitive signatures of the knowledge of a scientific theory, which ideally represent a significant component of the initial state of the learner. They are the understanding of, and the ability to use for descriptive, explanatory and problem-solving purposes
\begin{enumerate}[1.]
\item different public representations of relevant concepts: linguistic, mathematical, visual, etc. \cite{Arabatzis2020};
\item the exemplars of the theory: tasks and resolution strategies encountered in lectures, exercises, laboratory assignments, textbooks, etc. \cite[p. 134]{Hoyningen1993}.
\end{enumerate}
Theory change is always accompanied by change in exemplars and in relevant concepts at different representational levels (new formation, evolution, disappearance \cite{Arabatzis2020}). Therefore, we need to consider not only ontological change in concepts, but also change in constructs used by the scientific community to represent these concepts, as well as the change in tasks and in resolution strategies. These features mark important differences with CC processes at introductory level, since na\"{\i}ve science is neither socially shared nor mathematized.

Conceptual dynamics involved in theory change can give rise to various types of learning challenges. New formation may entail, e.g., coalescence or differentiation of familiar notion in nonintuitive terms. Evolution may determine difficulties to identify which aspects of a familiar entity can be productively used in the new theory and which not, to develop a consistent understanding of the new aspects, and to clearly discriminate between the old and the new version. Disappearance may deprive students of important resources in organizing scientific knowledge. Change in exemplars - that may be strongly context-dependent - is reasonably related to knowledge fragmentation. However, it is clear that each factor of change may have an influence on both challenges, and therefore that overcoming these challenges requires a coordination of knowledge revision and knowledge organization strategies.

As we have seen in Section \ref{Sec:Intro}, research shows that the trajectories of concepts and constructs from CM to QM (e.g., the concept of state and vector superposition), and the new nature of basic tasks (such as predictions on the results of measurement in different contexts) are often connected with deep learning challenges.

According to this account of the impact of theory change on cognitive challenges, the basis for addressing knowledge revision and its organization in the transition from CM and QM are respectively the educational analysis of change in concepts/constructs, and of change in exemplars. We present them in two separate subsections.

\subsubsection{Change in concepts and constructs: addressing the challenge} \label{Sec:2.1.1}

The analysis informing our design started by building a taxonomy of changes that may affect scientific concepts (as regards their ontology) and mathematical constructs (as regards their representational role). After a selection of basic notions of CM and QM and their characterization according to standard presentations of the theories, we used this taxonomy to trace their dynamics. Finally, we scanned the literature on student understanding of QM for finding connections between documented difficulties and aspects of these dynamics. The final result was the identification of educationally significant patterns of continuity and change and the development of pattern-dependent strategies to promote concept revision, with a special focus on the use of prior intuition. This analysis is reported in full in Zuccarini and Malgieri \cite{Zuccarini2022}.

Here we illustrate how we use it in the design of teaching strategies with the support of dynamic frames, a tool employed by philosophers of science to represent aspects of the categorical structure of a concept in a theory, and therefore to analyze its dynamics in theory change \cite{Andersen2006}. To this purpose, we adopt the format of Zuccarini and Malgieri, designed to directly visualize continuity and change, not only in concepts (ontological change) but also in constructs (representational change), linking specific conceptual content to the related pattern of change.

The first example we show is the trajectory of the concept of \textit{system quantity}.
\begin{figure*}[!htbp]
    \centering
\begin{tabular}{|r|l|} \hline
    \includegraphics[width=.48\textwidth]{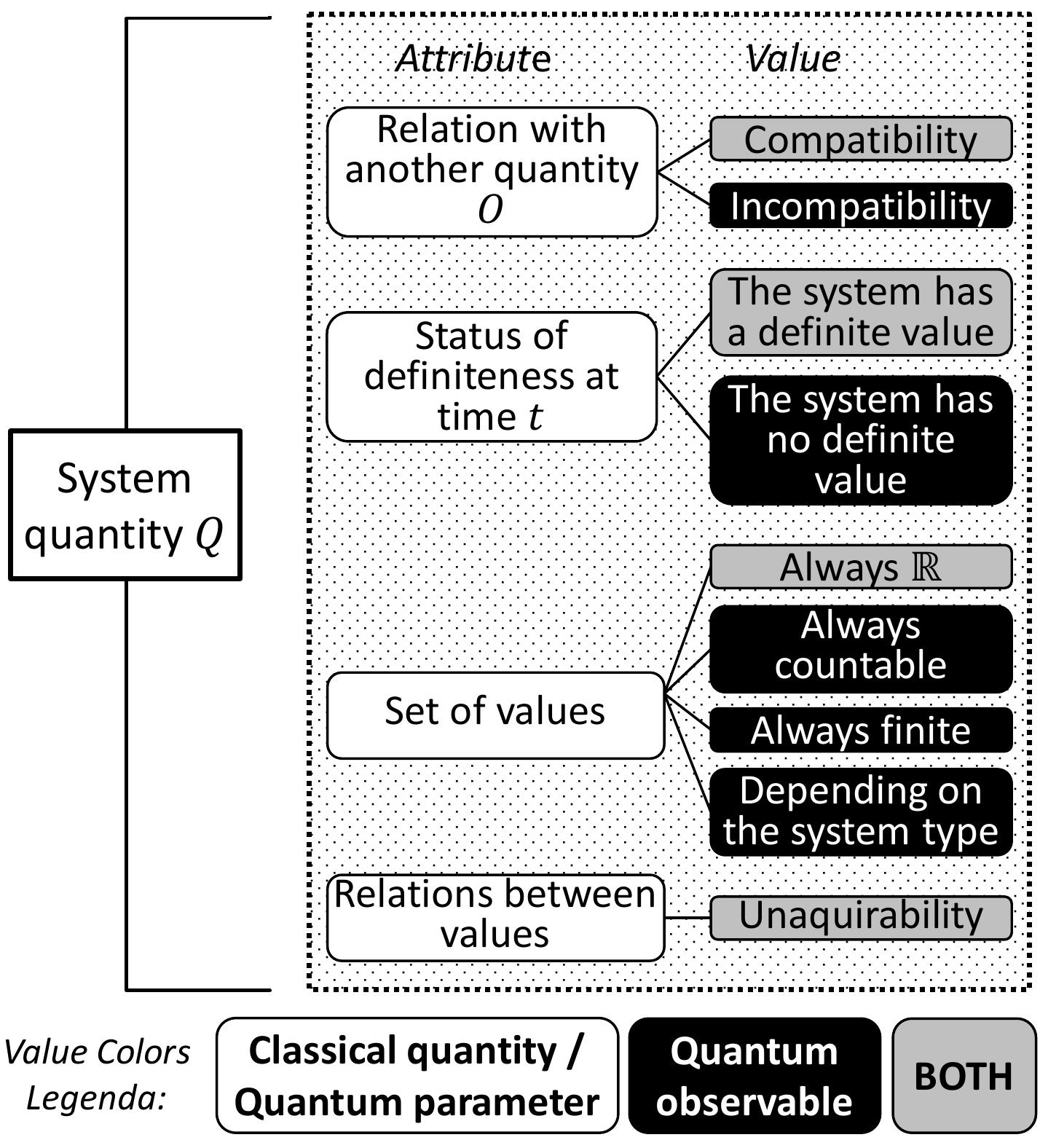} &
    \includegraphics[trim=0 0 0 -5, width=.52\textwidth]{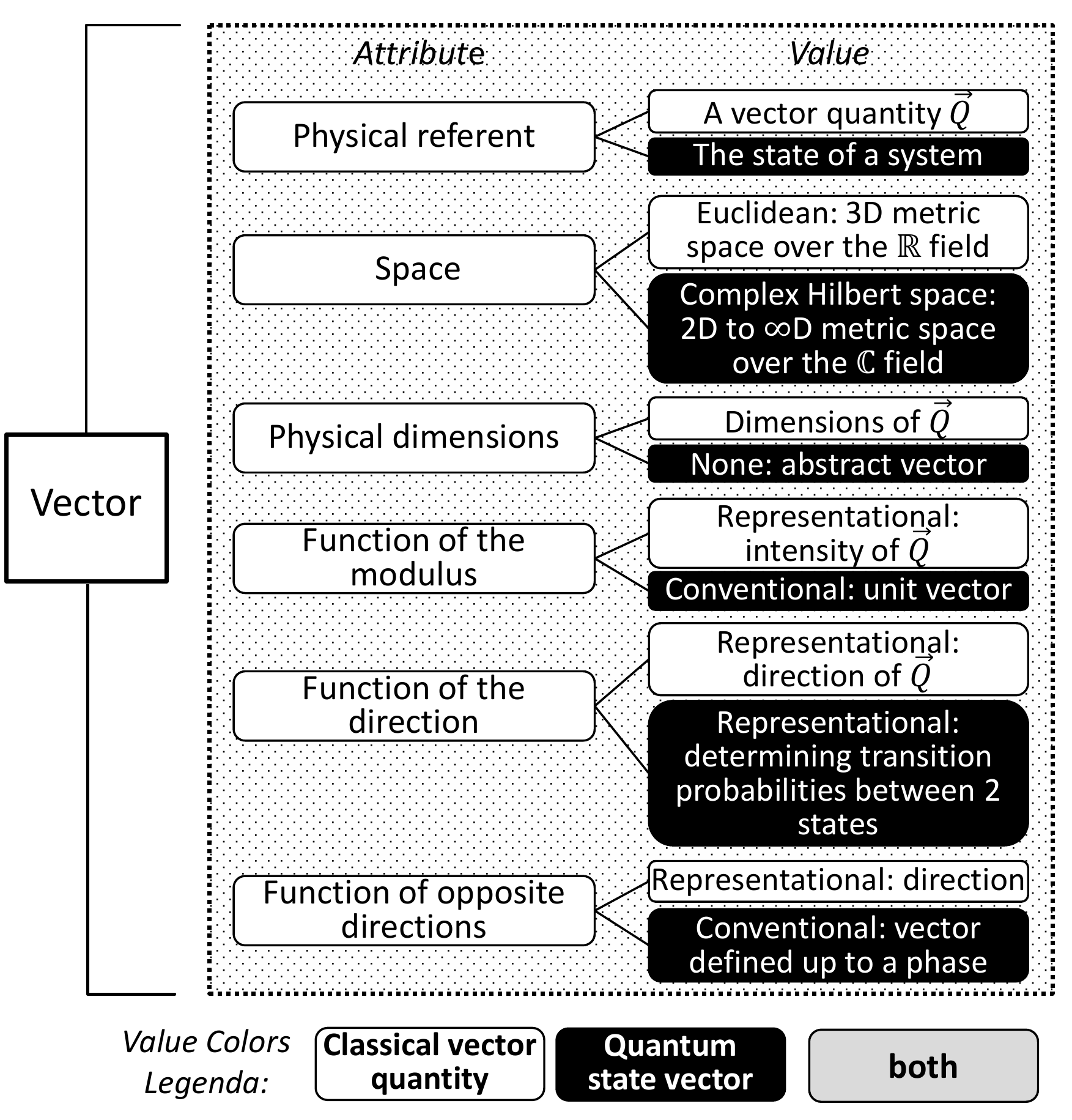} \\ \hline 
\end{tabular}
    \caption{Visualizing categorical change: (a) the concept of \emph{system quantity}; (b) the \emph{vector} construct.}
    \label{FIG:1}
\end{figure*}
This expression refers to physical quantities describing properties of systems and includes both dynamical variables, that in QM become observables, and parameters such as mass, that in non-relativistic QM behaves as a classical quantity.
Its main pattern of evolution is denoted as \textit{categorical generalization} (Fig. \ref{FIG:1}.a): for an aspect of the notion (attribute), some of the corresponding conceptual properties (values of the attribute) pertain to both CM and QM (gray boxes), the others only to QM (black boxes). In the light of this pattern, a strategy to promote a smooth revision may be the following: since the aspect involved includes at least a property that is valid both in CM and in QM, the instructor can start its discussion by fostering a productive use of classical intuition. Then, the new property(ies) can be made available to students by means of constructivist strategies (e.g., use of analogies for concepts, mathematical sense-making and/or embodied cognition for mathematical constructs). Finally, following the prevalence model of CC \cite{Potvin2015}, we rely on dissonance strategies to produce cognitive conflict. The frame is used for structuring an end-of-unit table containing interpretive tasks to discriminate between the old and the new properties of the notion, identifying the correct context of application of each one .

The second example is the trajectory of the \textit{vector} construct, that in CM is primarily used to represent physical quantities, while in QM it typically refers to the state of a system. Its conceptual dynamics presents only one pattern, denoted as \textit{value disjunction} (Fig. \ref{FIG:1}.b): for an aspect of the notion, there are properties valid only in CM (white box) and properties valid only in QM (black box). In this respect, the quantum version and the classical one are totally different. The main distinction between the strategy suggested for categorical generalization and the strategy we suggest here is the need to start by building an understanding of the new property(ies), and to use prior intuition at the end, as a contrast \cite{Henderson2017}.

Last, while other modern theories present a clear demarcation line between their phenomena and classical ones (a low $v/c$ ratio in special relativity), in QM the so-called ``classical limit'' is a deep and controversial issue \cite{Ballentine1994, Klein2012}. It appears that students need to bridge, at a conceptual and a formal level, the world of the new theory to that of the old one, in order to facilitate the transition between the two perspectives. End-of-unit tables based on the frames can also serve to visualize and interpret continuity and change in concepts and constructs, offering a bridge between the two theories, not in terms of limiting processes, but of categorical structure.

All this is summarized in the first design principle:\\

\centerline{\fbox{\begin{minipage}{\columnwidth}
  \centering{PRINCIPLE OF KNOWLEDGE REVISION}\\
  the analysis of continuity and change in concepts and constructs will be used for developing
  \begin{itemize}
  \item trajectory-dependent strategies for a smooth transition to their quantum versions
  \item end-of-unit tables containing interpretive tasks on selected aspects of their trajectory  $\Rightarrow$\\
  promoting the discrimination between the classical and the quantum version of a notion by identifying the correct context of application of each aspect
  \end{itemize}
  as a result, this approach to knowledge revision provides an opportunity to address student's need of comparability with CM
\end{minipage}}}

\subsubsection{Change in exemplars: addressing the challenge} \label{Sec:2.1.2}
The analysis of challenges related to knowledge fragmentation in QM has played a fundamental role in the development of the course. A difficulty was represented by the search for quantum exemplars at secondary school level. As a matter of fact, quantum formalism is among the less common curriculum content in traditional TLSs for secondary school students, as well as real lab assignments and simulated experiments \cite{Stadermann2019}. In upper-division courses, instead, students are exposed to the basic mathematical machinery of the non-relativistic QM and to plenty of exercises in lectures, recitations, homework and exams. As a result, analyzing the nature of these tasks and corresponding resolution strategies became the key for contrasting classical and quantum exemplars.

This work fed into a recent publication on the structure of quantum knowledge for instruction \cite{Zuccarini2020}. According to it, textbooks and educational research mainly focus on the following tasks and related subtasks: finding information (1) on the results of the measurement of an observable on a state, (2) on the time evolution of the state, and (3) on the time evolution of the probability distribution of an observable on a state.

\begin{figure*}[!htpb]
    \centering
       \fbox{\includegraphics[width=14cm]{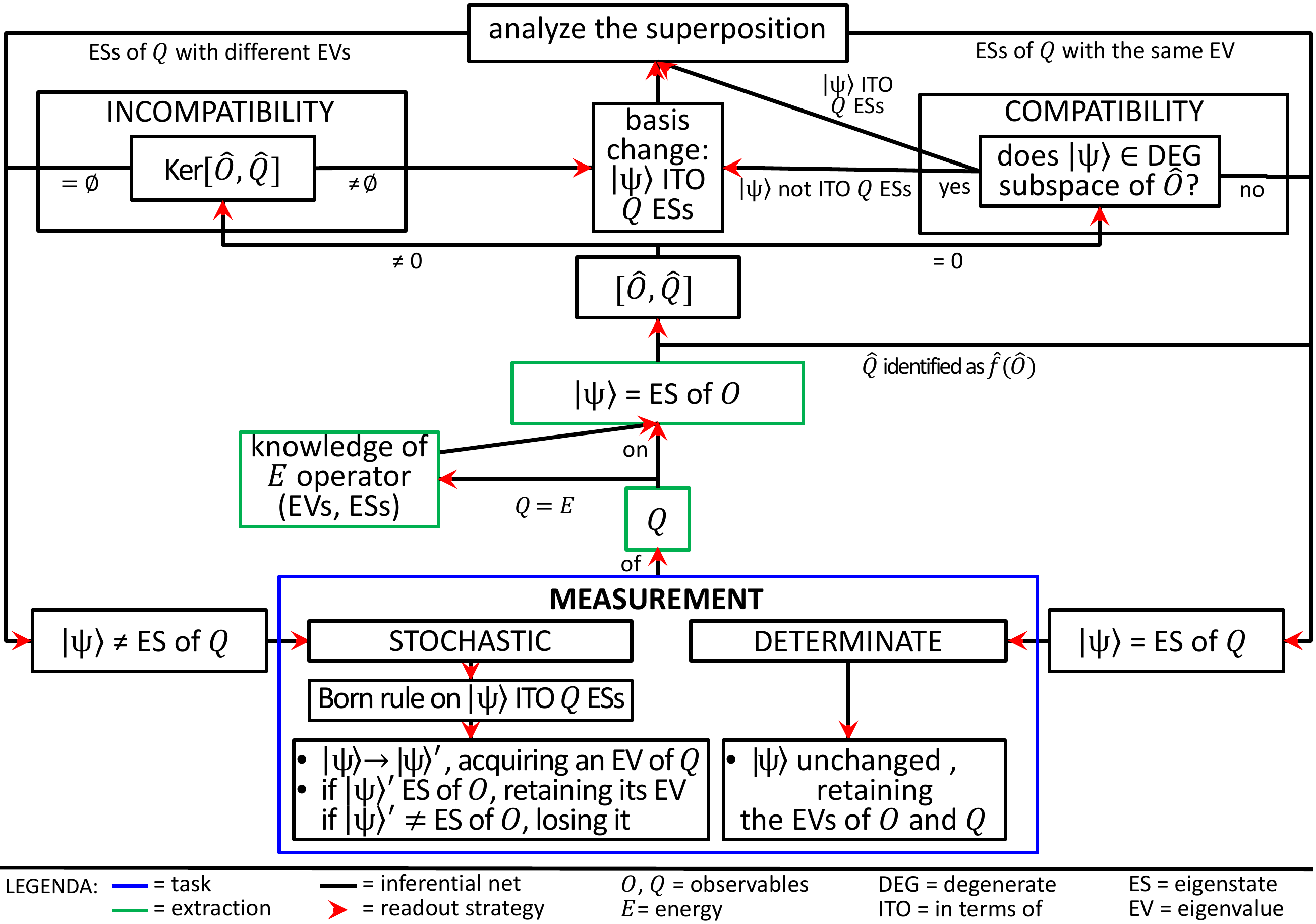}}
    \caption{Measurement on an eigenstate of an observable.}
    \label{FIG:2}
\end{figure*}

The strategies for accomplishing them are radically different from those used in solving CM problems. See, for instance, the structure of the task of getting information on the measurement of an observable on a state, which is written here as an eigenstate of a given observable (Fig. \ref{FIG:2}).
The task is described according to the language of coordination classes \cite{diSessa1998, diSessa2016}, a theoretical perspective suitable to understand how scientific concepts function in determining a particular class of information about the physical world. From the analysis of Fig. \ref{FIG:2}, it is immediate to identify at least two reasons behind the difficulty to build a global knowledge structure in QM. First, the context-specific elements of this structure are in turn complex objects. Second, the subtasks related to using prediction tools and procedures are also complex, unfamiliar, and highly variable from context to context.

If we aim to provide school students with valuable support to make predictions on quantum processes, we need to considerably simplify the picture: set aside time evolution to focus only on measurement; give priority to qualitative predictions; set aside operators, commutators, and eigenvalue equations. Despite these constraints, it is still possible to promote the building of a well-organized knowledge structure about quantum measurement, allowing students to identify, in a wide range of contexts: 1) the result of measurement (in terms of acquisition, loss, retention of definite values of observables); 2) the nature of the process (stochastic or determinate); 3) a consistent explanation of this nature; 4) quantitative predictions; 5) comparability with ideal classical measurement.

The key is represented by decomposing a basic element of the inferential net of Fig. \ref{FIG:2}, i.e., the knowledge of the relation between the relevant observables, into binary relations between their values, that arise in measurement. To this purpose, we borrow some language and elements from a specific perspective on the quantum theory, the Geneva-Brussels approach \cite[see, e.g.,][]{Debianchi2011}. In our course, a value of a \emph{system quantity} - classical or quantum - that can be said to be either possessed by a physical system (when the probability to measure it is $1$) or not, is denoted as physical ``property'', and relations existing between values are denominated as ``relations between properties''. The concept of property we use is a strongly restricted version of the original one, which includes not only values but also the union of disjoint intervals of values. Unless indicated otherwise, we will describe ideal measurements of discrete and continuous quantities only in terms of single values.

The relations between properties of interest to us are defined as follows: two different properties, $P_a$ and $P_b$, belonging respectively to the \emph{system quantities} $O$ and $Q$, not necessarily distinct from each other, are
\begin{itemize}
    \item \emph{mutually unacquirable}: if any system possessing one of them retains it and can never acquire the other in the measurement of the corresponding quantity. No system can ever possess $P_a$ and $P_b$ at the same time (mutual exclusivity);
    \item \emph{incompatible}: if any system possessing one of them loses it and may stochastically acquire the other in the measurement of the corresponding quantity. No system can ever possess $P_a$ and $P_b$ at the same time (mutual exclusivity);
    \item \emph{compatible}: if any system possessing only one of them retains it and may stochastically acquire the other in the measurement of the corresponding quantity. If the system possesses $P_a$ and $P_b$ at the same time, it retains them in the measurement of any of the corresponding quantities.
\end{itemize}

Given the abstract nature of these definitions, we illustrate some concrete consequences and examples. \emph{Mutually unacquirable} properties are, in the first place, different properties of the same quantity, but also properties of different quantities that are mutually unacquirable due to physical constraints. An instance of the latter case is the following: if the azimuthal quantum number of a system is $l=1$, it is not possible for this system either to possess $m=4$ or to acquire it in the measurement of $L_z$, and viceversa. An arbitrary value of position is always \emph{incompatible} with any value of its conjugate momentum. A property of spin is always \emph{compatible} with properties of spatial observables (position, momentum, kinetic energy, orbital angular momentum, etc.). As with the relations between observables, relations between properties are invariant across contexts except for those between energy properties and properties of other observables. For the latter, the validity of the relations is restricted to the type of system described by the Hamiltonian at hand.

This framework can be used as a conceptual basis for an educational reconstruction of quantum measurement across contexts and of the transition from the classical to the quantum measurement task. Since mutual unacquirability and incompatibility naturally arise in the exploration of spin or photon polarization measurements, which can also be addressed in a simple quantitative form (Malus's law for photon polarization and its equivalent for spin), these relations can be introduced and justified to students as empirical regularities that are specific to quantum systems. Moving on to the relations between \emph{system quantities} is almost immediate: two quantities are compatible if every property of each one is compatible with at least one property of the other, otherwise they are incompatible. Except in a limited number of cases (when two quantities are incompatible, but admit simultaneous eigenstates), relations between quantities can be qualitatively assessed in a similar way:\\
the \textit{system quantities} $O$ and $Q$ are
\begin{itemize}
        \item \emph{incompatible}: if any system possessing a property of one of them loses it in the measurement of the other quantity and stochastically acquires one property of the latter. No system can ever possess properties of $O$ and $Q$ at the same time;
        \item \emph{compatible}: if any system possessing only a property of one of them retains it in the measurement of the other quantity and stochastically acquires one property of the latter. If the system possesses properties of $O$ and $Q$ at the same time, it retains them in the measurement of these quantities.
\end{itemize}

Other contexts can be added to the picture either by using relations between properties or between observables. If we opt for the latter, by knowing which relations exist between the observables that initially have a definite value and the measured observable, we determine which of them are definite after the measurement and which not. In addition, the possession in advance or the lack of a property of the measured observable allows us to assess the nature of the process: respectively determinate or stochastic. This task can be accomplished also in the presence of degeneracy.

If the course includes the discussion of the vector representation of the state, quantitative predictions on discrete state systems are also possible. Measurements on systems such as harmonic oscillators, potential wells, bound states of hydrogen-like atoms can be studied by exploiting the conversion of the relations into algebraic constraints. With the introduction of the state vector, mutual unacquirability of properties becomes orthogonality of the corresponding states, and paves the way for examining the superposition of a finite number of vectors. A task within the reach of school mathematics.

Finally, the relations represent an instrument for addressing student's need of comparability with the classical measurement task. All of them except incompatibility can be identified also in the classical regime: all \emph{system quantities} are compatible with one another; properties of the same quantity are mutually unacquirable; every point particle possesses one property of each quantity.
Thus, the emergence of incompatibility represents an explanation of theory change with relation to the measurement task and the description of systems at a point in time.

We are now able to formulate the second principle:\\

\centerline{\fbox{\begin{minipage}{.9\columnwidth}
  \centering{PRINCIPLE\\ OF KNOWLEDGE ORGANIZATION}\\
  the framework of the relations between properties and then between observables will be developed together with students in the simple context of two-state systems, and will be used to
    \begin{itemize}
  \item promote the construction of a unifying picture of quantum measurement and the ability to manage it in problem-solving, allowing students to explore this process in other scientifically significant contexts
  \item  promote a smooth transition to a quantum perspective and help address student's need of comparability between CM an QM, since it constitutes a transtheoretical framework
  \end{itemize}
  \end{minipage}}}


\subsection{Transition to quantum mechanics: personal and scientific epistemology} \label{Sec:2.2}

\subsubsection{Personal epistemology: theoretical modelling cycles} \label{Sec:2.2.1}
Personal epistemology may be introduced as an individual's answers to questions such as ``how do you know?'' and ``why do you believe?'' \cite{Wittmann2020}. Recent reviews on CC and epistemic cognition report that there is a convincing body of research establishing a connection between more sophisticated epistemologies and deeper conceptual understanding in a particular domain \cite{Amin2014, Elby2016}. QM represents an ideal context for exploiting this synergy: a focus on epistemology may promote the learning of counterintuitive quantum content; conversely, a course of QM may be an opportunity for studying the practices of scientific modelling. Wittmann and Morgan, e.g., structured most of their course around activities in which students work to build new concepts and create new knowledge \cite{Wittmann2020}.

In order to put the aforementioned synergy in the service of learning QM, we chose to adopt a similar approach. However, given the wide range of possible knowledge-building activities, we endeavoured to identify the most appropriate ones for the content at hand. According to Sandoval \textit{et al.}, the conceptual, procedural, and epistemic expertise of a discipline is bound up in its specific practices \cite{Sandoval2016}. But what practices characterize the construction of QM as a knowledge domain? A peek at the history of physics in the early 20th century suggests that theory-building is at the core of these practices. We concluded that involving students in theoretical modelling activities could be a promising strategy for helping them accept the quantum description of the world as a plausible product of their own inquiry, developing theoretical reasoning skills in the process. However, educational research on the epistemic practices that characterize the work of theoretical physicists is currently lacking. In Fig. \ref{FIG:3}, we propose a list of historically significant practices of this kind.

The specification of these practices served as a starting point for the design of strategies to engage students in theoretical modelling cycles, which is performed by drawing on perspectives concerning the use of mathematics in physics \cite{Uhden2012, Redish2015}, the conduction of thought experiments \cite{Gilbert2000, Stephens2012}, on different forms of inquiry- and modelling-based approaches, e.g., the ISLE learning system \cite{Etkina2015}.\\

\begin{figure*}[!htpb]
\centerline{\fbox{\begin{minipage}{.9\textwidth}
\centering{EPISTEMIC PRACTICES OF THEORETICAL NATURE IN THE HISTORY OF PHYSICS}\\
FUNDAMENTAL: generating, extending and revising interpretive models that act as comprehensive systems of explanation with the aim to develop a unified picture.
\begin{enumerate}
 \item building new knowledge on a topic by means of thought experiments (e.g., Galileo's free fall experiment, Maxwell's demon, Einstein's elevator)
 \item interpreting already known laws within the framework of new models (e.g., Clausius, Maxwell and Boltzmann's interpretation of thermodynamic quantities and laws within the framework of atomistic models)
 \item deepening the theoretical investigation of a phenomenology by adopting multiple perspectives (e.g., Euler's two specifications of the flow field in fluid mechanics)
 \item identifying mathematical constructs suitable to describe features of physical objects and processes (e.g., Newton's adoption of constructs of infinitesimal calculus for describing gravity and mechanics at large)
 \item analyzing mathematical constructs already representing features of physical objects or processes to deduce results that have not been unveiled yet (e.g., Lagrange's laws of fluid dynamics: an application of one of Euler's specifications of the flow field to special cases with new mathematical methods)
 \item starting from results found in one context and extending or adapting them to other contexts (e.g., Maxwell's hydrodynamic model of the magnetic lines of force)
  \end{enumerate}
  \end{minipage}}}
  \caption{Practices of theoretical nature that have been historically used by physicists for building new knowledge.}
    \label{FIG:3}
\end{figure*}


\centerline{\fbox{\begin{minipage}{.9\columnwidth}
  \centering{EPISTEMIC PRINCIPLE }\\
  design the course around a modelling process that includes theoretical practices used by physicists
  in the historical development of the discipline, with the goal to help students
    \begin{itemize}
  \item accept the quantum description of the world as a plausible product of their own inquiry, thus promoting a smooth transition to a quantum perspective
   \item  build theoretical reasoning skills
    \end{itemize}
\end{minipage}}}


\subsubsection{Scientific epistemology: approach to interpretation} \label{Sec:2.2.2}
Research on students transitioning from classical to quantum thinking shows that when interpretive themes are deemphasized, interest in QM decreases, while learners still develop a variety of (sometimes scientifically undesirable) views about the interpretation of quantum phenomena \cite{Baily2015}. For this reason, we built our course around a \emph{clearly specified} form of standard approach \cite{Bub1997}, schematically set apart from other schools of thought by means of rules of correspondence between the structure of the theory and its physical referents in the world:
\begin{enumerate}
    \item a pure state provides complete information on the behavior of an individual quantum system (ruling out statistical interpretations);
    \item an observable of a system has a determinate value if and only if the quantum state of the system is an eigenstate of the operator representing the observable (ruling out modal interpretations);
    \item the quantum description of processes includes two different types of state evolution: in the absence of measurement, the unitary evolution governed by the Schr\"{o}dinger equation; in measurement, the evolution prescribed by the projection postulate (ruling out other no-collapse interpretations).
\end{enumerate}
As mentioned at the end of each statement, all have been questioned by part of the scientific community, with the third being the most unsatisfactory one for a variety of reasons \cite{Bub1997}, starting from the measurement problem \cite{Schlosshauer2007}.

An additional interpretive choice concerns the wave-particle duality. Baily and Finkelstein adopt a ``matter-wave perspective'' \cite{Baily2015}, that allows students to interpret without paradoxes how a system can ``know'' whether two paths are open or only one of them in a ``which-way'' experiment. However, if the system propagates as a wave, students may ask what kind of medium supports or, equivalently, is perturbed by this wave. For this reason, in the construction of a full quantum model of a system, we adopt a field ontology, a perspective put forward in education also in recent years \cite[e.g.,][]{Hobson2013}.\\

\centerline{\fbox{\begin{minipage}{.9\columnwidth}
  \centering{EPISTEMOLOGICAL PRINCIPLE }\\
  design the course around a clearly specified form of (standard) interpretation in order to
    \begin{itemize}
  \item build a coherent educational proposal
  \item identify which facets of the foundational debate are triggered by each interpretive choice, and how to discuss them  according to the educational level of the students, helping them develop an awareness of the cultural significance of the debate, of the limits the chosen stance, of the open issues
    \end{itemize}
\end{minipage}}}

\section{Implementing the principles: development of the path and of the activities}

In this section, we illustrate how the principles of design drove the development of the course and of individual activities within it. The content and the modelling process were informed by the interplay of the \textit{Principle of Knowledge Revision}, the \textit{Principle of Knowledge Organization} and the \textit{Epistemic Principle}.  The construction of the learning path is described in Section \ref{Sec:3.1}. In the subsequent sections, each principle is addressed separately, showing how it guided the design of individual activities aimed to implement it. The first and the second principles are discussed respectively in Section \ref{Sec:3.2} and \ref{Sec:3.3}. Converting the specific practices listed in Fig. \ref{FIG:3} into authentic inquiry activities has been a particularly complex task that required the examination of different research perspectives and approaches (Section \ref{Sec:3.4.1}). In Sections \ref{Sec:3.4.3} - \ref{Sec:3.4.5} we show how they were blended in the design of various types of activities. The impact of the \textit{Epistemological Principle} on the design was conditioned by the chosen contexts and topics, and is addressed in Section \ref{Sec:3.5}.

\subsection{Structuring the learning path} \label{Sec:3.1}
The main source of inspiration and materials for this course has been an educational path for the introduction of QM in the context of polarization developed and evaluated by the physics education research group of the University of Udine \cite[e.g.,][]{Ghirardi1996, Michelini2004, Michelini2019}. The Udine's path is focused on the superposition principle and its consequences, starting from the distinction between properties and states. It makes use of hands-on activities with cheap experimental tools (polarizing filters, birefringent crystals), quantitative measurements with light intensity sensors, and of JQM \cite{Michelini2002}, an open-ended environment for computer-simulated experiments on photon polarization. However, the two curricula are different with respect to their design principles, various strategies, physical situations included and sequence. The sequence of activities of our course will be presented in Section \ref{Sec:4}, and displayed in full in Fig. \ref{FIG:9}. Here, we describe the the construction of the learning path, showing how the interplay of the first three principles defined its shape.

As a matter of fact, starting with polarization is compatible with the implementation of each of the principles. Since the phenomenon can be experienced by means of classical light beams and explained both in classical and quantum terms, it easily lends itself to a gradual building of a quantum model of the physical situation (\textit{Epistemic Principle}) and to the revision of classical concepts and mathematical constructs (\textit{Principle of Knowledge Revision}). In addition, two relations between properties (mutual unacquirability and incompatibility) naturally arise in photon polarization measurements. Along with compatibility, they represent the conceptual tools needed for extending the examination of measurement to distant physical situations (in our case, the hydrogen-like atom), promoting the construction of a unifying picture across contexts (\textit{Principle of Knowledge Organization}).

The introductory phases of our modelling cycle are described according to the template of the \emph{Model of Modelling} \cite{Gilbert2002, Gilbert2016}, an artifactual view that ascribes particular importance to the process of the creation and expression of the model. In these stages, the artifact is denoted as `proto-model', since it will be complete only after its expression by means of external modes of representation:
\begin{enumerate}[1.]
    \item \textit{Creation of the proto-model}:
        \begin{enumerate}[(a)]
        \item experiences for supporting its creation: (1) exploration of the  phenomenology of the linear polarization of light (interaction of macroscopic beams with polarizing filters/birefringent crystals); (2) empirical determination of its quantitative laws (Malus's law for beams polarized at $\theta$ incident on a filter with axis at $\phi$: $I_{out}=I_{in}\cos^2{(\theta-\phi)}$, reduction to half for unpolarized ones: $I_{out}=I_{in}/2$); (3) presentation of fundamental experiments on the detection \cite{Grangier1986, Grangier2005} and polarization of single photons after passing a filter;
        \item sources: the heuristic criterion according to which the hypotheses on the behavior of individual photons must be compatible (1) with the experimental evidence on the detection and polarization of a photon, and (2) with the classical phenomenology and laws for macroscopic light beams.
        \end{enumerate}
    \item \textit{Expression of the proto-model}: a fundamental mode of representation used in this course is the iconic language of JQM for the depiction of idealized physical situations and experiments involving the polarization of single photons. The representation includes photons - visualized by means of their polarization property (Fig. \ref{FIG:4}) - and devices such as single photon sources, polarizing filters, calcite crystals, screens and counters (Fig. \ref{FIG:5}). This language will represent an essential support for the implementation of theoretical epistemic practices such as thought experiments and the interpretation of classical laws of polarization in terms of photons. Mathematical modes of representation accompany these activities (e.g., Malus's law) and support the implementation of mathematical modelling practices (e.g., hypothesizing a mathematical representation of the quantum state and interpreting the meaning of its properties);
\begin{figure}[!ht]
    \centering  
       \fbox{\includegraphics[width=\linewidth]{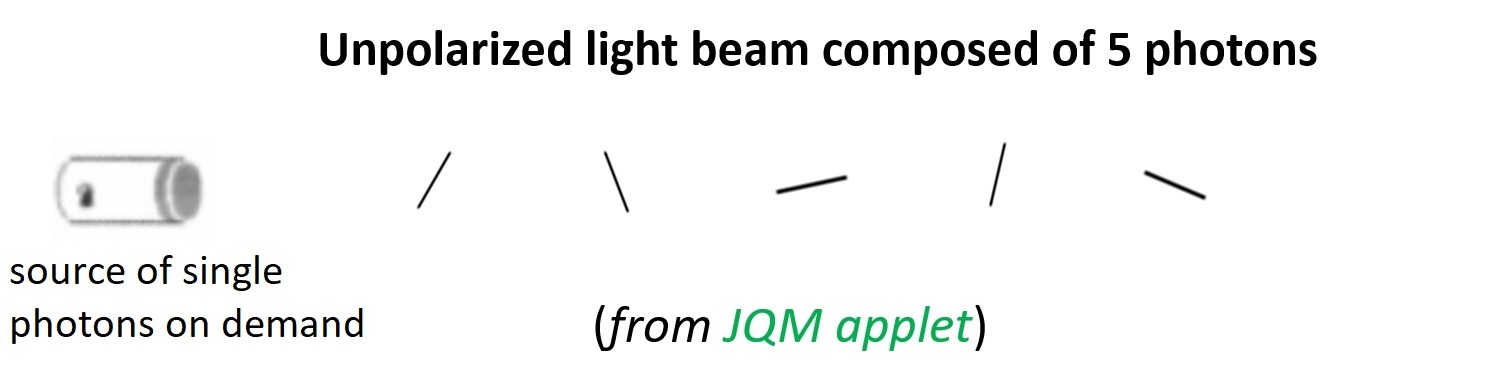}}
    \caption{The linear polarization of a photon is represented as a segment to highlight the fact that it is identified by a direction, and not a vector \protect\cite{Michelini2002}. Students are informed that segments are not to be intended as real physical representations of photons, but as a support for theoretical reasoning about photon polarization and related physical situations.}
    \label{FIG:4}
\end{figure}
\begin{figure}[!ht]
    \centering  
       \fbox{\includegraphics[width=\linewidth]{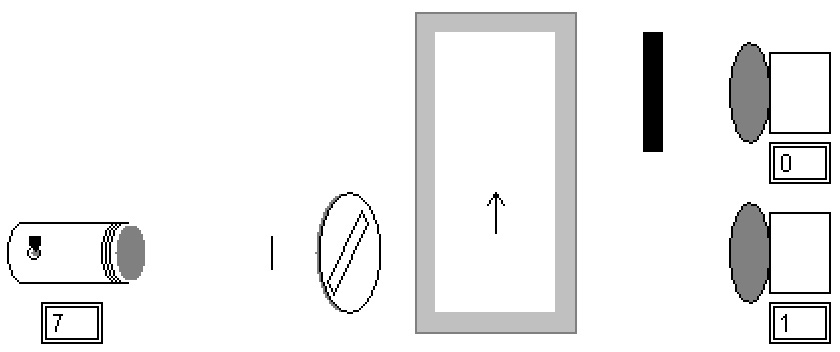}}
    \caption{Iconic representation of a single photon source with a predetermined polarization property (vertical, in this case), one vertically polarized photon, a polarizing filter with an arbitrary axis (here at $45^{\circ}$), a birefringent crystal with a $0^{\circ}$ and a $90^{\circ}$ channel, a screen placed on the extraordinary one, two photon counters.}
    \label{FIG:5}
\end{figure}
\end{enumerate}

After its creation and expression, the full-fledged model is ready to be developed and revised through a process conducted by means of theoretical epistemic activities (\textit{Epistemic Principle}), where students need to reinterpret, at a single-photon level, macro-phenomena and macro-laws which have already been explored by means of cheap experimental tools. It starts as the model of an object (the photon) for what concerns its detection and polarization. It soon grows to become a model of the interaction between photons and devices composed of filters/crystals followed by counters. The interaction with crystals and detectors is interpreted as an instance of the quantum measurement process, leading students to identify the relations between the initial property and those that correspond to possible outcomes of measurement. By means of this interpretive lens, the discussion can go beyond the scope of polarization: the relations are applied at a global level (incompatibility: position or velocity measurement on a system) and in the context of the hydrogen-like atom (compatibility: measurements of $E$, $L$, $L_z$, $S_z$). The relations between properties are then upgraded in terms of relations between observables. Next, the model can be embedded into the algebraic language of the polarization state vectors. Another inroad into the context of the hydrogen-like atom is made to introduce and discuss its state vector in terms of quantum numbers and calculate transition probabilities by means of vector superposition. The model is thus ready to undergo a major revision, incorporating also the propagation of photons - wave-like interference included - and their entanglement, therefore leading to the construction of a far-reaching model of radiation (the photon) and matter (the hydrogen-like atom).

A fundamental choice is addressing the quantum state, its vector and then quantum superposition only after the discussion of the concepts of measurement and observable. There are various reasons behind this choice. First, this sequence allows us to focus on the revision of a notion at a time (\textit{Principle of Knowledge Revision}). This would not be possible if we started directly with the superposition principle, that in QM is inextricably linked to all the aforementioned notions. The possibility to postpone the introduction of the state and superposition is granted by the \textit{Principle of Knowledge Organization}, which provides instruments for discussing quantum measurement and observables without resorting to the concept of state. Second, implementing the \textit{Epistemic Principle} involves structuring math modelling activities, e.g., related to the introduction of the state vector, that may cause a high cognitive load. In the context of polarization, building the mathematical representation of the state requires a consistent understanding of the single-photon interpretation of the Malus's law as probabilistic law of transition between different polarization properties: $p(\theta\mapsto\phi)=cos^2{(\theta-\phi)}$. Since this topic has been widely discussed in the unit on measurement, the state of polarization can be simply presented as a change of perspective on the same phenomena, without adding new physical content. This allows students to focus exclusively on the revision of the concept of state and on math modelling activities, thus reducing the cognitive load. One example is expressing the law of transition in terms of relations between state (ket) vectors\footnote{In the context of linear polarization, there is no need of complex numbers. Therefore, we do not introduce bra vectors and express the Born rule by using the square of a dot product.}: $(|\theta\rangle\cdot|\phi\rangle)^2=cos^2{(\theta-\phi)}$. Third, our course includes not only the context of photon polarization, but also of the hydrogen-like atom. An immediate examination of the concept and mathematical representation of the quantum state of the latter would be too challenging to our student population. Instead, the knowledge of measurement processes on this type of system together with the discussion of the polarization state vector represent a natural basis on which to build the state of a hydrogen-like atom and the corresponding (ket) vector in terms of quantum numbers\footnote{We restrict the mathematical discussion to superposition states with real coefficients: no need of bra and square moduli.}: $|n, l, m, s \rangle$.

The inclusion of the hydrogen-like atom offers various educational opportunities. In the discussion of the state, it allows us to break the one-to-one correspondence between properties and states that characterizes linear polarization (identifying the $|n, l, m, s \rangle$ state of a hydrogen-like atom requires the specification of four properties), as well as the identity of the angle between polarization properties and corresponding state vectors  (directions in the state space of the hydrogen-like atom are clearly unrelated to directions in the physical space). In the case of superposition, linear combinations of $|n, l, m, s \rangle$ vectors make it possible to generalize the discussion of measurement and observables to situations in which no known quantity is initially defined, and to address the normalization of the state vector after a measurement, that is trivial when the components of a superposition are limited to two terms, as in the context of polarization. Moreover, it allows us to introduce the concept of product state, since $|n, l, m, s \rangle$ is the composition of two states, $|n, l, m\rangle$ and $|s \rangle$, and the first expression is the contracted form of $|n, l, m\rangle|s\rangle$. In the course, we leave out any reference to the mathematical construct known as tensor product, but explain that the last expression is a way to denote a state (the global state of the atom) that depends on two component states (its spatial state and its spin state). Last, the context naturally lends itself to an interdisciplinary approach in collaboration with the chemistry teacher on topics such as orbitals and the atomic structure.

A solid understanding of the concept of state, of its vector, and of quantum superposition represent a strong basis for building a consistent interpretation of quantum interference and entanglement at a conceptual and mathematical level. Hence, the course ends with the discussion of propagation (``which-path'' experiments) and entanglement (first, of spatial and polarization modes of a photon, then of the polarization of different photons, and finally in the measurement problem).

To sum up, we identified the following path of learning and concept revision from CM to QM as potentially productive: linear polarization $\rightarrow$ measurement  $\rightarrow$ system quantity $\rightarrow$ state $\rightarrow$ vector $\rightarrow$ superposition $\rightarrow$ interference $\rightarrow$ general model (of a system) $\rightarrow$ correlation between internal components of the state (in QM, they can be entangled).

In the course, this learning path has been organized into four units: 1) Introduction to quantum measurement and observables, 2) The quantum state and its vector, 3) Quantum superposition, 4) Propagation and entanglement.

\subsection{Structuring activities according to the principle of knowledge revision} \label{Sec:3.2}

This section is devoted to the design of activities to support students in the revision of classical concepts and constructs according to the \emph{Principle of Knowledge Revision}. We describe two cases: the first concerning the ontological shift of a concept (measurement), the second the representational shift of a construct (vector superposition). Here we examine the path for the introduction of quantum measurement and of quantum superposition, with a special attention to end-of-unit table on superposition. Such tables are scheduled at the end of a unit and are designed to promote the discrimination between the classical and the quantum version of a notion with a birds's eye view on the revision process and address student's need of comparability with CM.

\subsubsection{Measurement} \label{Sec:3.2.1}
\begin{figure}[!htpb]
       \fbox{\includegraphics[width=\columnwidth]{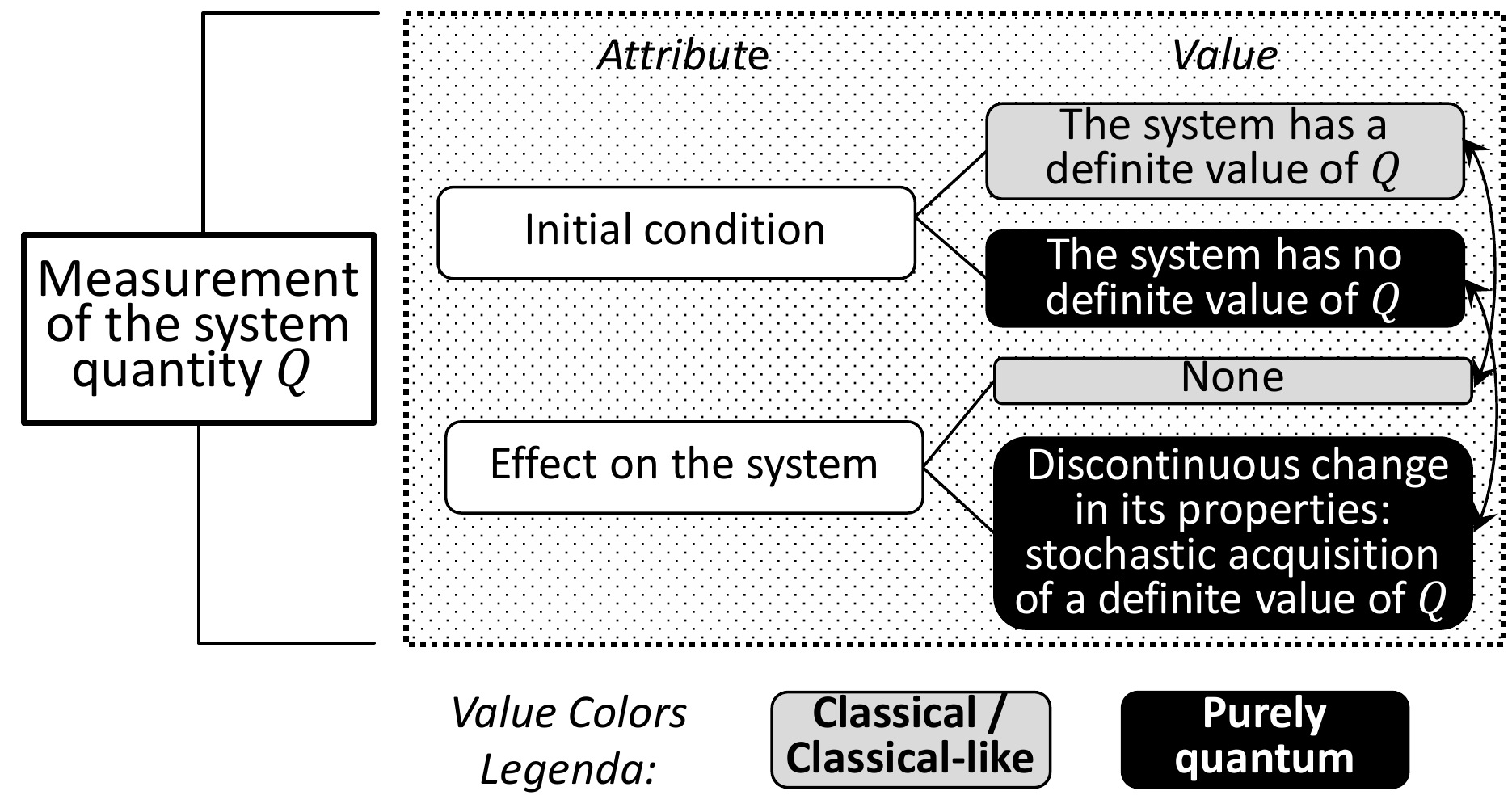}}
    \caption{Ideal measurement: concept trajectory from CM to QM \cite{Zuccarini2022}.}
    \label{FIG:12}
\end{figure}

In the transition to a quantum picture, the trajectory of the concept of \emph{ideal measurement} (see Fig. \ref{FIG:12})
and, as a consequence, its revision, are of crucial importance. In the context of polarization there is a specific challenge to take into account. While the linear polarization of macroscopic light beams can have any orientation in the plane of polarization and is identified by measuring its angle, the linear polarization of a photon can also have any orientation, but its measurement gives one of two angles that may differ from the initial one. Research found that students have difficulties in interpreting the quantum case as a two-state system \cite{Singh2015} and to frame the physical situation as a measurement: in our first design experiments, some of the students wondered how it was possible to describe it as such, because to them it was just ``a weird interaction altering the property''.

As can be deduced from Fig. \ref{FIG:12}, the trajectory of the concept of measurement in theory change is an instance of \emph{categorical generalization} (see Section \ref{Sec:2.1.1}). This type of dynamics suggests to start from the special case pertaining to both CM and QM, which is described by the gray boxes in the figure, i.e., the classical-like and determinate case. In general, it happens when the system possesses in advance a property of the measured quantity (in  the context of polarization: when the initial property of the photon coincides with one of the two possible outcomes of measurement). This situation is familiar to students, since it can be interpreted as an ideal classical measurement. Then, we move on to discuss its new feature, active and stochastic (when the photon does not possess in advance a property of the measured polarization quantity) in analogical terms, as a form of generalization of the first case. Finally, it is possible to focus on the conditional nature of measurement and the different physical situations corresponding to each case.

In the cycles of refinement of our course, we considered two possible devices for measuring photon polarization: 1) polarizing filter + detector; 2) birefringent crystal + detectors placed on the output channels. A prerequisite to discuss quantum measurement by means of these devices is framing their interaction with a photon in terms of information obtained on its polarization property as a result of the process. Some authors who introduce quantum measurement in the context of polarization use the analysis of vector superposition to support student reasoning \cite{Heyde2020, Michelini2019}: e.g., a photon prepared in $|45^{\circ}\rangle=\frac{1}{\sqrt{2}}|0^{\circ}\rangle+\frac{1}{\sqrt{2}}|90^{\circ}\rangle$ is absorbed by a filter with axis at $0^{\circ}$ as a result of a transition to $|90^{\circ}\rangle$, or equivalently, of the acquisition of the property at $90^{\circ}$. Since we explicitly renounce to start from the state and superposition, the discussion of polarization measurements must be preceded by an activity designed to support students in interpreting the absorption of a photon, either by a polarizing filter or by detectors placed after a crystal, as the result of a transition in polarization property or (in the classical-like case) of the retention of the initial property. At this stage, polarization observables have not been introduced yet. Therefore, we suggest students to use the expression ``outcome-property'', which denotes the properties associated with the outcomes of the polarization measurement at hand.

A possible outline of the revision of measurement, expressed as a sequence of goals, could be the following: supporting students in recognizing that
\begin{itemize}
    \item in the context of photon polarization, every possible outcome of the interaction of a photon with the device can be interpreted as the consequence of the retention/aquisition of a polarization property;
    \item the classical-like, passive and determinate interaction with the device can be interpreted as a measurement (familiar case, in which the system already possesses an outcome-property);
    \item the purely quantum, active and stochastic interaction with the device can be interpreted as a measurement (categorical generalization, corresponding to the new situation in which the system does not possess an outcome-property);
    \item the nature of measurement is conditional: passive and determinate when the system already possesses an outcome-property, active and stochastic when it does not.
\end{itemize}

\subsubsection{Superposition} \label{Sec:3.2.2}
\begin{figure}[!htpb]
       \fbox{\includegraphics[width=\columnwidth]{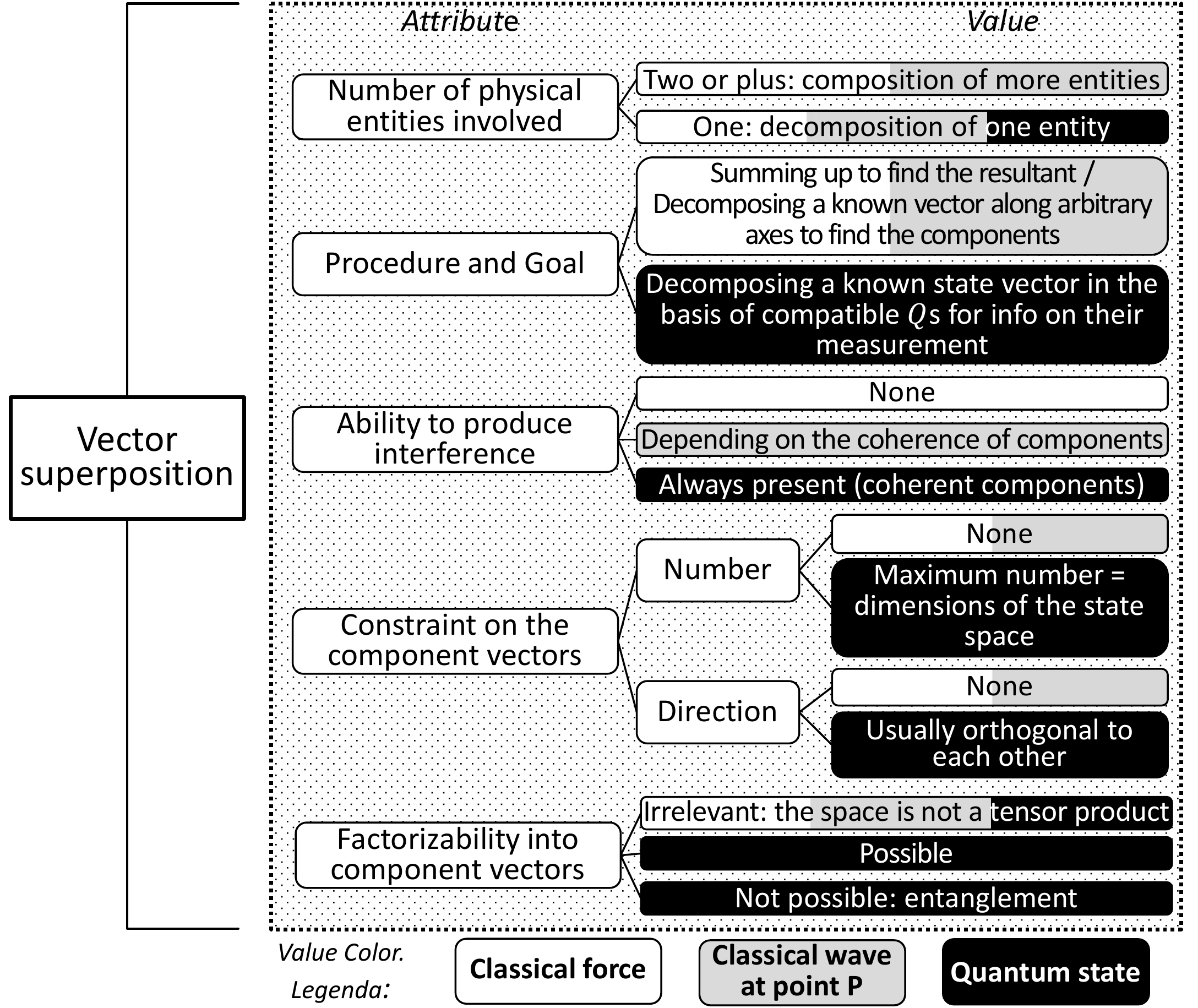}}
    \caption{Vector superposition: concept trajectory from CM to QM \cite{Zuccarini2022}.}
    \label{FIG:17}
\end{figure}

Quantum superposition is a fundamental topic in the learning path. Hence, the development of the course has been heavily influenced by the need to promote an effective revision of the representational properties of vector superposition. The design of these activities is based on a careful examination of the trajectory of this construct in theory change, which is displayed in Fig. \ref{FIG:17}.
The attributes included in the figure identify the basic representational features of vector superposition and the changes it undergoes in the transition to the new paradigm.

Interference and entanglement are not addressed in Unit 3 during the initial discussion of superposition, but in Unit 4 together with propagation. Therefore, the revision process related to these aspects (in the figure: \textit{Ability to produce interference} and \textit{Factorizability into component vectors}) was postponed to the final unit of the course.

In this section, we examine the remaining features, \textit{Procedure and goal}, \textit{Number of physical entities involved}, and \textit{Constraint on the component vectors}, where the pattern of \emph{value disjunction} (see Section \ref{Sec:2.1.1}) occurs in two out of three cases. This patterns suggests a different strategy to address the revision process: first building an understanding of the new features of the construct, and then contrasting them with those of its classical counterparts, in order to identify which of the familiar features lead to unproductive reasoning in a quantum context. Hence, prior intuition is used as a contrast at the end of the instructional sequence on the topic.

Since we are dealing with the revision of the representational features of a mathematical construct, which become evident in the transition from a physical situation to its mathematical representation (mathematization) and viceversa (interpretation), we need to focus on these processes in a quantum context. Mathematization has been already performed in the introduction of the state vector. Here we use interpretive tasks.

Procedure and the goal (decomposition of the state vector in a given basis to obtain info on the measurement of the corresponding observable) are introduced in worksheet items on polarization measurements of the standard observable (horizontal and vertical properties), and then of different polarization observables. In the latter, we go in depth on the issues of decomposition and basis change.

The last task can be used also to draw the attention to the physical referent and the goal of quantum superposition: one entity, the state vector, that can be decomposed in different bases depending on the information we need to derive. In later teaching experiments, additional support has been provided by means of embodied cognition, as described in Zuccarini and Malgieri \cite{Zuccarini2022}. In this activity, the perceptual experience of passive rotations (simulating the passage of the same state from superposition in the basis of one observable to that in the basis of another) was put in the service of promoting the awareness that quantum superposition involves only one physical entity.

The new constraints on the number of component vectors (maximum number equal to the dimensions of the state space) and on their directions (orthogonal to one another) are dealt with by means of interpretive tasks on the meaning of quantum superposition in measurement. In the context of polarization, we have two mutually unaquirable results, and therefore only superpositions of two orthogonal vectors. For helping students make this connection, we administer tasks on the examination of various mathematical expressions: superpositions of three polarization state vectors and of two non-orthogonal vectors, asking to determine whether the formulas have physical meaning.

The addition of the hydrogen-like atom provided a further context with more general and significant features. By discussing measurement on a superposition of its eigenstates in terms of quantum numbers (see Section \ref{Sec:3.1}), it is possible, e.g., to generalize the physical meaning of the constraint on the number of component vectors.

The end of the instructional sequence is represented by the end-of-unit table on superposition. Here we guide students to compare the features of familiar forms of vector superposition (of forces and waves) represented in Fig. \ref{FIG:17} with quantum superposition, using prior intuition as a contrast. As a matter of fact, the representational role of this mathematical process in QM is very different from that of the most commonly used forms of superposition in CM. In particular, a consistent interpretation of its referent (only one vector) and its goal (decompose it to obtain info on measurement)  is a prerequisite for addressing the quantum notion of interference, which is totally internal to the individual system. From this derives the seemingly contradictory statement of Dirac: ``Each photon then interferes only with itself'' \cite[p. 9]{Dirac1967}. In such respect, the structure of the frame of Fig. \ref{FIG:17} represents a useful template that can be converted into a comparison table, where students are required to identify whether a given statement (e.g.: a goal of superposition is to find the resultant) pertains to one or more forms of superposition (forces, waves, state vectors).

\subsection{Structuring activities according to the principle of knowledge organization} \label{Sec:3.3}
Implementing the principle entails the construction of activities supporting students in identifying the relations between properties and using them to manage the measurement process in multiple contexts. Here we describe the activation of the principle: first, the design of activities for the introduction of the relations in the context of polarization; then, for the use of the relations in the qualitative assessment of measurement at a global level (position and velocity) and in the context of the hydrogen-like atom ($E$, $L$, $L_z$, $S_z$, $x$).

\subsubsection{Introducing the relations between properties} \label{Sec:3.3.1}
As we said in Section \ref{Sec:2.1.2}, the relations of mutual unaquirability and incompatibility naturally arise in polarization measurements. After the revision of the concept of measurement both at a qualitative and quantitative level (according to the probabilistic interpretation of Malus's law), and after the introduction of polarization observables, which are identified by their two outcome-properties ($\phi$, $\phi+90^{\circ}$), students have all they need to determine whether there exists a measurement in which a photon loses its initial property $\theta$ and may stochastically acquire a property $\phi$ (incompatibility) or not (mutual unacquirability, when $\phi=\theta+90^{\circ}$).

\begin{figure}[!htpb]
       \includegraphics[width=\columnwidth]{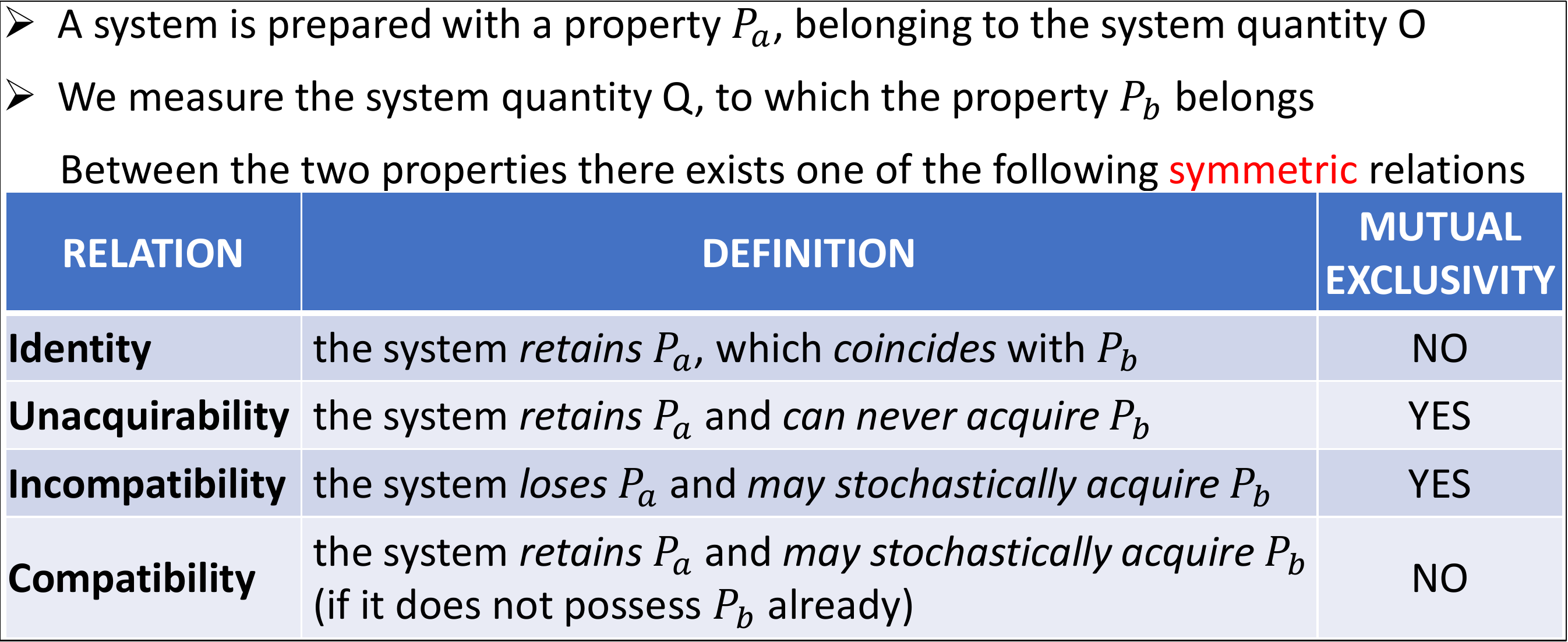}
    \caption{Definition of the relations between properties: summary table.}
    \label{FIG:22}
\end{figure}

Therefore, instead of lecturing on the definition of the relations, we can support students in acquiring the appropriate perspective for their introduction, by asking them to assess different statements on the retention, loss, and acquisition of properties in measurement. The template on which to build the task is provided by a table containing a concise definition of the possible relations between two properties, $P_a$ and $P_b$ (Fig. \ref{FIG:22}).
The statements are four because, to complete the picture, we consider a further relation: identity, embodying the fact that if the initial property ($P_a$) is also a possible outcome of measurement, the system retains this property. The statement may seem trivial at first, but corresponds to an important feature of quantum systems: when the system has a property of an observable $O$ at an instant (either as a result of preparation or acquired in a previous measurement), sufficiently rapid measurements of $O$ will provide the same property with certainty.

The core of the task is the following: given a photon prepared with an arbitrary polarization property $P_a$, we ask, for each of the four statements corresponding to the definition of different relations, whether the event occurs in measurement and, if it does, under which conditions. The item can be totally abstract or set in the usual context of calcite crystals and counters, as a support for student reasoning about measurement. In any case, this task corresponds to an increase in the level of abstraction of the activities, since the properties at hand are totally arbitrary. We assume that the work done on transition probabilities and on the revision of measurement represents an adequate basis for the analysis of the statements. After the activity, each statement can be identified by the instructor as the definition of a relation between properties.

Finally, we observe that this interpretive activity corresponds to a practice of the theoretical physicists: ``deepening the theoretical investigation of a phenomenology by adopting multiple perspectives'', and thus concurs to implement also the \emph{Epistemic Principle}. As a matter of fact, we are dealing with a change in perspective on the same phenomenon: from the revision of the concept of measurement to the analysis of what relation is established by measurement between two given properties. In accordance with the basic goals of the practice, such change of perspective significantly extends the scope of the student journey through the quantum realm.

\subsubsection{Extending the use of the relations between properties to other contexts} \label{Sec:3.3.2}
Since the only transition law students know at this stage is the probabilistic interpretation of Malus's law for polarization, and quantum superposition is not yet available, the discussion of measurement of different observables in other contexts has to start at a qualitative level. Here, the relations between properties can act as an organizing principle for managing the process.

Students may address quantum exemplars by relying on the revision of measurement and on the definition of the relations between observables. They include: 1) deducing what relation there can be between properties of the same observable and of different ones based on their behaviour in measurement; 2) deriving which observables will have a definite value after a measurement and the nature of the process (stochastic or determinate) based the knowledge of the relations between the properties of the measured observable and of the initially definite observables.

The cases we consider are
\begin{itemize}
    \item measurements of $x$ and $v$ at a global level: for discussing mutual unaquirability of the properties of the same quantity, and the measurable consequences of incompatibility between the properties of different observables;
    \item measurements of $E$, $L$, $L_z$, $S_z$, $x$ in the context of the hydrogen-like atom: the first four observables, for discussing compatibility between the properties of different observables and its measurable consequences; position, for discussing what happens when the properties of the measured observable are incompatible with those of some initially definite observables ($E$, $L$, $L_z$), and compatible with others ($S_z$).
\end{itemize}

As before, also this activity is an epistemic practice of the theoretical physicist: ``starting from results found in one context and extending or adapting them to other contexts.'' In order to support students in running this practice, the tasks can be presented in terms of a \emph{structured inquiry} (see Section \ref{Sec:3.4.1}). This means that the instructor defines the problem (extending the use of the relations) and the procedure (here: a sequence of inferential and interpretive questions), while students generate an explanation based on the theoretical knowledge at their disposal.

\subsection{Structuring activities according to the epistemic principle}  \label{Sec:3.4}

\subsubsection{Approaches and research perspectives for running theoretical epistemic practices} \label{Sec:3.4.1}

\emph{Inquiry-based learning and Modelling-based teaching}. The first expression denotes a set of active-learning approaches devised to support students in the design and conduction of investigations that mirror the processes
used in science for building its body of knowledge. In this work, we describe the forms of inquiry on which we drew to develop epistemic activities by referring to Llewellyn \cite{Llewellyn2012}. They differ in the amount of information and direct instruction provided to students during the activity:
\begin{itemize}
    \item Demonstrated inquiry: the instructor poses the research questions or generates the hypotheses to test, plans the inquiry procedure and communicates the results. A demonstrated inquiry differs from conventional demonstrations in the way the instructor poses questions to the students during the presentation, soliciting input in the design of the experiment;
    \item Structured inquiry: questions/hypotheses and procedure are still provided by the instructor; however, students generate an explanation supported by the evidence they have, possibly drawing implications;
    \item Guided inquiry: in this case, the instructor provides the questions/hypotheses, but the student is responsible for designing and conducting the investigation, as well as for drawing conclusions.
    \item Self-directed inquiry: the instructor provides the initial orientation, but students raise their own questions/hypotheses, design their own procedures, and organize and analyze their own results.
\end{itemize}
In all cases, the instructor offers content and process support during all phases of the inquiry, providing extensive scaffolding. Since in our course, the goal of inquiry is not to ‘test variables’ but to develop and refine a scientific explanation in the form of a theoretical model, it will not be conducted by means of controlled experiments, but of hypothetico-deductive reasoning (a mental process in which students progress from a hypothesis to specific conclusions by means of logical inferences), thought experiments, etc. The requirements for running a specific inquiry cycle (experiences, sources) will be identified - when needed - by using the template of the model of modelling, which has been presented in Section \ref{Sec:3.1}.

\begin{figure}[!htpb]
       \fbox{\includegraphics[width=\columnwidth]{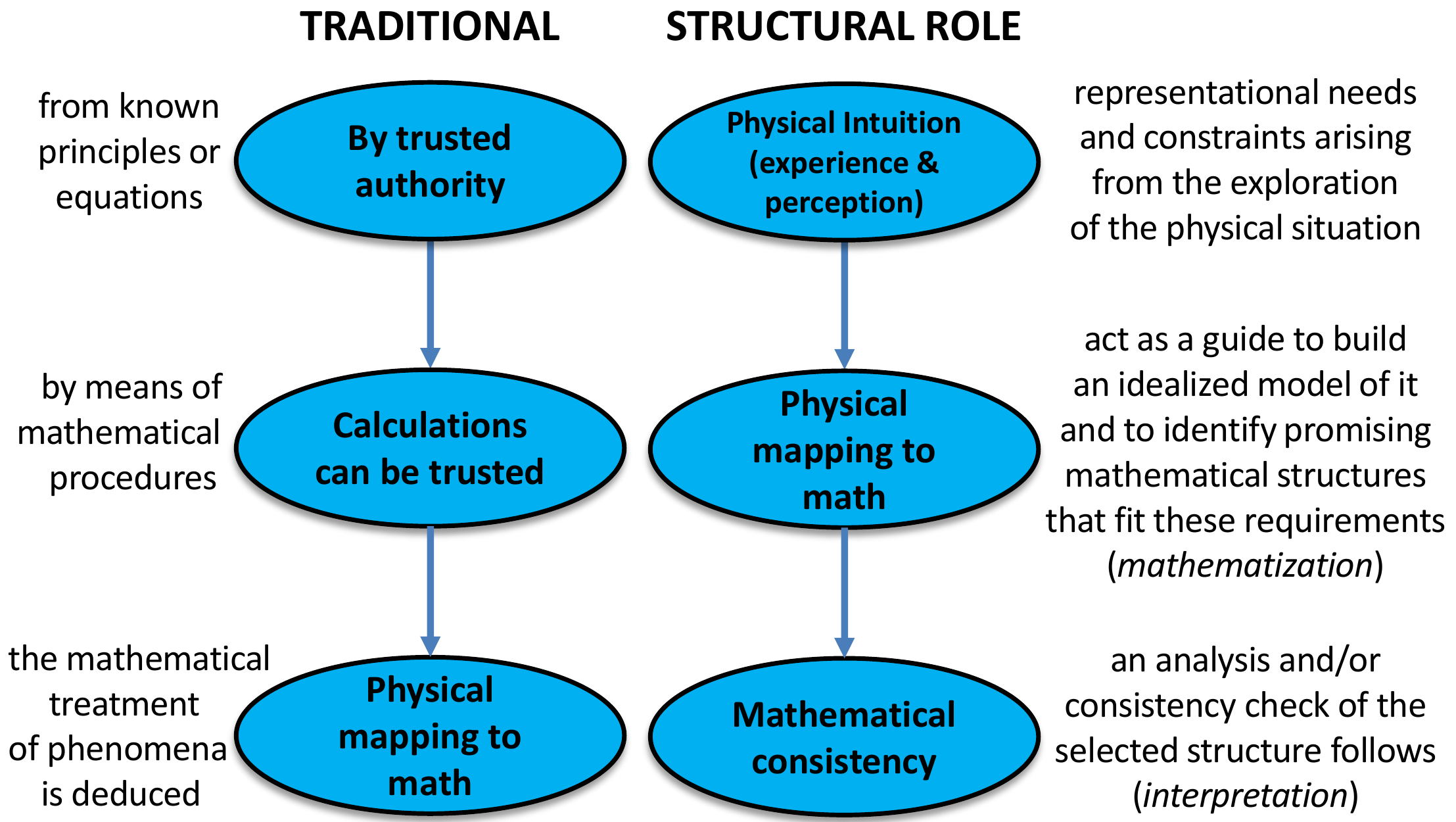}}
    \caption{Contrasting a traditional chain of activation and one highlighting the structural role of mathematics in physics.}
    \label{FIG:6}
\end{figure}

\emph{Mathematical modelling strategies}. In order to provide insight on the ways in which mathematics can be put in the service of physical modelling, we drew on theoretical studies on the role and the language of mathematics in physics. Uhden \textit{et al.} identified two fundamental aspects to consider \cite{Uhden2012}: the deeply tangled unity of mathematical and physical models, and the distinction between technical and structural role of mathematics in physics. The latter is the role of math in structuring physical concepts and situations that is made explicit in the processes of mathematization and interpretation. Another perspective is provided by Redish and Kuo, who analyze the language of mathematics in physics by means of cognitive linguistics in a resources framework \cite{Redish2015}, and suggest to initially focus on physical intuition and embodied experience rather than equations and principles. A common theme of both studies is the line of development of the discourse: from concrete to abstract (from physical issues to mathematics) followed by a new interpretive activity, aimed at clarifying further physical implications of the newly introduced structure (Fig. \ref{FIG:6}).
Variations on this theme are used in the design of inquiry activities, highlighting the structural role of math in the modelling of physical concepts and situations. See, for instance, the design of activities on the revision of superposition that has been outlined in Section \ref{Sec:3.2.2}. Since the construction of an idealized representation of the physical situations discussed in our course has been performed in the creation and initial expression of the proto-model, we can skip this step in the structural chain of activation described in Fig. \ref{FIG:6}.

\emph{Operationalizing thought experiments}. Thought experiments may play a significant role in the presentation of modern physics, opening ``a unique window to the strange and unknown world of super-large and super-small scales''\cite{Galili2007}, where real experiments are excluded from regular classroom activity. This practice is employed for a variety of purposes \cite{Brown1991}: facilitating a conclusion drawn from available experiences and sources, rejecting a given explanation (Galileo's leaning tower of Pisa experiment), illustrating some of the counter-intuitive or unsatisfying aspects of a theory (Maxwell's demon, Schr\"{o}dinger's cat), finding new constraints that help guide positive modifications of a theory (Einstein's elevator).

A definition of thought experiment that is potentially productive in education has been proposed by Stephens and Clement \cite{Stephens2012}, who emphasize the process rather than the product: performing an untested thought experiment [..] ``is the act of considering an untested, concrete system (the `experiment' or case) and attempting to predict aspects of its behavior. Those aspects of behavior must be new and untested in the sense that the subject has not observed them before nor been informed about them.'' This emphasis on the relationship between the agent and the process allows us to widen the scope of thought experiments in educational practice: students making a prediction for an unfamiliar analogy, running a model for the first time, or applying a model to an unfamiliar transfer problem, are performing an untested thought experiment.

As regards creating and running a thought experiment, Gilbert and Reiner \cite{Gilbert2000} propose an analytical schema that consists of six steps, intended as component elements, which could represent a conceptual basis for the design of educational activities:
\begin{enumerate}
    \item posing a question or a hypothesis;
    \item creating an imaginary world, consisting of entities (objects, or mental creations which can be treated as objects) relating to each other in a regulated manner;
    \item designing the thought experiment;
    \item performing the thought experiment mentally;
    \item producing an outcome of the thought experiment with the use of the laws of logic;
    \item drawing a conclusion.
\end{enumerate}
While thought experiments may play a variety of roles, we focus on their use as a form of testing procedure. We qualify these procedures as \emph{humble thought experiments}, because they are not meant to achieve the purposes of historically significant thought experiments (e.g., Einstein's elevator); yet, their structure corresponds to that described by Gilbert and Reiner \cite{Gilbert2000} for a thought experiment, and their conduction may be within the reach of secondary school students. In order to design and analyze tasks in which students are actively engaged in running a thought experiment, we refer to the ISLE learning framework \cite{Etkina2015}, in which the phases of testing experiments are clearly identified and associated with scientific abilities. These abilities are described in rubrics introduced in Etkina et al. \cite{Etkina2006}, which are to be used as self-assessment instruments. For scientific abilities related to testing experiments, see Etkina \cite[Appendix B]{Etkina2015}. While the rubrics have been developed as a support in the design and running of experiments in the lab, they may be easily adapted to activities in which the instructor
\begin{itemize}
    \item provides the issue to explore, encouraging students to generate different hypotheses and to test them by running - step by step - a thought experiment specifically designed by the instructor (Section \ref{Sec:3.4.3});
    \item provides a hypothesis and asks students to design a thought experiment to test it, to run the thought experiment, and to draw appropriate conclusions on the initial hypothesis (Section \ref{Sec:3.4.4}).
\end{itemize}


\subsubsection{Thought experiment: Description of an unpolarized beam in terms of photons} \label{Sec:3.4.3}

This activity marks the start of student work in modelling and of the use of worksheets. At this point of the course, students have explored the polarization of macroscopic light beams by means of cheap experimental materials and discussed evidence on the detection of a single photon and on its polarization property after a filter, visualized as a segment in JQM screeshots.

The task is extending the model, describing unpolarized beams in terms of photons. Since this is the beginning of the theoretical modelling cycle, its purposes are manyfold: a) inviting students to engage in theoretical modelling tasks; b) discussing the basic heuristic principle that drives the modelling process: our hypotheses on the behavior of individual photons must be compatible with the classical quantitative laws for macroscopic beams; c) minimizing the axiomatic basis of the course; d) putting prior intuition in the service of learning QM.

\begin{figure*}[!htpb]
       \fbox{\includegraphics[width=\textwidth]{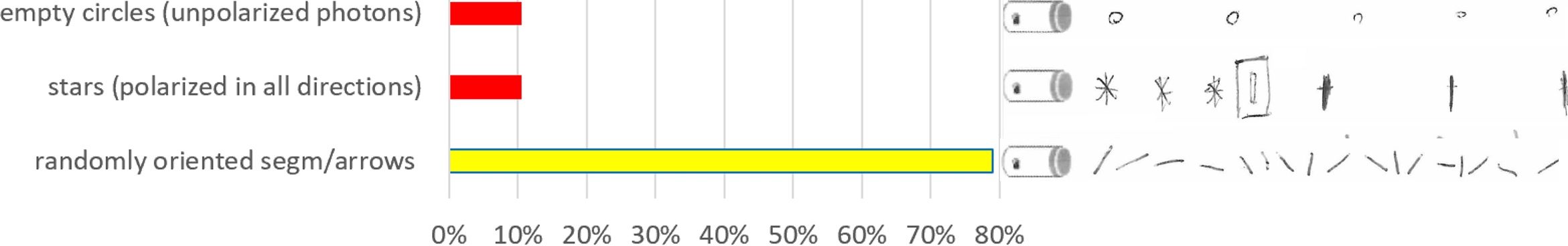}}
    \caption{Spontaneous models of unpolarized light in terms of photons: Summer School of Excellence, Udine, 2015 (41 students).}
    \label{FIG:39}
\end{figure*}

The last purpose leads us to the genesis of the activity. In the first teaching experiments, we investigated spontaneous models of photon polarization, identifying a strong tendency to interpret unpolarized light as made of
\begin{itemize}
    \item photons polarized in different directions (sketching differently oriented segments: correct model);
    \item unpolarized photons (empty balls);
    \item photons polarized at all angles (stars).
\end{itemize}
See Fig. \ref{FIG:39} for student sketches.

As a result, we saw the possibility of settling the matter by means of a modelling activity in which all three hypotheses are elicited in a whole-class discussion, and the unproductive ones are ruled out by asking a prediction on the number of photons transmitted by a filter (possible reduction according to the first hypothesis, no reduction for the other ones) and comparing it with the classical law for unpolarized light (reduction to half). This kind of activity can be described as a testing experiment of theoretical nature or, in short, as a thought experiment.

In order to identify the correct hypothesis and discard the others, students need the following experiences: 1) Classical law for unpolarized light; 2) Classical law for polarized light: the intensity of a beam of polarized light passing through a filter with a different axis is reduced; 3)  polarization is a property of the single photon; 4) photons of a polarized beam are all polarized in the same direction;  5) light intensity grows with the number of photons emitted at a given instant. The source is the compatibility of the hypothesis on the single photons with the classical laws, that apply to beams of many photons. All is available to students except 5), which is intuitive and shortly discussed by the instructor, and the source, that is made explicit in the last task asked by the activity.

The relevant scientific abilities are based on a subset of those described by Etkina. They are tested for each hypothesis and can be used in the analysis of student answers as target performance of the main phases:
\begin{enumerate}[(a)]
    \item able to make an assumption based on the hypothesis about the action of the experimental device (filter with vertical axis) on the photon beam;
    \item able to make a reasonable prediction (on the number of transmitted photons) based on the assumption;
    \item able to decide whether the prediction is compatible with the outcome prescribed by the macroscopic laws (reduction to half).
\end{enumerate}

\begin{figure}[!hb]
    \centering  
       \fbox{\includegraphics[width=\linewidth]{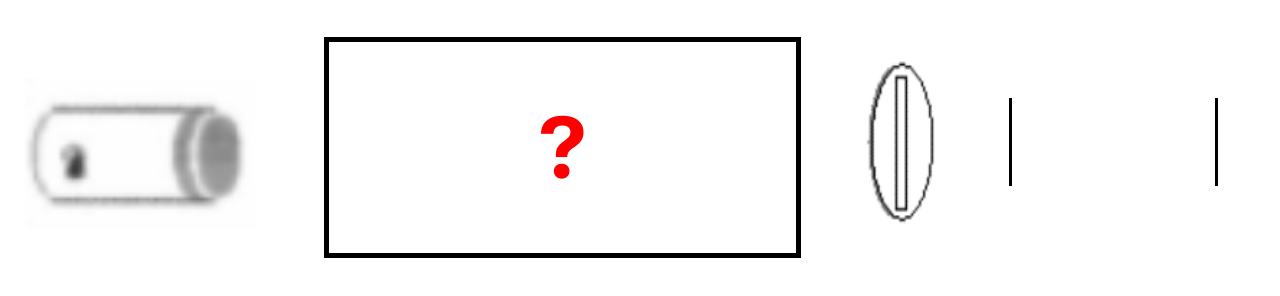}}
    \caption{An unpolarized beam is incident on a filter with vertical axis. How can it be described in terms of photons?}
    \label{FIG:48}
\end{figure}

The proposed task is a form of inquiry with elements of the \emph{structured inquiry} and of the \emph{self-directed inquiry} \cite{Llewellyn2012}. The phases of the activity are reported in Table \ref{table:1}, its iconic representation in Fig. \ref{FIG:48}.

\begin{table}[!ht]
\caption{Description of an unpolarized
beam in terms of photons: Phases of the task} 
\centering 
\begin{tabularx}{\columnwidth}{cX}
\hline
1 &  the instructor provides the issue to explore: how to describe an unpolarized beam of four photons \\
\hline
2 &  students generate different hypotheses in a whole class discussion\newline (1. randomly oriented photons; 2. photons polarized in all directions at the same time; 3. unpolarized photons)\\
\hline 
3 & based on each hypothesis, students are asked to make an assumption on the action of a polarizing filter with vertical axis on the photon beam\newline (1. absorbing some photons \& changing the polarization of the transmitted ones; 2. removing all polarization properties except $90^{\circ}$; 3. adding a polarization property at $90^{\circ}$) \\
\hline
4 & for each assumption, students are asked to make a prediction on the transmission process\newline (1. a possible reduction in the number of photons; 2. \& 3. transmission with certainty) \\
\hline
5 & students draw an appropriate conclusion on each hypothesis by comparing the corresponding prediction with the prescription of the macroscopic law, i.e., reduction to half\newline (since for hypotheses 2. \& 3.
 no photon is absorbed, the only hypothesis left is 1.) \\
\end{tabularx}
\label{table:1} 
\end{table}


\subsubsection{Interpreting already known laws within the framework of new models: Malus's law} \label{Sec:3.4.2}

This activity is scheduled immediately after the thought experiment. The experiences and sources needed for reinterpreting Malus's law in terms of photons as a stochastic law of transition are all available. A support for this task is represented by the thought experiment itself, which draws student attention to the absorption of individual photons in the interaction with a filter.

\begin{figure}[!hb]
    \centering  
       \includegraphics[width=\linewidth]{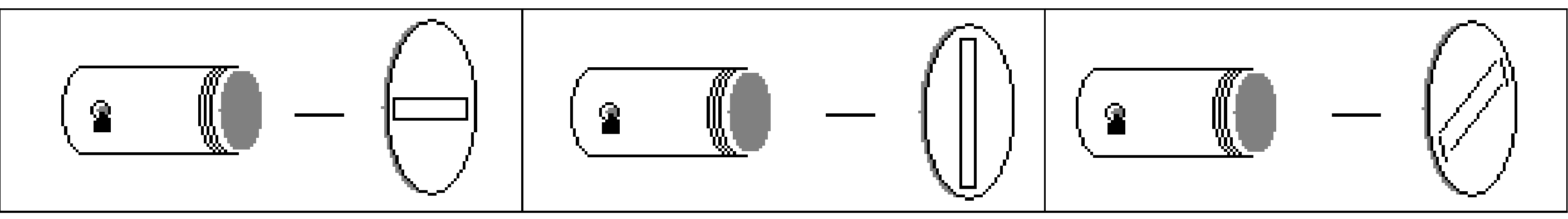}
    \caption{Students are asked to make qualitative and then quantitative predictions on the future prospects of a photon meeting a filter.}
    \label{FIG:49}
\end{figure}

Therefore, in order to help students develop a consistent interpretation of Malus's law, we found it sufficient to design a \emph{structured inquiry} \cite{Llewellyn2012}. The phases of the activity are reported in Table \ref{table:2}, its iconic representation in Fig. \ref{FIG:49}.

\begin{table}[!hb]
\caption{Interpreting Malus's law in terms of photons: Phases of the task} 
\centering 
\begin{tabularx}{\columnwidth}{cX}
\hline
1 &  students are engaged in an analysis of possible interactions between a photon prepared with polarization at $0^{\circ}$ and a filter, in the cases displayed in Fig. \ref{FIG:49}, to highlight the dichotomy between the determinate case (axis at $0^{\circ}$ or $90^{\circ}$) and the uncertain one (axis at $45^{\circ}$). \newline In each situation, students are asked to predict whether a photon is transmitted or not, knowing that by definition it is not possible to transmit half photon. After that, the instructor shows the possible results by means of JQM screenshots.\\
\hline
2 & in the light of Malus's law (and implicitly of the heuristic principle), students are asked to make a quantitative prediction on the future prospects of a photon meeting a filter (denoted as ``Horoscope of the photon''), taking into account the polarization property of the photon and the axis of the filter.\\
\hline
3 & students are asked to extend their predictions to arbitrary angles of polarization and axes of the filter.\\
\hline
\end{tabularx}
\label{table:2} 
\end{table}


\subsubsection{Thought experiment: Superposition as statistical mixture of component states?} \label{Sec:3.4.4}
While the thought experiment described in Section \ref{Sec:3.4.3} plays a platonic role \cite{Brown1991}, both destructive and constructive (ruling out some hypotheses and identifying the one that is compatible with available evidence), this thought experiment plays only a destructive role: excluding the possibility that quantum superposition can be interpreted as a statistical mixture.

The activity has two purposes: 1) helping students distinguish between superposition states and mixed states; 2) launching an epistemological discussion on quantum uncertainty and the completeness of QM (the Einstein-Bohr debate). The first one corresponds to addressing a persistent issue in the learning of the theory (see Section \ref{Sec:Intro}). For the second purpose, we need to draw students' attention to the consequences of interpreting a superposition as a mixture of the component states: it entails that a measurement on a single system is deterministic, that the observable to be measured is definite, and that the use of probability is due to lack of knowledge about the state of the system. By analyzing and rejecting the claim, we pave the way for introducing one of the main problems with the standard interpretation of QM: how is it possible that identical systems interacting with the same measurement device in the same conditions may give different and unpredictable results?

For discussing the topic, we propose to students a \emph{guided inquiry} \cite{Llewellyn2012}. The hypothesis is provided by the instructor, pretending it has been advanced by students of the previous years: ``the expression $|30^{\circ}\rangle = \sqrt{3}/2|0^{\circ}\rangle + 1/2|90^{\circ}\rangle$ means that the state $|30^{\circ}\rangle$ is composed of a random mixture of photons in the states $|0^{\circ}\rangle$ and $|90^{\circ}\rangle$, in a proportion corresponding to the square of the coefficients.'' In order to make the most of it for our second purpose, we guide students to unfold the physical consequences of the hypothesis with a series of questions. Then, we ask them to design a thought experiment to test the hypothesis, to run the thought experiment, and to draw the conclusions on the hypothesis.

The prerequisites for running the activity are the following experiences: 1) awareness of the probabilistic meaning of superposition in measurement, which is the first topic addressed in the introduction to superposition, 2) awareness that if we measure the observable to which the initial property of the system belongs ($30^{\circ}$, in this case), we find this property with certainty, which has been addressed in the revision of measurement. Source of the mixture hypothesis is the alternative interpretation of the referent of superposition: not an individual entity that is decomposed, but a composition of different entities. The strength of this hypothesis is that it has the same goal of the consistent interpretation (finding information on measurement), and that both interpretations link component states to the possible results and the square of their coefficients to the corresponding probabilities.

The thought experiment is elementary: all we need is to direct the beam to a filter with axis at $30^{\circ}$. The prediction associated with the mixture hypothesis is that some of the photons will be absorbed (on average: 3/8). However, we know that at a macroscopic level, all the light polarized at $30^{\circ}$ will be transmitted by the filter. The hypothesis is false.

The scientific abilities that are associated with the conduction of the thought experiment are again a subset of those described by Etkina, and are expressed as in her article \cite[Appendix B]{Etkina2015}:
\begin{enumerate}[a.]
\item able to design a reliable experiment that tests the hypothesis;
\item able to make a reasonable prediction based on the hypothesis;
\item able to decide whether the prediction and the outcome disagree.
\end{enumerate}

\subsubsection{Identifying and interpreting mathematical constructs for describing physical situations and deriving new results: State vector and Entangled superposition} \label{Sec:3.4.5}

\emph{State vector}. The refinement of the first mathematical modelling activity has been briefly described in a previous work \cite{Pospiech2021}. The main issue concerned the common difficulty to discriminate between quantum states and measurable properties in the context of polarization. While it is tempting to introduce the polarization state vector by using its correspondence with polarization properties, i.e. directions on the polarization plane (for instance, $0^{\circ}$ $\rightarrow$ $|0^{\circ}\rangle$, $90^{\circ}$ $\rightarrow$ $|90^{\circ}\rangle$, $|0^{\circ}\rangle\cdot|90^{\circ}\rangle=0$), this approach suggests students that state vectors and polarization properties are different representations of the same thing. Therefore, we tried an alternative strategy: introducing the state vector by fostering an algebraic form of reasoning. After reminding students that the quantum state describes the behavior of the system in measurement, we asked which mathematical objects are suitable for describing the transition rule from one state to another by means of simple operations (sums/products) between the entities representing each state. Among algebraic entities, students chose vectors, since the transition rule depends on directions. However, even after these activities, we found that only half of the students consistently discriminated the two notions. The matter was settled only when we asked about the nature of the state vector after introducing the vector of the hydrogen-like atom, which is unequivocal in this respect.

\begin{figure}[!hb]
       \fbox{\includegraphics[width=.98\linewidth]{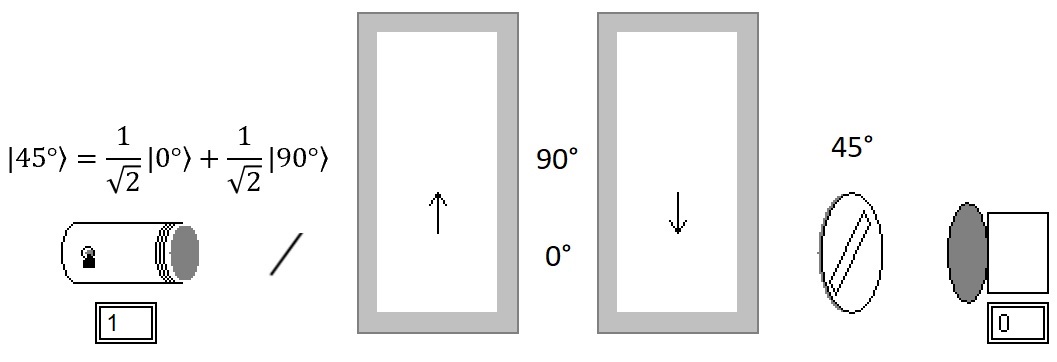}}
    \caption{Iconic representation of a ``which-way" experiment.}
    \label{FIG:8}
\end{figure}

\emph{Entangled superposition of modes}. This task is addressed after the work on propagation, where students are led to conclude that the position of a photon between a direct and reversed calcite crystal (see Fig. \ref{FIG:8})
is indefinite, to identify a new form of interference, and to build a full quantum model of a system for measurement and propagation. They are also expected to know the concept of component states ($|n, l, m \rangle$ and $|s\rangle$) and of product state, which coincides with the global state of the atom ($|n, l, m \rangle |s\rangle$).

\begin{figure} [!h]
\centering
\begin{tabular}{|r|l|} \hline
    \includegraphics[width=.48\linewidth]{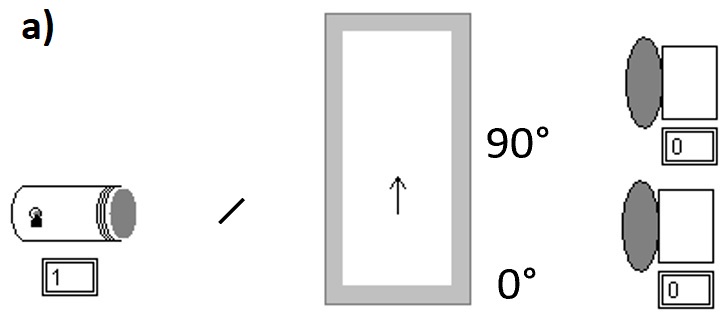} &
    \includegraphics[width=.48\linewidth]{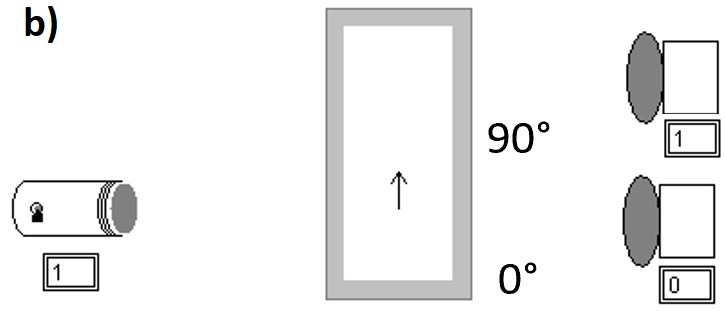} \\ \hline
\end{tabular}
    \caption{(a) Preparation; (b) Measurement.}
    \label{FIG:7}
\end{figure}

In the context of the course, the ideal physical situation for a smooth and compact discussion of entanglement is the usual case of a photon incident on a calcite crystal followed by two detectors, which has been used since Unit 1 to introduce quantum measurement. The only difference is that here we do not focus on preparation or measurement (see Fig. \ref{FIG:7}),
but on the properties of the particle beyond the crystal, just before the absorption. It is worth noting the richness of this simple context: if we choose to discuss position, we address the wave-particle duality, if we discuss the connection between position and polarization, we are led to the entanglement of modes.

The learning path we propose to students is a \emph{structured inquiry} \cite{Llewellyn2012}, in which we guide them to identify a new mathematical construct and its representational role. For this purpose, we need first to specify the relevant experiences - concerning the description of photon polarization after the crystal and of its measurement - and sources of the model - concerning the basic ingredients of the mathematical representation. Then, these elements are embedded into a sequence based on the template of the structural chain of activation (see Fig. \ref{FIG:6}): exploration of empirical constraints and mathematical prerequisites, mathematization, interpretation. Finally, we describe the design of the individual activities used to build the components of the sequence.

Experiences and sources:
\begin{itemize}
\item Experience 1: the knowledge of the fact that polarization is indefinite after the crystal;
\item Experience 2: the knowledge of the fact that by measuring position you also measure polarization and vice versa;
\item Source 1: the mathematical representation of the spatial state of the photon in an elementary form;
\item Source 2: the product of spatial and polarization states;
\item Source 3: the physical interpretation of the component vectors and of the coefficients of a superposition state.
\end{itemize}
The two experiences are not yet available to students. Source 3 is already known from the study of superposition, sources 1 and 2 consist in applying to the photon what we already know about the global state of the hydrogen-like atom, replacing spin with polarization. The modes of representation are the iconic language of JQM (available to students after the discussion of measurement in Unit 1) and ket representation of product vectors (available after the discussion of the state of the hydrogen-like atom in Unit 2).

Now we are ready to propose a structural chain of activation:
\begin{enumerate}
    \item exploration of the physical situation, highlighting those aspects that are relevant to the issue at hand (experience 1 and 2);
    \item introduction of the mathematical ingredients needed to derive the new construct (sources 1 and 2);
    \item mathematization: task for supporting the identification of the construct (source 3);
    \item interpretation: task for analyzing the new construct (rediscovering and deepening the content of the qualitative experiences).
\end{enumerate}

The activities designed to build the components of the sequence are the following:

Experience 1. Just before starting the part on entanglement, we use the definition of the possession of a property (a system possesses a property if and only if the probability to measure it is 1) to guide students to determine that, after the crystal, not only the position of a photon prepared in the superposition state $|\psi\rangle=1/\sqrt{2}|0^{\circ}\rangle+1/\sqrt{2}|90^{\circ}\rangle$  is indefinite, but also its polarization. As a matter of fact, there is 1/2 probability that the photon is collected by the detector in the $0^{\circ}$ channel or by that in the $90^{\circ}$ channel.

Experience 2. We direct student attention to a fact known from the revision of measurement, but that had not been emphasized until now: a measurement of position after the crystal is a polarization measurement and vice versa. In particular, each position property is correlated to a specific polarization property and vice versa.

Sources 1 and 2. After the second experience, position has clearly come into play. Therefore, we propose students to analyze the global state of the photon, using as a reference the description of the hydrogen-like atom. For this purpose, we introduce the spatial state of the photon in terms of three position (eigen)states:
\begin{itemize}
    \item localized immediately after the source: $|x\rangle$
    \item localized at the entrance of the detector on the ordinary channel at $0^{\circ}$: $|x_1\rangle$
    \item localized at the entrance of the detector on the extraordinary channel at $90^{\circ}$: $|x_2\rangle$
\end{itemize}

Mathematization.\\
\emph{Special cases (determinate)}: we ask students about the global state of the photon at the time of its absorption by a detector, if it is prepared in $|x\rangle|0^{\circ}\rangle$ (answer: $|x_1\rangle|0^{\circ}\rangle$), and if it is prepared in $|x\rangle|90^{\circ}\rangle$ (answer: $|x_2\rangle|90^{\circ}\rangle$).\\
\emph{General case (stochastic)}: we ask students how to write the global state, an instant before the absorption, of a photon prepared in an arbitrary state of polarization, i.e., in the global state $|x\rangle(a|0^{\circ}\rangle + b|90^{\circ}\rangle)$. Based on the results of the previous questions and on the quantum interpretation of component vectors and coefficients (Source 3), a candidate can be proposed: $a|x_1\rangle|0^{\circ}\rangle + b|x_2\rangle|90^{\circ}\rangle$.

Interpretation. We ask students about properties of this state. Specifically, those corresponding to the results of Experience 1 and 2.

We conclude the section with two comments. First, in QM it is not possible to univocally reconstruct a superposition state by measuring only one observable, as we do not get information on the phases \cite{Michelini2014}. Given that we focus exclusively on the modeling of entangled superposition, this issue was considered negligible. Second, these activities allow us to immediately apply the conceptual and mathematical discussion of the entanglement of modes to a new physical situation: the purely quantum entanglement of different systems. A new mathematical modelling activity can be implemented by describing the physical situation of two photons emitted by parametric down-conversion. Information provided to students is the following: the possible results of polarization measurements on one of the photons, the effect of this measurement on the other photon, and the transition probability. Based on these elements, students can pass from an expression like $a|x_1\rangle|0^{\circ}\rangle \pm b|x_2\rangle|90^{\circ}\rangle$ to a structurally identical formula such as $a|0^{\circ}_1\rangle|90^{\circ}_2\rangle \pm b|90^{\circ}_1\rangle|0^{\circ}_2\rangle$.

\subsection{Implementing the epistemological principle} \label{Sec:3.5}

\subsubsection{Impact of the interpretive choices on the coherence of the design}  \label{Sec:3.5.1}
Here we describe how the interpretive choices are used to strengthen the internal coherence of the course. In what follows, first we recall the individual choice, then we explain how it affects the design.

\begin{itemize}
    \item a pure state provides complete information on the behavior of an individual quantum system (ruling out statistical interpretations);
\end{itemize}

In the course, we adopt a single system ontology. Therefore, we always refer to individual systems, favoring a probabilistic language over a statistical one. Ensembles of systems, identically prepared or not, are treated on a probabilistic basis, making use of the law of large numbers when appropriate. The implementation of this language choice played a productive role in the running of epistemic practices such as the interpretation of Malus's law in terms of photons, where a gradual transition from a discussion of quantum objects in terms oscillating between ensembles and individual systems, to a language carefully centered on the latter, was accompanied by an increasing rate of success (see Section \ref{Sec:5.4.1}).

\begin{itemize}
    \item an observable of a system has a determinate value if and only if the quantum state of the system is an eigenstate of the operator representing the observable (ruling out modal interpretations);
\end{itemize}

Since we do not use operators in the course, we do not introduce the terms ``eigenstate'' and ``eigenvalue''. However, the definition of the possession of a property by a system stands for the eigenstate-eigenvalue link: a system possesses a property if and only if the probability to measure it is $1$. The language of properties also helps suggest students a coherent interpretation of quantum superposition. As a matter of fact, while the superposition of linear polarization states is usually interpreted as a ``neither, nor'' situation (the system is in neither of the component states, and has neither of the corresponding properties), a superposition of two position eigenstates is sometimes interpreted as the system being ``in both places.'' However, the link between possessing a property and measuring it with certainty allows us to reconcile this case with the general frame: the system has neither of the component position properties. Its position is indefinite. For a productive use of this language in the development of an activity, see Section \ref{Sec:3.4.5}: after the passage of a photon through a calcite crystal, it is possible to prove that both position and polarization of the system are indefinite by using the same criterion.

\begin{itemize}
    \item the quantum description of processes includes two different types of state evolution: in the absence of measurement, the unitary evolution governed by the Schr\"{o}dinger equation; in measurement, the evolution prescribed by the projection postulate (ruling out no-collapse interpretations);
\end{itemize}

In the course, we always promote a clear distinction between measurement and propagation. While dealing with transitions in measurement, we make use of iconic representations showing an initial situation, e.g., in which a photon has just been emitted by a single-photon source (Fig. \ref{FIG:7}.a), and a final one, e.g., in which the photon has been absorbed and counted by a detector (Fig. \ref{FIG:7}.b). In these situations, we always direct student attention to the preparation and the measurement process.
The only exception occurs near the end of the course, when we discuss the ``which-way'' experiment by means of a photon beam directed to a device composed of a sequence of two calcite crystals, one reversed with respect to the other, followed by a filter and a detector (Fig. \ref{FIG:8}). This shift in focus is essential both in the discussion of the wave-particle duality and in that of entanglement (see Section \ref{Sec:3.4.5}).\\

\begin{itemize}
    \item in the construction of a full quantum model for propagation and measurement, we adopt a field ontology;
\end{itemize}

While we feel that this perspective can be perceived as plausible by students, who can make a connection with already familiar classical fields (especially in the case of a photon), ascribing a quantum field ontology to physical systems is a controversial operation \cite[e.g.,][pp. 133-135]{Passon2019, Norsen2017}. For this reason, we adhere to a cautious approach, suggesting students to model the system as a ``field of actual and potential properties.'' This expression means that the field describes the system in terms of properties it possesses (e.g., one might be a property of a spin component) and of ``potential properties that can possibly be actualized [...] through measurement processes'' \cite{Debianchi2013}. For more information on the concept of potential property and its transition to actuality, see also C. J. Isham \cite{Isham1995}. Based on the examination of the ``which-way" experiment, students are led to identify two further elements of revision in the concept of field: differently from a classical field, a quantum one displays a punctual interaction with detectors (we can identify the detector with which the interaction takes place), but this interaction affects the entire field at the same instant, i.e., in a non-local way.\\

\subsubsection{Structuring the discussion of epistemological themes} \label{Sec:3.5.2}

The three rules of correspondence naturally lend themselves to a discussion, respectively, of the completeness of the theoretical description, of indefiniteness and uncertainty, and of the measurement problem. Based on the examination of the ``which-way'' experiment and entanglement, it is also possible to add to the picture a discussion of the problem of locality.

Format, content and placement of the activities on the foundational debate need not only be instrumental to the implementation of the \textit{Epistemological Principle}, but also compatible with the educational level of the student population at hand, the structure of the learning path, and the duration of the course (12 hours).
The chosen format consists in a short introductory lecture given by the instructor, followed by a whole class discussion of the topic, which can be supported by pre-class reading assignments. This structure is in accordance with the method used to conduct worksheet-based activities (see Section \ref{Sec:4.3}), which represent the common thread of the course, and is adapted to the complexity of foundational topics by extending the time for the whole class debate, that needs to be at center stage, and by providing multiple forms of support. The texts are preferably selected among those works of leading scientists whose understanding does not require sophisticated mathematical or physical knowledge \cite[e.g., excerpts from][]{Schilpp1998, Bell2010}. Since the first three units of the course concern preparation, measurement, and their formalism, while propagation, wave-particle duality, and entanglement are addressed in Unit 4, it is natural to discuss first the debates on indefiniteness, uncertainty, and completeness.

The first occasion to introduce the problem of indefiniteness and uncertainty may be the extension of the relations between properties to the case of position and velocity, in Unit 1, where students deal with the loss of the property of one observable in the measurement of the other (a limiting case of the uncertainty principle). Another occasion is offered by an activity of the third unit, which is designed to promote the distinction between a superposition state such as $|\psi\rangle=a|0^{\circ}\rangle+b|90^{\circ}\rangle$ and a mixture of a fraction of $a^2$ photons prepared in $|0^{\circ}\rangle$ and $b^2$ in $|90^{\circ}\rangle$, and to launch the discussion of related epistemological issues (see Section \ref{Sec:5.4.3} for a description of the goals and of the development of this activity). Completeness may be examined in Unit 2, during the discussion of the quantum state, or together with the other issues in the third unit.

In the initial versions of the course, we discussed the Heisenberg's microscope thought-experiment and Bohr's criticism of it in Unit 1, in order to contrast a disturbance interpretation of the principle - where system properties are well-defined but it is not possible to measure them simultaneously with an arbitrary precision - with an interpretation in which they are not well-defined \cite{Tanona2004}. For the revision of this activity, see Section \ref{Sec:5.5}. In Unit 3, instead, we discussed the debate between Einstein and Bohr on the completeness of quantum mechanics (e.g., hidden variables) and - again - the uncertainty principle, leaving out the part on the EPR paradox, which will be taken up when dealing with entanglement \cite{Schilpp1998}. The discussion of these issues was concluded in Unit 4, where we proposed students, as a plausible interpretation of the wave-particle duality, an ontology based on the ``field of actual and potential properties.''

After the conceptual and mathematical examination of the polarization entanglement of two photons produced by parametric down-conversion, students are presented with the problem of non-locality, which is discussed only at a qualitative level. This discussion allows us to emphasize the importance of the foundational debate in the development of scientific knowledge. In the words of Alain Aspect, ``there was a lesson to be drawn: questioning the `orthodox' views, including the famous `Copenhagen interpretation', might lead to an improved understanding of the quantum mechanics formalism, even though that formalism remained impeccably accurate'' \cite[][p. xix]{Bell2010}. As regards technological development, it is possible to illustrate to students that a deeper understanding of entanglement is at the root of a second quantum revolution that is now unfolding \cite[e.g.,][]{House2018}, and that John Bell has been its prophet \cite{Bell2010}.

By having students work on the modelling and interpretation of the mathematical description of entanglement, we gain a further opportunity: ending the course with the discussion of the measurement problem. We illustrate the Schr\"{o}dinger's cat thought experiment and more in general the measurement problem,
indicating three lines of solution proposed by members of the scientific community: 1) accept the standard interpretation and modify the dynamics of the theory; 2) accept the dynamics and modify the standard interpretation; 3) accept both the standard interpretation and the dynamics, and try to show that their conflict can be ignored for all practical purposes \cite{Bub1998}.
As an instance of the first, we mention the Ghirardi-Rimini-Weber's theory \cite{Norsen2017}. The second is illustrated by hinting at the Everett's ``many
worlds'' interpretation \cite{Norsen2017}. The third is represented by the decoherence research program \cite{Schlosshauer2007}. We explain that decoherence provides an answer to the nonobservability of interference effects on macroscopic scales. However, outside the scope of decoherence remains the explanation of why only a particular outcome is realized in each measurement \cite{Schlosshauer2007}: one of the most significant open issues in modern physics, that affects also our proposed interpretation of the wave-particle duality.

\section{The course} \label{Sec:4}

\subsection{Structure of the course and types of activities} \label{Sec:4.1}
The course is designed for an optimal duration of 12 hours, even if some design experiments lasted only ten. The time devoted to each topic is organized as follows: four hours for Unit 1, two for Unit 2, two for Unit 3, four for Unit 4. Lessons are divided into two-hour blocks, that represent a compromise between the time required to engage secondary school students in a series of inquiry- and modelling-based activities they are not accustomed to, and the need to limit the cognitive load associated with the discussion of non-intuitive and novel content.

\begin{figure*}[!htpb]
       \fbox{\includegraphics[width=\textwidth]{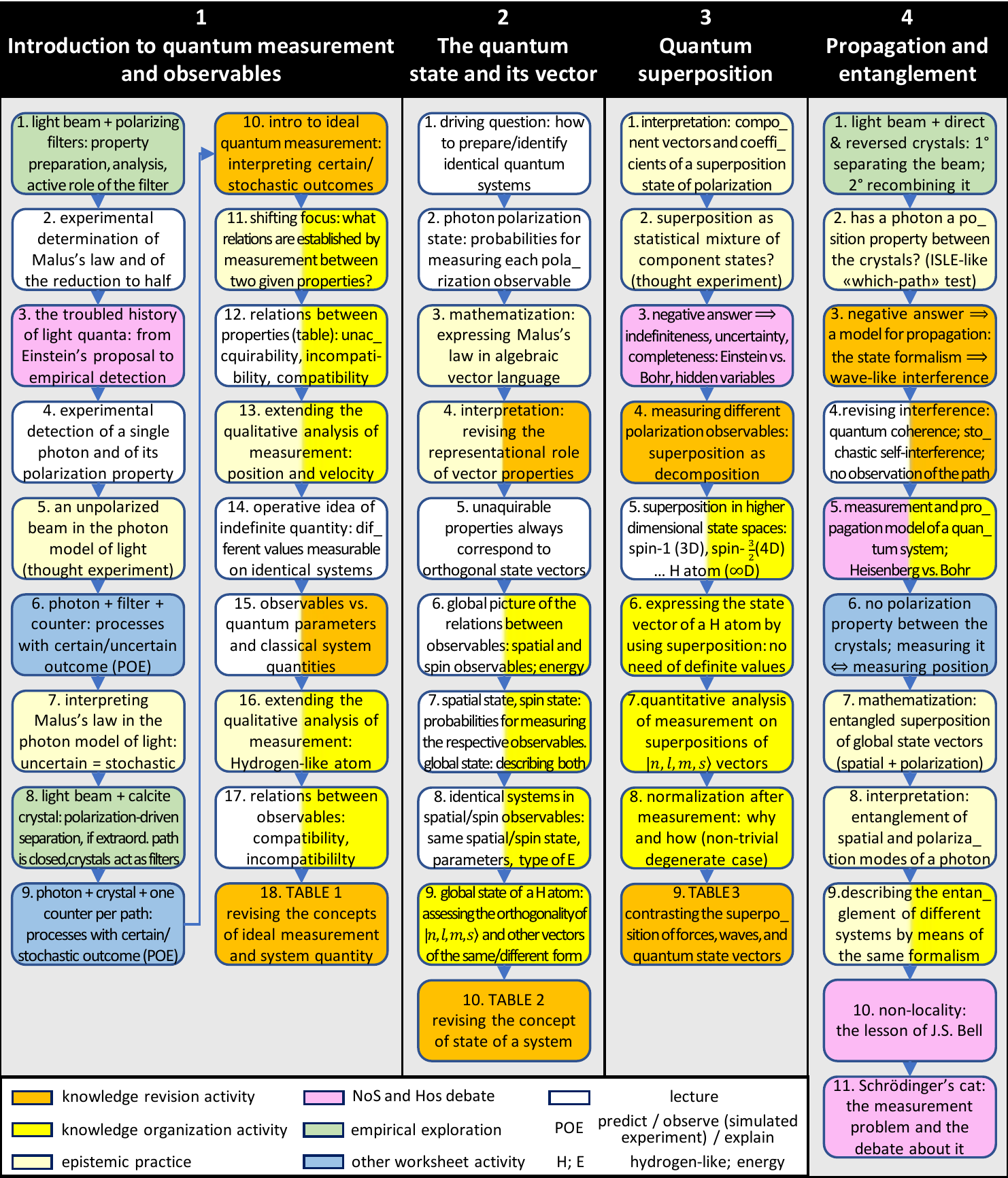}}
    \caption{The structure of the sequence as a composition of building blocks: units and individual activities. Two-colour boxes with a white half represent lectures aimed to implement the design principle associated to the other color. The other two-colour boxes represent active-learning strategies that play more than one role.}
    \label{FIG:9}
\end{figure*}

The structure of the path in terms of units, individual activities and their typology, is displayed in Fig. \ref{FIG:9}.
The figure is intended as a guide to design and instruction. As regards the former, it allows us to monitor the implementation of the principles and their interplay, step by step. As regards instruction, it gives guidance on the crucial aspects to focus on at each step, which could be, knowledge revision, knowledge organization, epistemic cognition, epistemological debates, etc. By examining the figure, it is possible to see that, while each unit builds on the previous ones, individual units can be described as self-contained. In accordance with the implementation of the first and the second design principles, each unit is concluded with a bird's-eye view across contexts on the revision, due to theory change, of the basic concepts and constructs addressed in it. However, except for Unit 1, all the others can be introduced by means of a driving question or a need emerging from previous units, that suggests students the importance to acquire further knowledge \cite{Wittmann2020}. For Unit 2 on the quantum state, the driving question is an issue implicitly raised in Unit 1: how to prepare/identify identical quantum systems if some of the observables are necessarily indefinite (activity explicitly displayed in figure). Unit 3 on superposition is associated with the need to quantitatively determine the possible results of measurements on hydrogen-like atoms, which have been qualitatively explored in Units 1 and 2. For Unit 4, the question is how to describe propagation with the mathematical tools introduced in Unit 3 for describing measurement.

\begin{figure}[!htpb]
       \fbox{\includegraphics[width=\columnwidth]{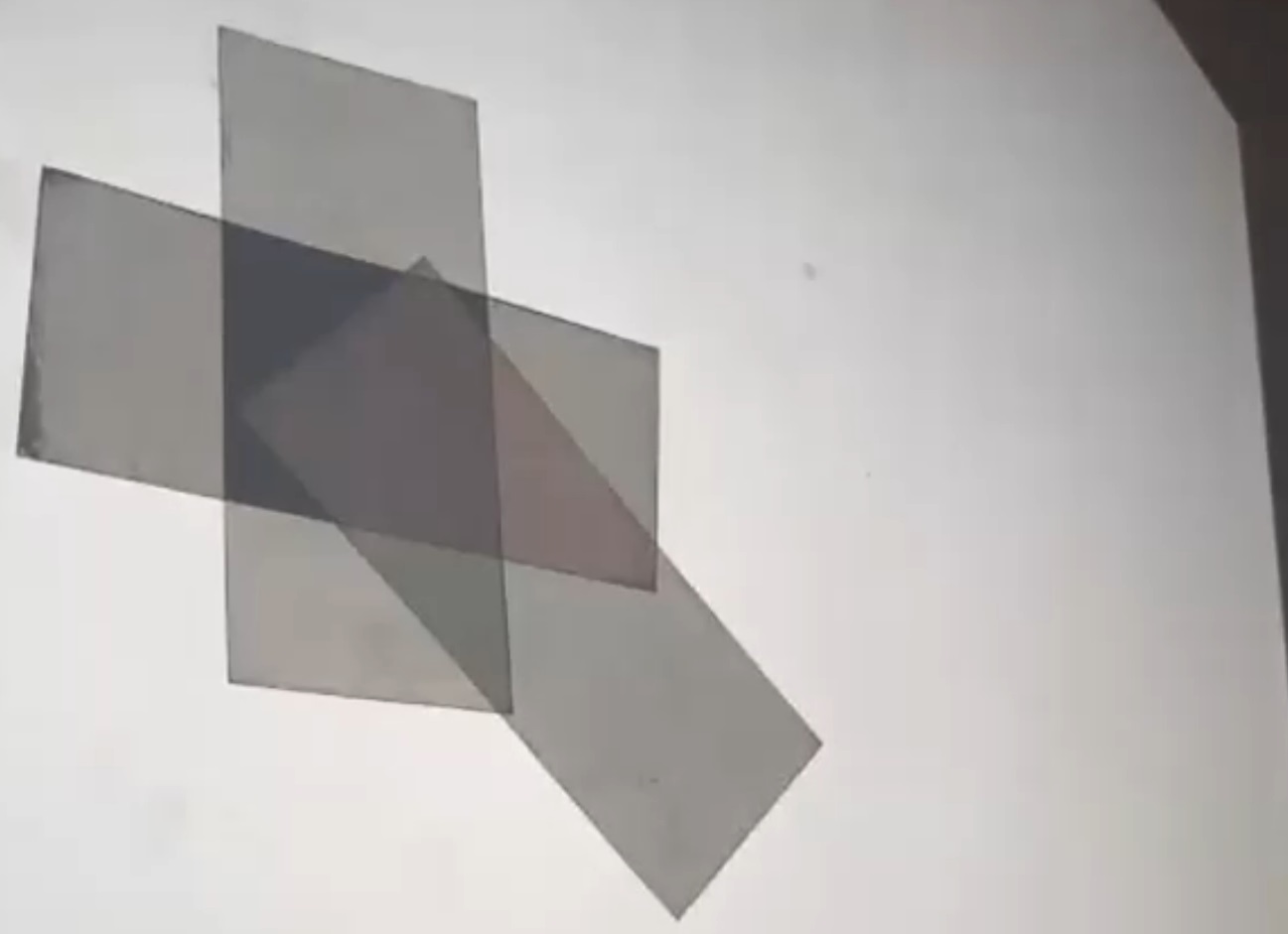}}
    \caption{The active nature of polarizing filters: by inserting a third filter between two filters with perpendicular axes, we observe an increase in transmitted intensity.}
    \label{FIG:10}
\end{figure}

\begin{figure} \centering
\begin{tabular}{|r|l|} \hline
    \includegraphics[width=.42\columnwidth]{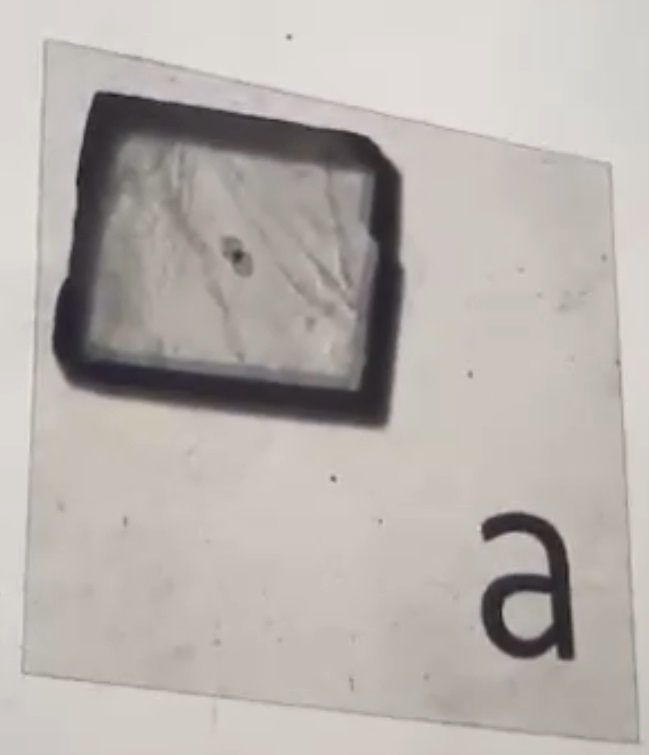} &
    \includegraphics[width=.58\columnwidth]{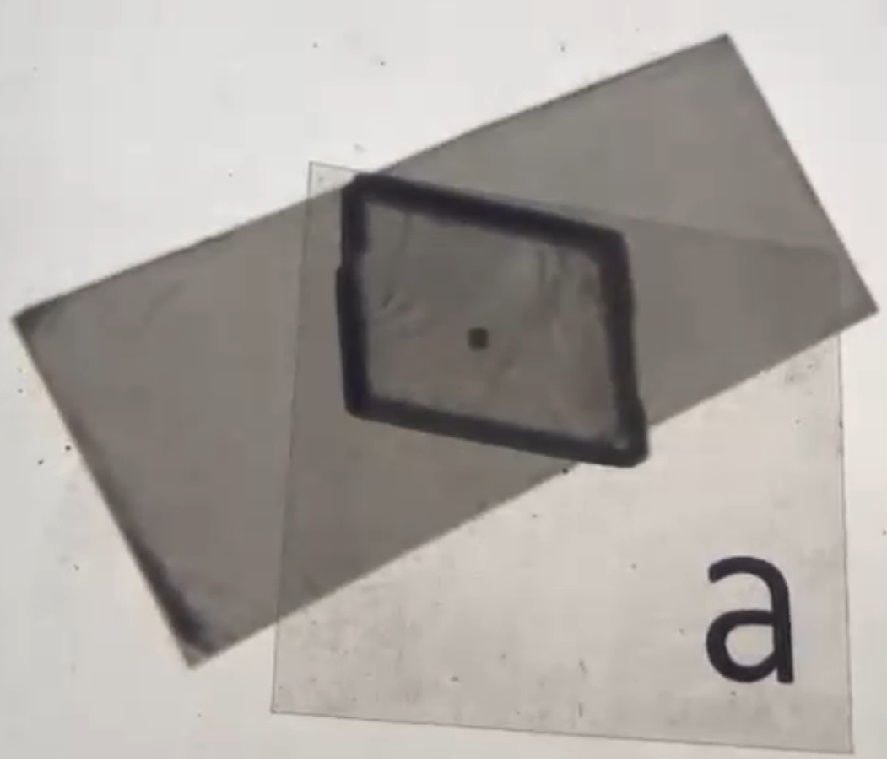} \\ \hline
\end{tabular}
    \caption{(a) The phenomenon of birefringence; (b) The outgoing light beams are polarized, as shown by adding a filter on the crystal.}
    \label{FIG:11}
\end{figure}

The typology of each activity has been displayed in Fig. \ref{FIG:9} by means of a color code. By looking at the color distribution, it is evident that a large majority of the activities are linked to the implementation of the four principles. The following is a synthetic description of each type of activity:
\begin{itemize}
    \item \textit{Knowledge revision activity}: relying on the representation of the conceptual trajectory of a classical notion \cite{Zuccarini2022} with the aim to promote a consistent interpretation of its quantum counterpart (often structured in terms of interpretive tasks);
    \item \textit{Knowledge organization activity}: relying on the relations between properties/observables in order to build a coherent body of knowledge by using the same conceptual tools for the analysis of different physical situations;
    \item \textit{Epistemic practice}: inquiry- and modelling-based activity that mirrors the processes used in theoretical physics for building new scientific knowledge. We remind the reader that, by running this kind of activities, students build knowledge that is \emph{new for the learner}. In order to promote an awareness of the nature of each practice and of its significance in the development of the discipline, the activity is followed (less frequently: preceded) by what we call ``a historical snapshot.'' It consists of a two-minute lecture on the practice and on a historically significant example of how it has been used by theoretical physicists in the building of classical physics knowledge (Fig. \ref{FIG:3} includes the summary of a historical snapshot on each practice).
    \item \textit{NoS and HoS debate}:  discussion of issues concerning the scientific epistemology of QM and the historical development of the discipline. The activity involves a ten-minute lecture with the aid of the slide presentation followed by a whole class discussion. Except for ``the troubled history of light quanta'' (Fig. \ref{FIG:9}, activity 1.3), which is instrumental to introduce the discrete nature of electromagnatic radiation, and to highlight the tangled and non-linear relation between experiment and theory in scientific development \cite{Kragh1992}, the other activities of this kind have been already described in Section \ref{Sec:3.5.2};
    \item \textit{Empirical exploration}: of the polarization of macroscopic light beams by using cheap experimental materials such as polarizing filters and calcite crystals. During the exploration, their action on the beams is visualized on the wall by means of a overhead projector (see Fig. \ref{FIG:10} and Fig. \ref{FIG:11}). The activity is conducted as a form of \emph{demonstrated inquiry} \cite{Llewellyn2012}: the instructor poses questions to the students, soliciting input in the design of the exploration, encouraging them to form hypotheses, to make predictions, and to explain the results. Three empirical explorations are scheduled at different points of the course: right at the start of the learning path, to introduce the phenomenology of the interaction of light with polarizing filters (Fig. \ref{FIG:9}, activity 1.1); after the probabilistic interpretation of the Malus's law, to present the phenomenology of birefringence, thus providing the experience needed for the modelling of quantum measurement at a microscopic scale (Fig. \ref{FIG:9}, activity 1.8); at the beginning of Unit 4, to present a simple form of ``which-way'' experiment, paving the way to the discussion of propagation and entanglement (Fig. \ref{FIG:9}, activity 4.1).
\end{itemize}

\subsection{Instruments}
The instruments we use in the course are the following: 1) worksheets, 2) cheap experimental tools, 3) the JQM environment for simulated experiments, 4) a specific use of language, 5) a slide presentation, 6) homework: reinforcement exercises, reading assignments, and slides used in the previous lessons.

Worksheets are designed to emphasize written explanations of student reasoning. In this course, they represent the common thread underpinning the development of learning from beginning to end, and the main instrument for collecting data on student learning. For each unit, we designed a worksheet of two-three pages of tasks. Each worksheet is divided into blocks with a general goal that is split into conceptual micro-steps addressed in different questions \cite{McDermott2013}. With the exception of lectures and of former worksheet questions that have been converted into oral ones, the sequence of activities displayed in Fig. \ref{FIG:9} mirrors the structure of the worksheets. All worksheets but the last one end with a block containing one or more concept revision tables (Fig. \ref{FIG:9}, activities 1.18, 2.10, and 3.9).

As we have seen in the last section, the exploration of the phenomenology of light polarization at a macro-level is performed
thanks to kits including an overhead projector, passive filters, polarizing filters (Fig. \ref{FIG:10}), calcite crystals and tracing paper with a black dot in order to examine the phenomenon of birefringence (Fig. \ref{FIG:11}).

We already introduced the JQM environment for simulated experiments in Section \ref{Sec:3.1}. With one notable exception, in the current version of the course, the adoption of this instrument is limited to its visual code, which is used both in the slide presentation and in the tasks proposed to students (see, e.g., Fig. \ref{FIG:15}). This code is instrumental to building a highly idealized environment designed to help students focus on essential theoretical aspects. The exception concerns the ISLE-like ``which-path'' activity, in which the simulation plays the role of a testing experiment  (Fig. \ref{FIG:9}, activity 4.2).

Language in the slide presentation and in questions has been structured according to the following guidelines: first, the adoption of the language of ``properties'' and of their relations in order to provide a unified framework for describing measurement, state, and superposition at a point in time; second, the use of colloquial language and student sketches in whole class discussions (as in Fig. \ref{FIG:39}) and, when possible, in questions (e.g., describing activity 1.7 in terms of a ``horoscope of the photon'', as illustrated in Section \ref{Sec:5.4.1}).

Every aspect of the lessons (lectures, worksheet questions, correct answers, discussion of the results of empirical explorations) is supported by slides on the multimedia board or on a projector. At the end of each lesson, the slide presentation used in the classroom is made available to students in the form of a pdf file.

As regards homework, we already discussed reading assignments in Section \ref{Sec:3.5.2}. Homework exercises contain further interpretive questions (e.g., on the physical meaning of the sign of a superposition) and questions for deepening the development of specific aspects of the model (e.g., mathematically deriving the reduction to half of the intensity of an unpolarized beam of photons passing a filter).

The combined use of worksheets, slides, and of an instructor diary reporting student comments and reactions, offered us the possibility to monitor their learning paths during design experiments, identifying unsolved difficulties in the specific question or slide in which they were elicited. As a result, these instruments helped the researchers in their investigation of student ideas and in the refinement of the course.

\subsection{Methods: conduction of worksheet-based and oral activities} \label{Sec:4.3}
During the whole course, students are actively engaged in a modelling process. Therefore their role and that of the instructor are defined in accordance with inquiry-based learning approaches. Worksheet activities represent the bulk of the lesson time, and are conducted in the following way:
\begin{itemize}
    \item \textit{Phases of the activity}: 1) preparation: the instructor displays the slide containing the worksheet items, reads them, and specifies the number of minutes allotted for the writing assignment (depending on its difficulty); 2) writing assignment: each student on her/his individual worksheet, discussion with the deskmate is allowed and encouraged; 3) whole class discussion; 4) answer of the instructor: displayed on the slide and, if necessary, discussed.
    \item \textit{Role of the instructor during the writing assignment}: walking through the class, listening, observing, checking the progress of each student, answering clarification requests and posing stimulus questions (when realizing that some students are stuck) to help them overcome difficulties and to support their reasoning.
    \item \textit{Role of the instructor during the whole class discussion}: facilitating the discussion, e.g., asking a student to share her/his answer, inviting those who have given different answers to express their point of view in the attempt to convince their peers, asking further clarifications if the explanation is not fully clear to the other students, and going on in this process until a consensus has been reached.
\end{itemize}

Oral questions are displayed on a slide and directly addressed in a whole class discussion, after which the answer of the instructor is shown.

\section{Cycles of refinement} \label{Sec:5}

\subsection{Design-Based Research: data collection and analysis}

The course has been refined in cycles of testing and revision conducted in the framework of Design-Based Research (DBR). This framework is a collection of approaches devised for ``engineering'' teaching and learning sequences, and systematically studying them within the context defined by practices, activities and materials - in short, by the means - that are designed to support that learning \cite{Bakker2015}. DBR consists of cycles composed of three phases: preparation, design experiment, retrospective analysis. The results of a retrospective analysis feed a new design phase. When patterns stabilize after a few cycles, the instructional sequence at hand can become part of an emerging instruction theory.

The course has been experimented in classroom contexts of various nature.

The first one is the Summer School of Excellence on Modern Physics, held every year at the University of Udine, Italy. It consists of a one-week full immersion program in modern physics topics. The course was held in the years 2014-2018. Participant students ranged from a minimum of 29 in 2014 to a maximum of 41 in 2015. They were selected among a large number of applicants from a wide range of Italian regions. All of them had just completed the penultimate year of secondary school.

The second context consists of regular classrooms from Italian secondary schools. The course was held in Liceo Statale Corradini, in the city of Thiene, in November 2018 and in Liceo Scientifico Statale Alessi, in the city of Perugia, in February 2019. In the Italian system, Liceo is a type of school attended by students who intend to continue their studies in university. The design experiment involved three classes of the final year from Liceo Corradini, for a total of 61 students, and two classes of the same year from Liceo Alessi, for a total of 39 students.

The third context concerned self selected students from Liceo Scientifico Galilei, in the city of Trieste, at the end of March 2019. The course was offered as an optional study program, and was attended by 18 students.

In this work, we do not test the effectiveness of the course, but the refinement of individual activities. For this purpose, the differences between the three kinds of student population did not represent an issue. Future directions include the analysis of a pre-post-test administered in regular classrooms. Here we report on the path of refinement and the final design and learning outcomes of a set of activities chosen to illustrate the implementation of each of the four principles of design. Except for a limited number of recently added activities, cycles were iterated until patterns stabilized.

Data sources consist of written answers to worksheet questions, occasionally enriched by notes reported in the instructor diary during design experiments. Data were analyzed for correctness and for student lines of reasoning, since both informed the revision of the activities. The second type of analysis was conducted according to qualitative research methods \cite{Erickson2012}: the identification of crucial conceptual content and the examination of literature on learning difficulties in QM guided the building of a-priori categories. Then, based on conceptual elements introduced by student answers, the categories were revised. This process led to the identification of clusters and coherence elements in student reasoning.

Since the sample changed from experiments to experiment, in order to improve readability and to enable comparison, the rates of answers as regards both correctness and student reasoning are reported in percentages.

\subsection{Knowledge revision activities}  \label{Sec:5.2}

\subsubsection{Measurement} \label{Sec:5.2.1}

\emph{Goals}. Revision of the concept of measurement, intended as consistent interpretation of the interaction of a quantum system with a measurement device, and recognition of the conditional nature of the process: passive and determinate if the system possesses in advance a property of the measured quantity, active and stochastic if it does not. In the context of polarization, this requires interpreting the absorption of a photon by a filter or by a counter following a crystal in terms of a transition in property (in the passive case: of its retention).

\emph{Path of refinement}. In the initial version of the course (2014 Summer School), the introduction of measurement was scheduled right after an extensive work with polarizing filters, both at a macroscopic level and in terms of photons (see Section \ref{Sec:5.4.1}). Therefore, we used, as a measurement device, a filter followed by a counter. However, the design experiment showed that only a small number of students spontaneously interpreted the absorption of a photon by a filter in terms of acquisition or retention of a property in the direction perpendicular to its axis and, as a consequence, that most did not recognize the conditional nature of measurement. The addition of an iconic support in 2015 (two diagrams depicting the possible transitions of the initial property in case of transmission and of absorption) was not effective. In the stochastic case, only a minority interpreted absorption as a result of the acquisition of a property. Some students even wondered how the situation could be described as a measurement, and not as ``just a weird interaction altering the property''. Given the need to provide students with a context where the results of the interaction with the measurement device can always be visualized as the acquisition/retention of outcome-properties, we resolved to replace the filter with a birefringent crystal and two counters.

\begin{figure*}[!htpb]
       \includegraphics[width=\textwidth]{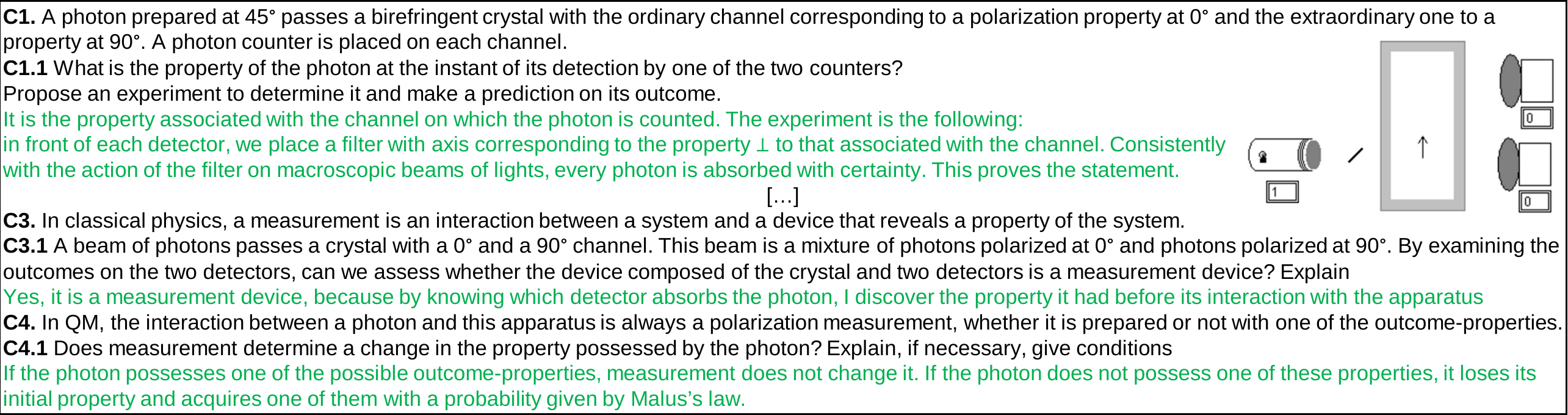}
    \caption{Worksheet block on quantum measurement: 2016 version. In black the item's text, in green the correct answer.}
    \label{FIG:15}
\end{figure*}

\emph{Final Version - Description of the activity}. In the 2016 Summer School (27 students),  we designed as preparatory task an empirical exploration of the interaction of light with crystals both at the macroscopic scale, with real instruments (Fig. \ref{FIG:9}, activity 1.8), and at the single photon level, with the aid of JQM screenshots. The activity was a predict-observe-explain sequence \cite{White1992} (Fig. \ref{FIG:9}, activity 1.9) in which single photons were shown just after the emission (properties at $0^{\circ}$, later $90^{\circ}$, and finally $45^{\circ}$) and then just after the absorption by one of two photon counters placed on the $0^{\circ}$ and $90^{\circ}$ channels of a calcite crystal (see Fig. \ref{FIG:7}).  After that, students were administered a worksheet block on measurement (see Fig. \ref{FIG:15}).
Item \textbf{C1} is an elementary form of thought experiment designed to promote a consistent interpretation of the absorption of the photon by the counters in terms of a transition in polarization property. Item \textbf{C2} is not shown in the figure, since it is not related to the issue at hand, and will be discussed in Section \ref{Sec:5.4.1}. \textbf{C3} is the most important item, promoting a consistent framing of the classical-like measurement, in which random mixtures of photons with $0^{\circ}$ or $90^{\circ}$ polarization are sent to a birefringent crystal with channels corresponding to the same properties. In \textbf{C4}, after giving a definition of quantum measurement as a form of generalization of classical measurement, in which the system does not possess an outcome-property in advance, we asked about the conditional nature of measurement.

\emph{Final Version - Results}. As regards \textbf{C1}, 85\% of the students identified the outcome-properties of a photon prepared at $45^{\circ}$ (probabilistic case) as $0^{\circ}$  and $90^{\circ}$. Most of them designed consistent experiments to prove their statement, using one filter on each channel, one with axis at $0^{\circ}$ and the other at $90^{\circ}$, either corresponding to the polarization associated with the channel (all photons pass the filters), or the opposite case (all absorbed by the filters).
All students but one interpreted the situation described in item \textbf{C3} (determinate interaction) as a classical measurement, half of them explicitly adding that we get to know the initial property according to the detector which collected the photon. As regards the revision of the concept of measurement (\textbf{C4}), 78\% gave consistent answers, recognizing the conditions for passive and determinate measurement, and active and stochastic one, e.g. ``if the property does not coincide with the outcome-properties, the system collapses into one of them. If it coincides, measurement does not change it.'' Even more important, due to the productive interpretation promoted by item \textbf{C3}, no student showed a reluctance to interpret the interaction as a measurement.

\begin{figure}[!htpb]
       \includegraphics[width=\columnwidth]{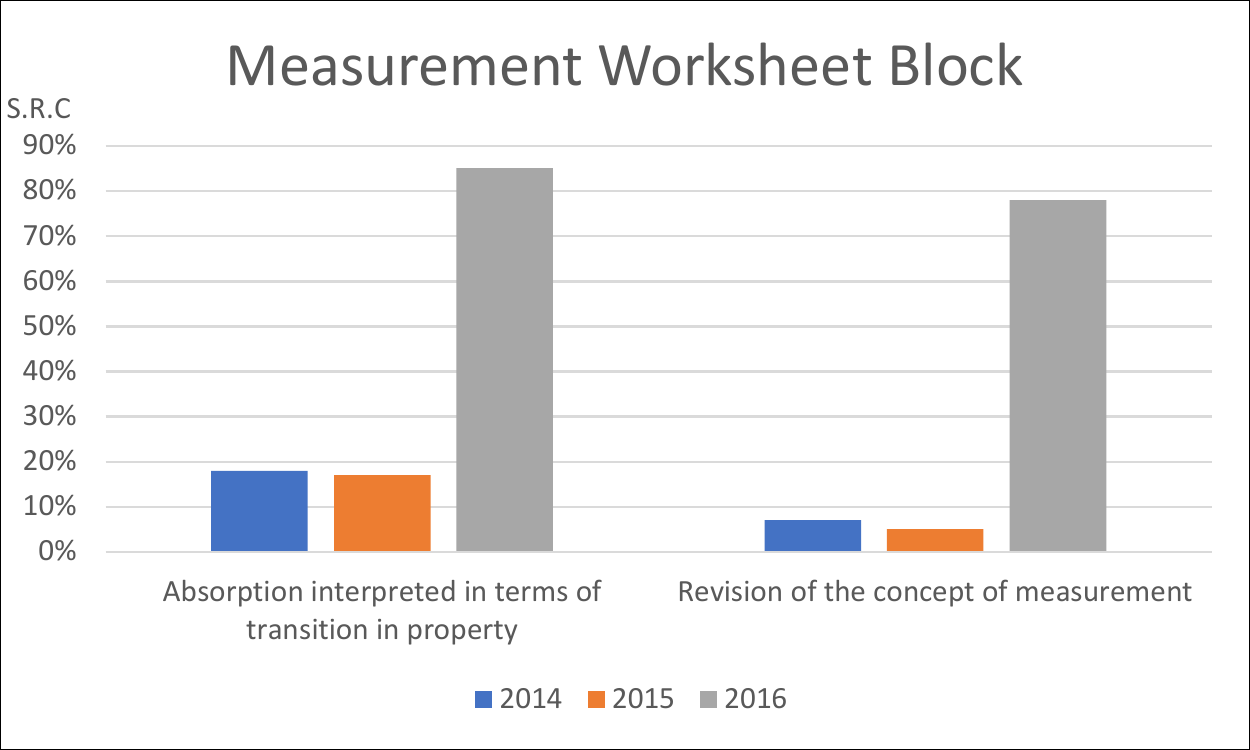}
    \caption{Worksheet results in 2014-2016. In this graph and in the next ones, S.R.C. means student's response correctness.}
    \label{FIG:16}
\end{figure}

Given the high rate of success (see the progression in Fig. \ref{FIG:16}), in the following design experiments the items were converted into oral questions.

\subsubsection{Superposition} \label{Sec:5.2.2}
\emph{Goals of the end-of-unit table}. Consistently contrasting the features of the familiar forms of vector superposition (of forces and waves) with quantum superposition.

\emph{Path of refinement}. The first version of the table was administered in the 2017 Summer School. At that time, the course focused only on polarization.
The table included four statements and asked students to assess whether each of them applied to the superposition of forces, waves, and state vectors, and to explain their reasoning in the last context. The statements concerned the aspects of superposition examined in Section \ref{Sec:3.2.2}: referent, goal, procedure, and constraints. The task on the referent did not address the number of entities involved, but was aimed to reinforce the discrimination between the abstract nature of state vectors and the physical nature of the classical ones, which proved to be a deep issue in the introduction of the state vector (see Section \ref{Sec:3.4.5}). While a majority of students consistently answered this item and the item on the constraints, the tasks on the goal and the procedure gave very poor results, due to difficulties with the number of entities involved in quantum superposition. This outcome prompted us to revise both the previous activities of Unit 3 and the table. In order to support students in recognizing that quantum superposition refers only to one entity - the state of the system -, we added an activity relying on embodied cognition (see Zuccarini and Malgieri \cite{Zuccarini2022}). We also added a more general context: the superposition of eigenstates of the hydrogen-like atom in terms of quantum numbers.

\begin{figure*}[!htpb]
       \includegraphics[width=\textwidth]{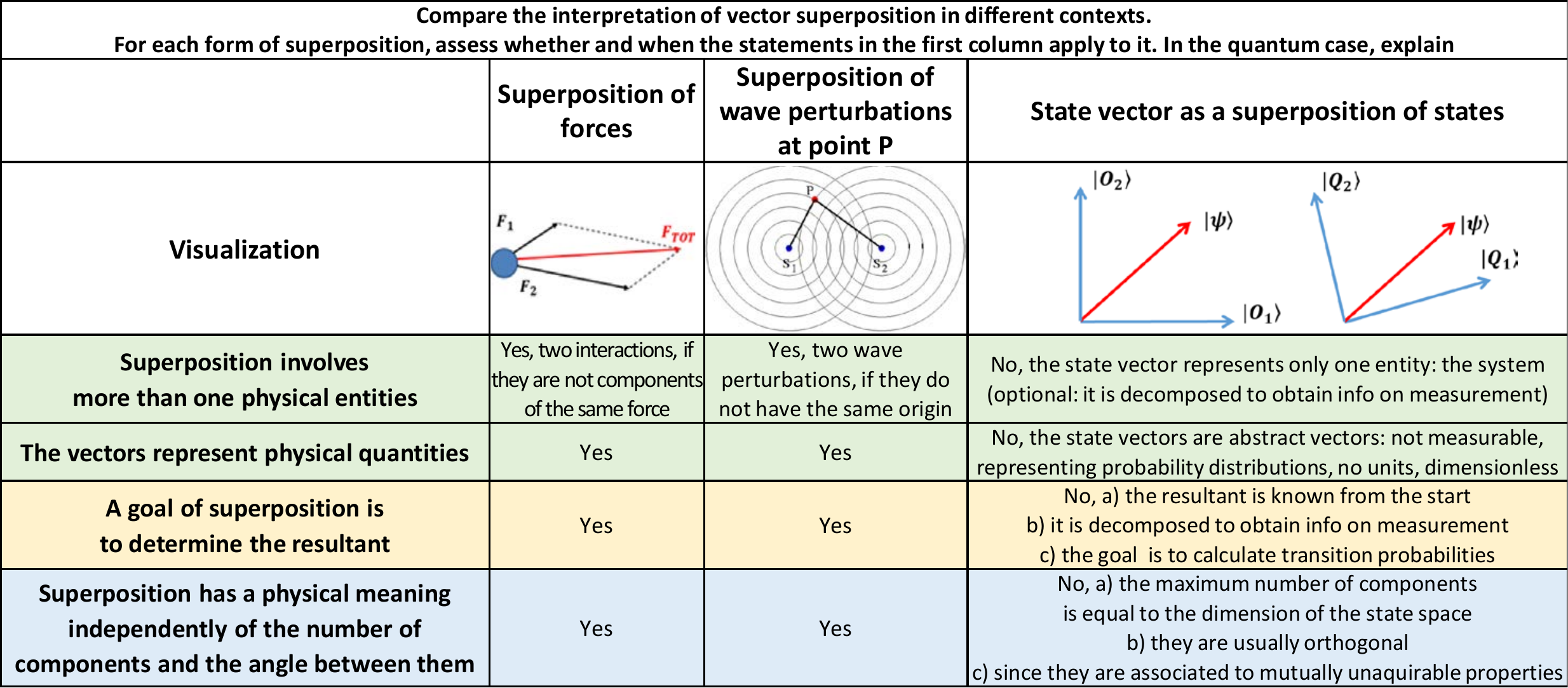}
    \caption{End-of-unit table on superposition filled with correct answers, final version. The nature of the items is highlighted by a color code: green refers to the referent, yellow to goal/procedure, blue to constraints.}
    \label{FIG:18}
\end{figure*}

\emph{Final Version - Description of the activity}. The end-of-unit table corresponds to activity 3.9 of Fig. \ref{FIG:9}, and is displayed in Fig. \ref{FIG:18}.
In the first row, we included a statement on the number of entities involved, which was considered as preparatory to the discussion of the goal. We left unchanged the two productive items on the nature of the vector and on the constraints. The item on the procedure was removed, since both the first and the third item already invited a discussion of this aspect.

\begin{figure}[!htpb]
       \includegraphics[width=\columnwidth]{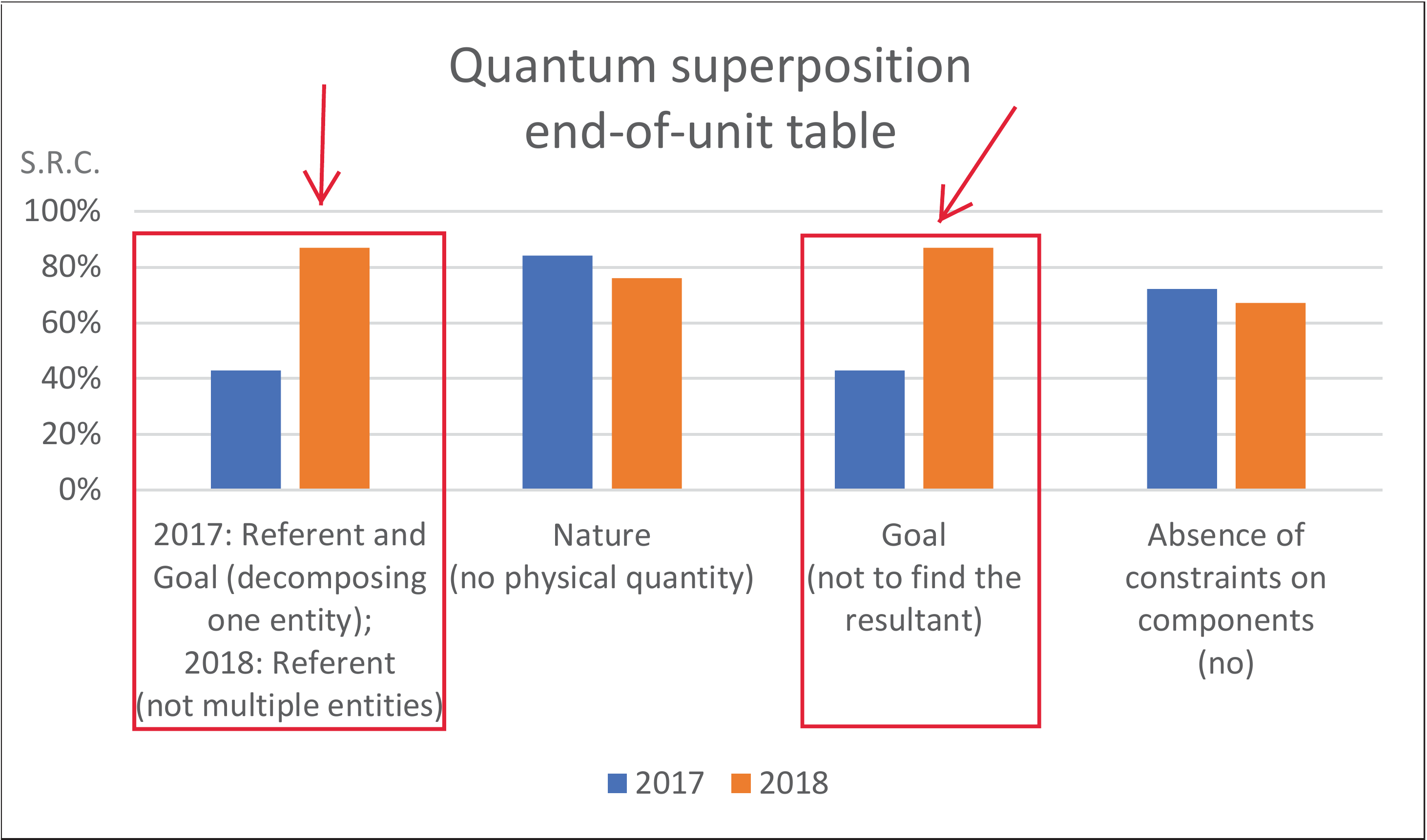}
    \caption{Answers with correct explanation in 2017 and 2018.}
    \label{FIG:20}
\end{figure}

\emph{Final Version - Results}. The new design was experimented in the 2018 Summer School (30 students). The classical boxes did not represent a challenge to students: the great majority consistently answered all of them, although without considering the cases in which only one entity is involved in the superposition of forces and waves. This shows that students spontaneously focus on superposition as a composition of multiple entities. As regards the quantum boxes, a comparison of the rate of success obtained in 2017 and 2018 is displayed in Fig. \ref{FIG:20}.
A strong improvement is evident in the answers to the first and the third item, with 87\% of the students providing consistent explanations. As regards the number of entities involved, 53\% did not limit themselves to state that quantum superposition concerns one entity, but interpreted the process as a decomposition of this entity. As to the goal, beside recognizing that determining the resultant is not among the goals of this form of superposition, most students identified its objectives as ``calculating [transition] probabilities'' (37\%) or ``decomposing the state vector'' (30\%). In both design experiments, more than 65\% consistently answered the other two items on the nature of the vector and on the constraints, with a slight decrease in performance in 2018. Students justified their answer about the nature of the vector by adding consistent explanations such as ``state vectors are unit vectors with no measurement units'', ``they express probability''. In the statement on the constraints, we need to consider the greater complexity introduced by the context of the hydrogen-like atom (where, for bound states, we can have a countably infinite number of components, all orthogonal to one another). Most students discussed the statement only in the context of polarization, e.g. ``In polarization, for instance, we can have up to two components $\rightarrow$ two values'' (47\%). However, some students gave global explanations by connecting the number of vectors to the spectrum of the measured observable: ``No, it depends on the number of values that a property can assume'' (13\%), which is true in the absence of degeneracy.

\subsection{Knowledge organization activities} \label{Sec:5.3}

\subsubsection{Introducing the relations between properties} \label{Sec:5.3.1}
\emph{Goals}. Supporting students in acquiring the appropriate perspective for the introduction of the relations between properties: analyzing polarization measurements in terms of loss or retention of the initial property and possible acquisition of a different one.

\emph{Path of refinement}. The task was initially administered in the 2018 Summer School. Its format is described in the penultimate paragraph of Section \ref{Sec:3.3.1}. Students were directed to refer to Malus's law as a reasoning tool, leaving out any details on the measurement device. This abstract, global formulation of the task proved ineffective. Only a small minority of students identified the conditions for two properties to be incompatible (the system loses its property and may stochastically acquire the other one). Many interpreted the task differently from what we intended, e.g., focusing on the mathematical description of Malus's law instead of specifying conditions on the two properties involved. In addition, since Malus's law had been discussed in the context of the photon-filter interaction, most students unproductively used this situation as a support for their reasoning. An old issue with filters reappeared: the difficulty to associate the absorption of a photon with a transition in property. On the contrary, those students who used birefringent crystals as a context for the task, were generally successful.

\begin{figure*}[!htpb]
       \includegraphics[width=\textwidth]{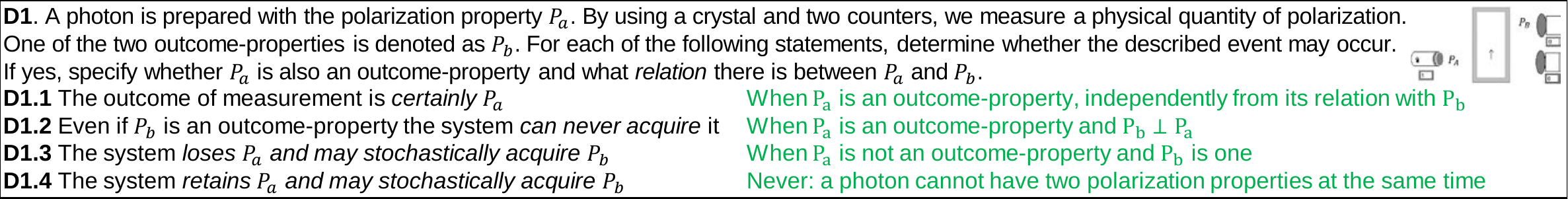}
    \caption{Identifying the possible relations between properties: Liceo Corradini, Thiene, final version.}
    \label{FIG:21}
\end{figure*}

\emph{Final Version - Description of the activity}. The task corresponds to activity 1.11 of Fig. \ref{FIG:9}, and is displayed in Fig. \ref{FIG:21}.
The introduction was radically modified, using calcite crystals as a concrete context for reasoning and visualizing the roles of $P_a$ and $P_b$ by means of a figure. Any reference to Malus's law was removed. The four statements were left unchanged. The activity was experimented in Liceo Corradini, Thiene, November 2018. It was administered in two of the three classrooms involved in the course, for a total of 40 students. In the third one, the activity was discussed orally for lack of time.

\begin{figure}[!ht]
       \includegraphics[width=\columnwidth]{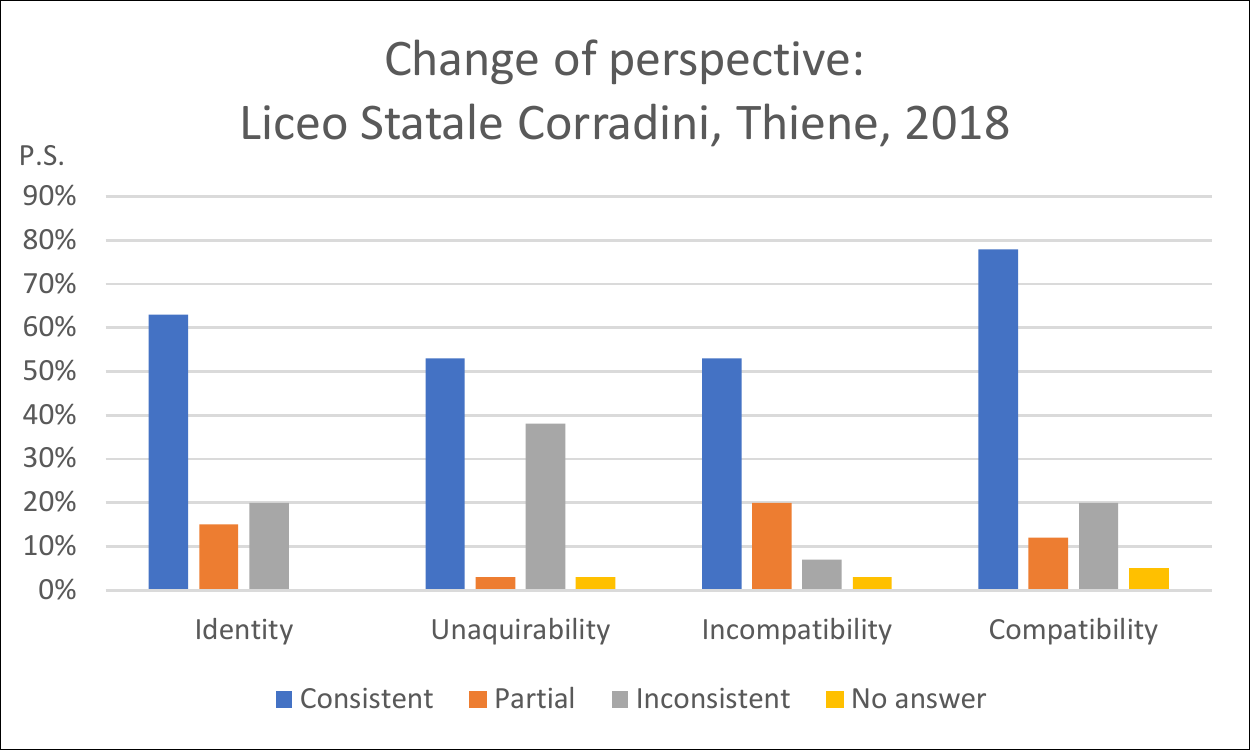}
    \caption{Identifying the relations. Results of the task. In this graph and in the next, P.S. means percentage of students.}
    \label{FIG:25}
\end{figure}

\emph{Final Version - Results}. Data are displayed in Fig. \ref{FIG:25}.
An answer is considered \emph{consistent} if its content matches that of the correct answer reported in Fig. \ref{FIG:21}, both in terms of outcome-properties and (if needed) of angular relations between $P_a$ and $P_b$. \emph{Partial} means that the student has identified one of the conditions for the occurrence of the event described in the statement but not all of them, or that has added unneeded conditions.

The rate of consistent answers is almost the same as in the 2018 Summer School, except for statement 3, on incompatibility, where it increases from 30\% to 53\%. If we consider that Summer School students are selected among a large number of applicants from all over Italy, while the design experiment at Liceo Corradini involved regular classrooms, this indicates a remarkable improvement.

\begin{figure}[!htpb]
       \includegraphics[width=\columnwidth]{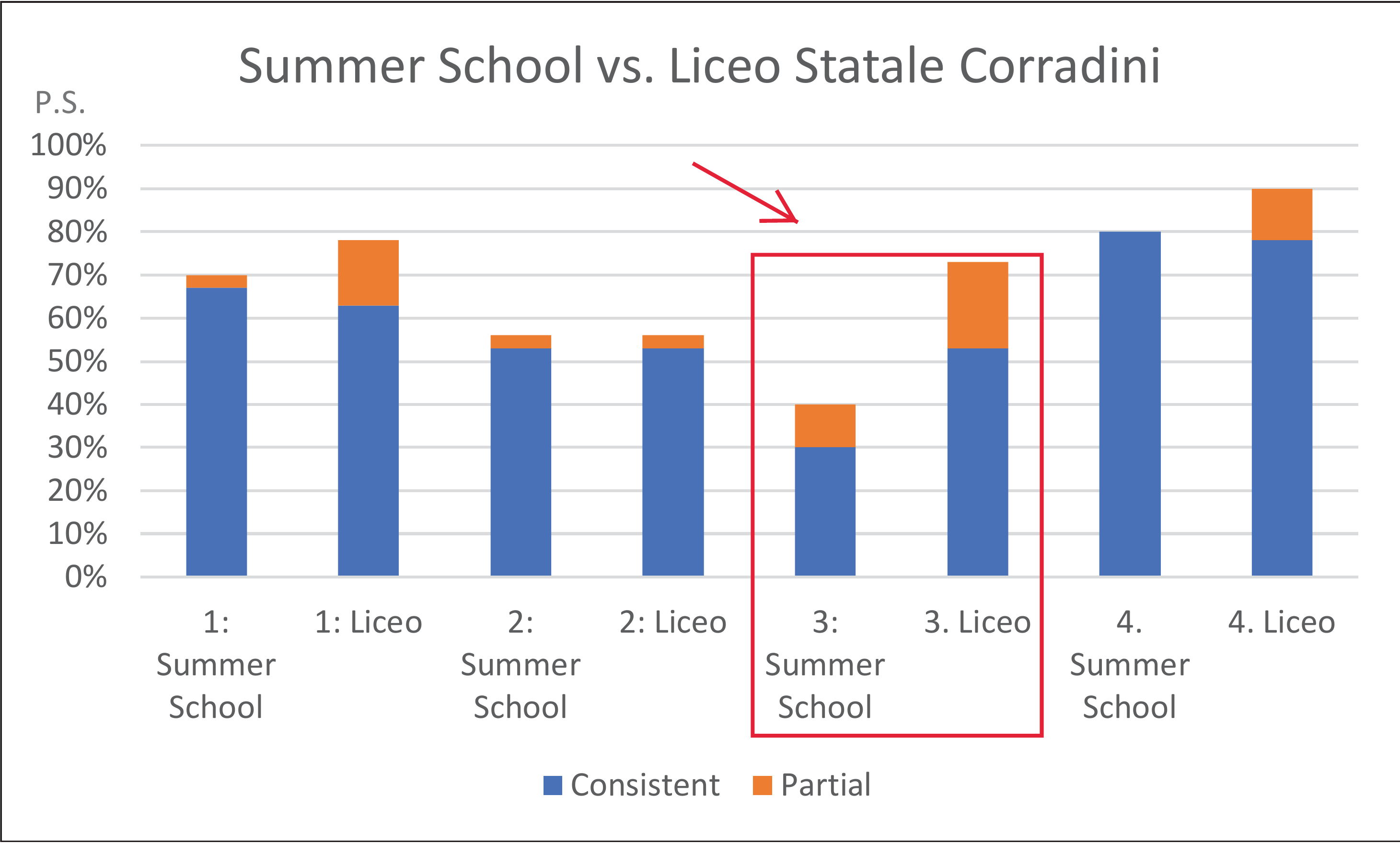}
    \caption{Identifying the relations. Comparison of consistent + partial answers.  1: Identity; 2: Unacquirability; 3: Incompatibility; 4: Compatibility.}
    \label{FIG:26}
\end{figure}

In addition, students from Liceo Corradini generally interpreted the task as intended by the researchers, which is mirrored in the much higher rate of partially correct answers, and practically all students adopted a global perspective. Reasons of failure were the difficulty to identify all the conditions for the occurrence of an event, both as regards the identification of the outcome properties and of the possible angle between $P_a$ and $P_b$. See, for instance, this answer to statement 4: ``Yes, $P_a$ is an outcome-property, $P_a \perp P_b$.'' The improvement is even more evident if we compare the sum of consistent and partial answers (Fig. \ref{FIG:26}).

\subsubsection{Extending the use of the relations between properties to other contexts} \label{Sec:5.3.2}
\emph{Goals}. Qualitatively managing measurement at a global level (position and velocity of a quantum system) and in the context of the hydrogen-like atom.

\emph{General description of the activities}. The tasks correspond to activity 1.13 of Fig \ref{FIG:9}, and are organized into two separate worksheet blocks, one on position and velocity, the other on the hydrogen-like atom. Here, we compare the results obtained in the 2018 Summer School (30 students) with those of the three classrooms of Liceo Corradini (61 students). The worksheet blocks used in the two design experiments are identical, except for a question added to the second block in the design experiment at Liceo Corradini.

\begin{figure*}[!htpb]
       \includegraphics[width=\textwidth]{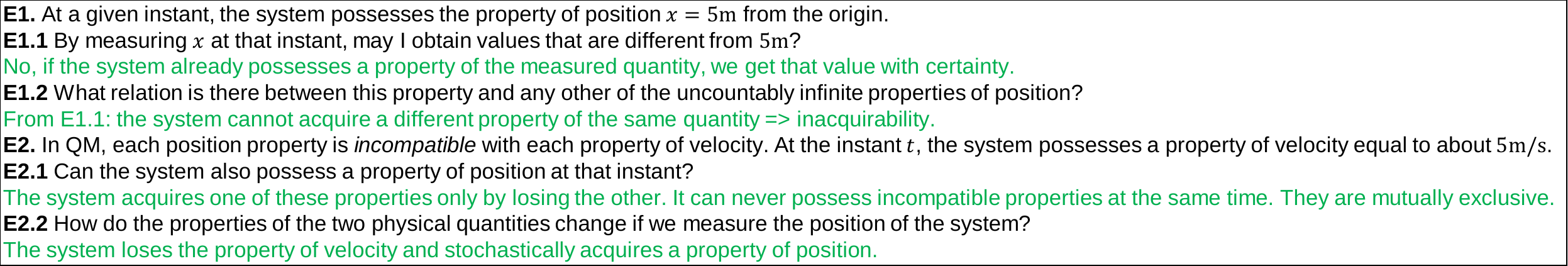}
    \caption{Using the relations between properties to discuss ideal measurement at a global level (position and velocity): worksheet block administered in the 2018 Summer School and in Liceo Corradini, Thiene, November 2018.}
    \label{FIG:27}
\end{figure*}

\emph{Description of the activity on position and velocity}. The worksheet block is displayed in Fig. \ref{FIG:27}.
From the definition of the relations, students are required to infer the mutual unacquirability of all the properties of the same quantity (\textbf{E1.1}, \textbf{E1.2}), to deduce that a system possessing a property of $v$ cannot have a property of an incompatible observable ($x$) at the same time (\textbf{E2.1}), and to qualitatively determine the results of the measurement of $x$ on that system in terms of change in properties: loss and acquisition, either determinate or stochastic (\textbf{E2.2}). For the last task, the definition of incompatibility is not sufficient, since it only mentions the \emph{possible} acquisition of a given property of $x$. In order to conclude that measurement always provides a single property of the measured quantity, also in QM, students need to activate resources on the revision of measurement.

\begin{figure}[!htpb]
       \includegraphics[width=\columnwidth]{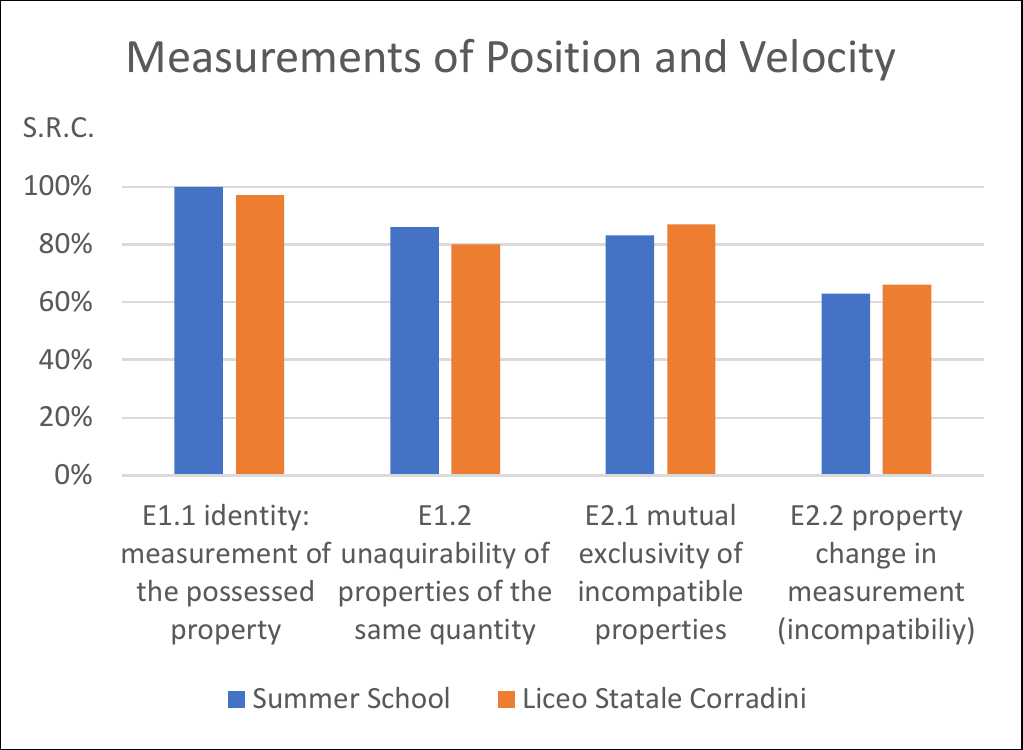}
    \caption{Comparison of the results of the task number 1.13 in Fig. \ref{FIG:9}.}
    \label{FIG:28}
\end{figure}

\emph{Results of the activity on position and velocity}. The comparison of the rate of consistent answers given by students in the two design experiments is presented in Fig. \ref{FIG:28}.
As we can see, while the student populations are very different from each other, in both cases a large majority consistently answered the items. Notably, regular classrooms outperformed summer school students in the most difficult questions (\textbf{E2.1} and \textbf{E2.2}, on incompatibility). A possible explanation is the improvement of the activity on the introduction of the relations between properties, where regular classroom had a much higher rate of success (see Fig. \ref{FIG:26}, statement 3 on incompatibility). However, assigning a physical meaning to their own answers was a totally different matter. Students belonging to regular classrooms were puzzled by a phenomenon they had not encountered in the context of polarization: the absence of any property of a familiar physical quantity ($x$ or $v$). Summer school students suggested explanations in terms of a ``perturbation'' introduced by the measurement device (20\%), a semiclassical model \cite{Ayene2011} that will be addressed by discussing Heisenberg's microscope thought-experiment and Bohr's criticism of it (see Section \ref{Sec:5.5}).

\begin{figure*}[!htpb]
       \includegraphics[width=\textwidth]{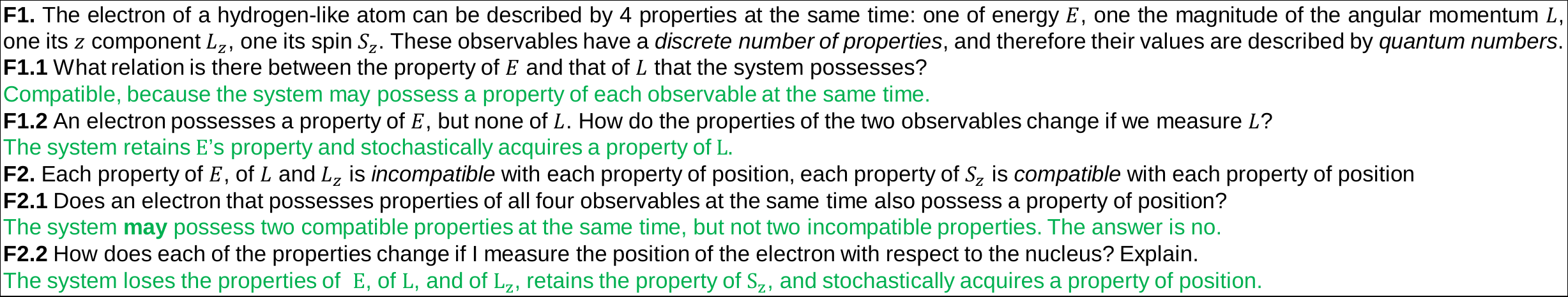}
    \caption{Using the relations between properties to discuss ideal measurement in the context of the hydrogen-like atom: the worksheet blocks administered in the 2018 Summer School and in Liceo Statale Corradini are identical, except for item F1.2, which was added in the Liceo version.}
    \label{FIG:29}
\end{figure*}

\emph{Description of the activity on the hydrogen-like atom}. The worksheet block is displayed in Fig. \ref{FIG:29}.
In the activities on the hydrogen-like atom, students are asked to perform the following tasks: \textbf{F1.1} recognizing that, if a system can have properties of different observables at the same time, these properties need to be compatible; \textbf{F1.2}  qualitatively determining the results of the measurement of an observable ($L$) on a system which has no property of that observable, but possesses a property of a compatible observable ($E$); \textbf{F2.1} in the case of multiple properties possessed by the system at the same time ($E$, $L$, $L_z$, $S_z$), some of which are compatible and some incompatible with the observable to measure ($x$), determining whether compatibility or incompatibility prevails (actually the latter: the system cannot have also a position property); \textbf{F2.2} qualitatively determining the results of the measurement of $x$ on the bound state at hand.

Question \textbf{F1.2} was added in the version for Liceo Corradini. Since the same case will be discussed also in Unit 3 by analyzing a superposition of bound states of the hydrogen-like atom, we meant to tighten the coherence of the course by proposing the same task also at a qualitative level. Moreover, we intended to investigate whether students could answer a ``compatible'' analogue to \textbf{E2.2}.

\begin{figure}[!htpb]
       \includegraphics[width=\columnwidth]{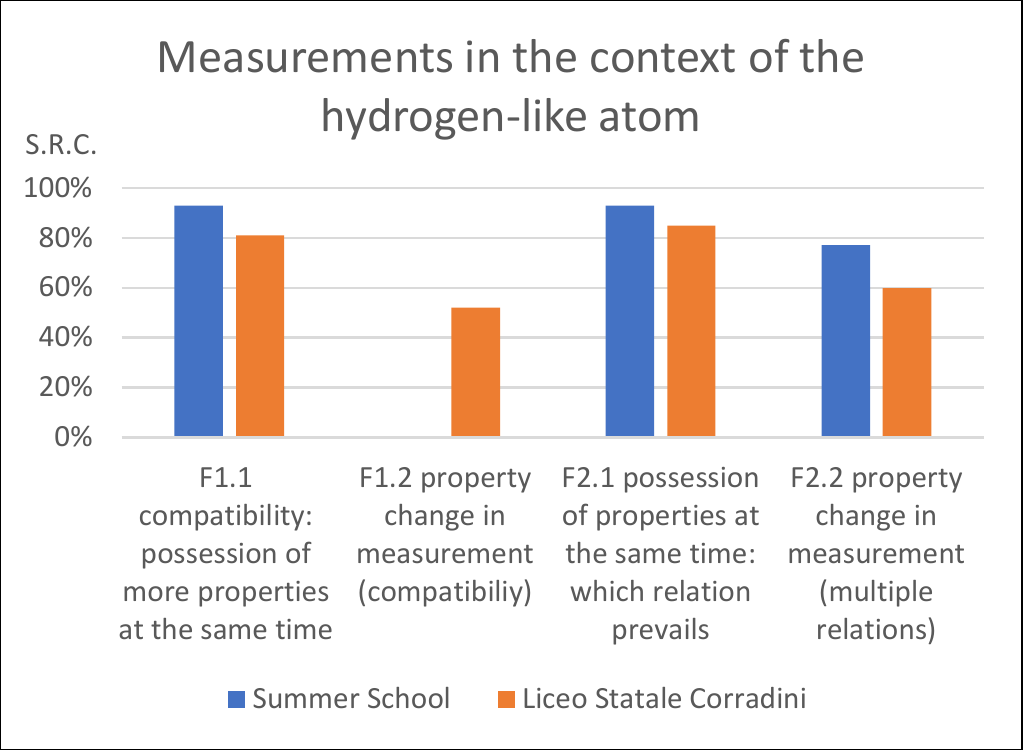}
    \caption{Comparison of the results of the task number 1.16 in Fig. \ref{FIG:9}.}
    \label{FIG:30}
\end{figure}

\emph{Results of the activity on the hydrogen-like atom}. Data on both design experiments are shown in terms of rate of consistent answers in Fig. \ref{FIG:30}.
While in both experiments, more than half of the students consistently answered all of the questions, their results were generally worse than that obtained in the first block, with summer school students outperforming regular classrooms. As to the latter, we still get a very high percentage of consistent answers to \textbf{F1.1} (85\%) and \textbf{F2.1} (81\%), while 60\% consistently answered \textbf{F2.2}. Question \textbf{F1.2} turned out to be the most difficult for them, with 53\% of consistent answers. Many students did not connect the situation described in the item with the relation of compatibility
Others said that if the properties are compatible, nothing changes in measurement
In question \textbf{F2.2}, an analogous but more complex version of the fourth item on $x$ and $v$, student explanations were quite similar to those reported for the latter. In general, the physical situations presented in question \textbf{F1.2} and \textbf{F2.2} were totally new to students, who had never encountered phenomena related to compatible observables in previous parts of the course. Here, the carefully selected and motivated students of the summer school were quicker to respond to the challenge posed by the new context.

\subsection{Epistemic practices} \label{Sec:5.4}

In the previous sections on the DBR cycles, we have already presented activities corresponding to theoretical practices for the construction of scientific knowledge: the elementary thought experiment in Section \ref{Sec:5.2.1}, the change in perspective described in Section \ref{Sec:5.3.1} and the extension of results found in one context to other contexts in \ref{Sec:5.3.2}. These activities are instrumental to the implementation of more than one principle of design at a time. Here we discuss activities that are exclusively designed to implement the \emph{Epistemic Principle}, with a focus on thought experiments and mathematical modelling.

\subsubsection{Thought experiment: Description of an unpolarized beam in terms of photons} \label{Sec:5.4.2}

\emph{Goals}. Identifying a consistent description of unpolarized beams in terms of photons by running a thought experiment specifically designed by the instructor.

\emph{Path of Refinement}. The first version of this activity was experimented in Liceo Alessi, Perugia, 2019. All students chose to discuss the following hypotheses: 1)  photons polarized in different directions (differently oriented segments) and 2) photons polarized at all angles (stars), that were proposed by them during the class discussion. The possibility that photons are not polarized was added by the instructor at the end of the discussion, but no student considered it. Since the discussion of quantum uncertainty and the stochastic interpretation of Malus's law were scheduled only after the task, assessing \textit{hypothesis 1} was much more difficult than assessing \textit{hypothesis 2}. With relation to the latter, some students associated the intensity not with the number of photons, but with its polarization. Therefore, we revised the previous part of the course on the introduction of light quanta, adding that in all considered cases the intensity of a light beam is not dependent on the polarization of its photons. Only two out of 39 students correctly completed the activity. This was not surprising, as it represented the first instance of theoretical testing experiment, an unfamiliar task for students.

\begin{figure}[!htpb]
    \includegraphics[width=\columnwidth]{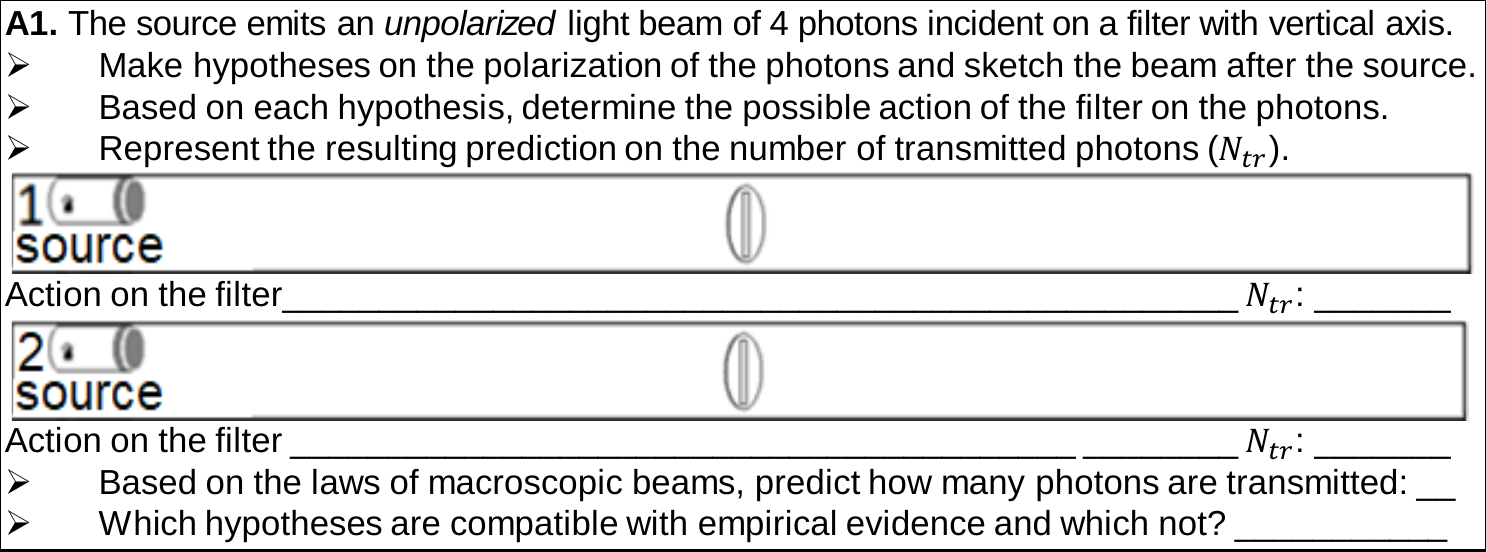}
    \caption{Worksheet block: unpolarized light in terms of photons, Liceo Galilei.}
    \label{FIG:42}
\end{figure}

\emph{Final Version - Description of the activity}. It corresponds to activity 1.5 in Fig. \ref{FIG:9}, and is displayed in Fig. \ref{FIG:42}.
In view of need to show students how to run a theoretical testing experiment, we revised the text by clearly articulating the phases of such a procedure. In addition, we weakened the conditions of acceptance of a hypothesis, replacing the mathematical expression ``satisfy the experimental results'' with a more qualitative one (``are compatible with empirical evidence'').  The activity was experimented in Liceo Galilei, Trieste, 2019 (18 attending the lesson).

\emph{Final Version - Results}.
The analysis of the assessment of each hypothesis is based on the use of the scientific abilities described in Section \ref{Sec:3.4.3}. Ability (a) is successfully used if the assumption on the action of the filter with vertical axis is compatible with the hypothesis: for photons polarized in all directions, elimination of all polarization properties that differ from $90^{\circ}$; for unpolarized photons, addition of a polarization property at $90^{\circ}$; for photons polarized in different directions, if the answer is compatible with Malus's law for polarized beams. Ability (b) is successfully used if the prediction on the number of transmitted photons it is coherent with the assumption on the role of the filter (regardless of its consistency). Ability (c) is successfully used if the conclusion on each hypothesis is coherent with the assumption and the prediction, and uses as empirical term of comparison either the reduction to half or its qualitative version (reduction of the number of photons).

\begin{figure}[!ht]
    \includegraphics[width=\columnwidth]{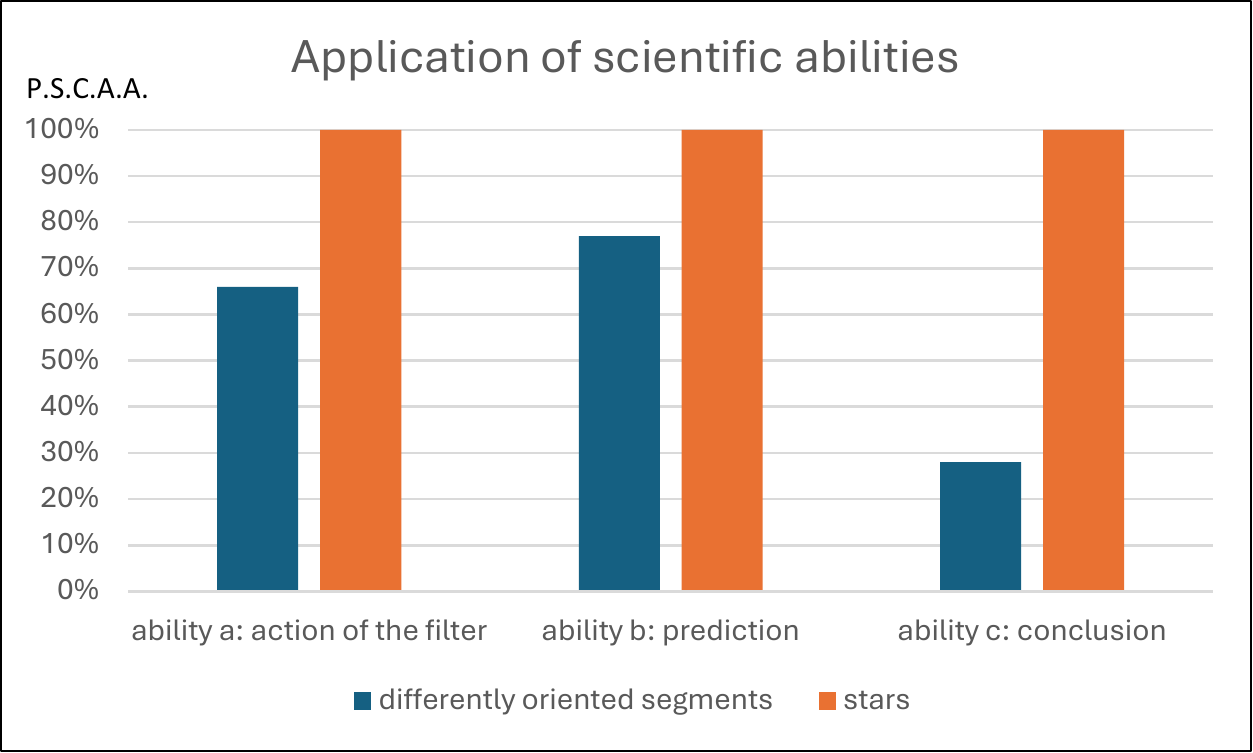}
    \caption{Consistent use of the abilities, Liceo Galilei. In this graph and in the next ones, P.S.C.A.A. means percentage of students consistently applying an ability.}
    \label{FIG:41}
\end{figure}

Results are shown in Fig. \ref{FIG:41}.
Also in this case, all students opted for discussing the same two hypotheses as in Perugia. However, the self-selected students of Liceo Galilei achieved much better results than regular classrooms.  With the information on the lack of correlation between intensity and polarization, the issue with \textit{hypothesis 2} was removed: all students consistently assessed the hypothesis. As to \textit{hypothesis 1}, 66\% of them successfully used ability (a) vs. 21\% in Perugia. Notably, while we had not mentioned uncertainty and probability before in the course, half of them spontaneously adopted a probabilistic approach to Malus's law: ``if they are not vertical, there is a certain probability'', ``photons may be stochastically transmitted or not.'' Others, while not mentioning probability, still displayed a global approach to the application of the law in terms of photons: ``photons pass or not based on the angular difference.'' The only issue with ability (a) was the idea that photons are transmitted exclusively if their polarization is identical to the axis of the filter. Such a condition is way too restrictive (infinitesimal), but might explain why some of these students wrote that no photon is transmitted by the filter. In general, around 30\% of the students correctly completed the activity. An additional 11\% successfully used ability (a) and (b), but left the section on the conclusion blank.

\emph{Future improvements}. Given the difficulties in using ability (b) and (c) to assess \textit{hypothesis 1}, we plan to focus only on qualitative aspects: we ask if, based on each hypothesis, there can be or not a reduction in the number of photons as a result of the interaction with the filter. Since we want students to assess all three possible hypotheses, we plan to add to the worksheet block another space for \textit{hypothesis 3}: unpolarized photons.

\subsubsection{Interpreting already known laws within the framework of new models: Malus's law} \label{Sec:5.4.1}
\emph{Goals}. a) Helping students develop a single photon interpretation of the law of Malus as a stochastic law of transition from the initial polarization property to that coinciding with the axis of the filter; b) since in the subsequent activities on the revision of measurement, filters have been replaced by crystal $+$ counters, supporting students in transferring the calculation of the transition probability from the first context to the second one.

\emph{Path of Refinement}. The revision of this activity spanned from 2014 to 2017, and was performed in each edition of the Summer School of Excellence. The activity started as a predict-observe-explain sequence \cite{White1992} with simulated experiments in the JQM environment, followed by an interpretive question on its results in terms of single photons. As the sequence asked predictions about the interaction of beams of ten photons polarized at $45^{\circ}$ with a filter with axis at $0^{\circ}$, a large majority of students interpreted the law as statistical and strictly related to the specific case at hand. The wording of the interpretive question had not been effective 1) in directing student attention to the single photon and 2) in helping them develop a general perspective on the law. We decided to strictly adhere to a single-photon ontology (see Section \ref{Sec:3.5.1}), directing student attention exclusively to interactions of one photon with a filter (``horoscope of the photon''), and structuring a clear path of generalization. In 2016, the replacement of filters with crystals described in Section \ref{Sec:5.2.1} highlighted a further issue that had not come to light before: the difficulty to transfer the calculation of the transition probability from the interaction of a photon with a filter to that with a crystal $+$ counters. The key to overcome this difficulty was identified with the following approach: encouraging students to focus only on the unifying features of the interactions between the photon and both devices (filter $+$ counter and crystal $+$ counters). These are: 1) the fact that most interactions have uncertain outcome, but in two special cases the outcome is determinate; 2) the existence of a transition depending only on the difference between the initial angle of polarization and the final one. The revision of the worksheet blocks was performed accordingly. Most importantly, we eliminated the use of simulated experiments in JQM, converting the activity a into a purely theoretical form of inquiry.

\begin{figure*}[!htpb]
       \includegraphics[width=\textwidth]{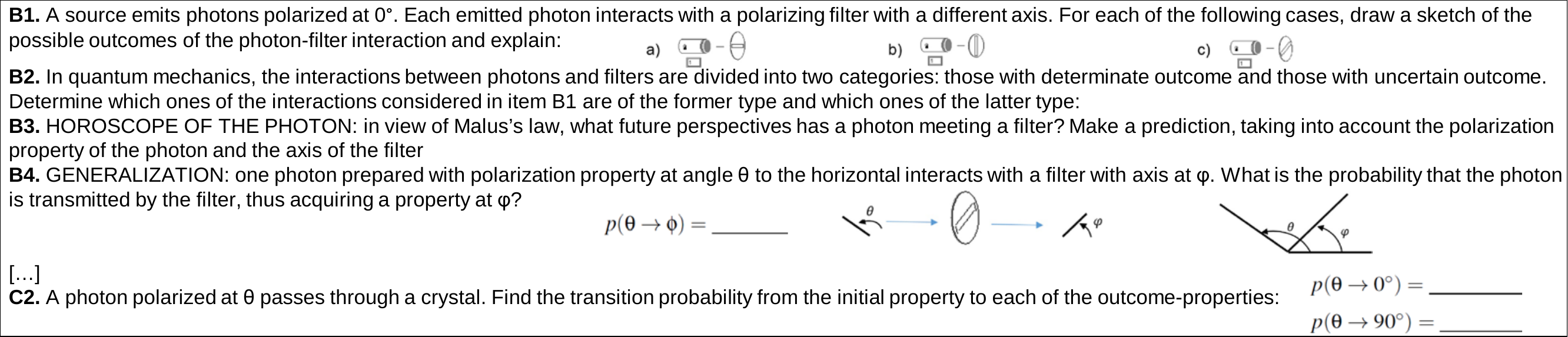}
    \caption{Worksheet block on the interpretation of Malus's law, and on its transfer to the context of calcite crystals.}
    \label{FIG:36}
\end{figure*}

\emph{Final Version - Description of the activity}. The tasks correspond to activities 1.6-1.7, and part of 1.9 in Fig. \ref{FIG:9}, and are displayed in Fig. \ref{FIG:36}.
The introductory items (\textbf{B1}-\textbf{B2}) examine the interaction of a single photon with a filter, discussing from the start the fundamental dichotomy between ``interactions with a determinate outcome'' and ``interactions with an uncertain outcome.'' After the Horoscope of the photon (\textbf{B3}), we added a specific item (\textbf{B4}) designed to generalize its results to the case of an arbitrary initial angle ($\theta$) and final angle ($\phi$) after the interaction, focusing exclusively on angles and not on the device. As regards the calculation of transition probabilities in the context of calcite crystals, in item \textbf{C2}, beams were replaced by a single photon.

\begin{figure}[!htpb]
    \includegraphics[width=\columnwidth]{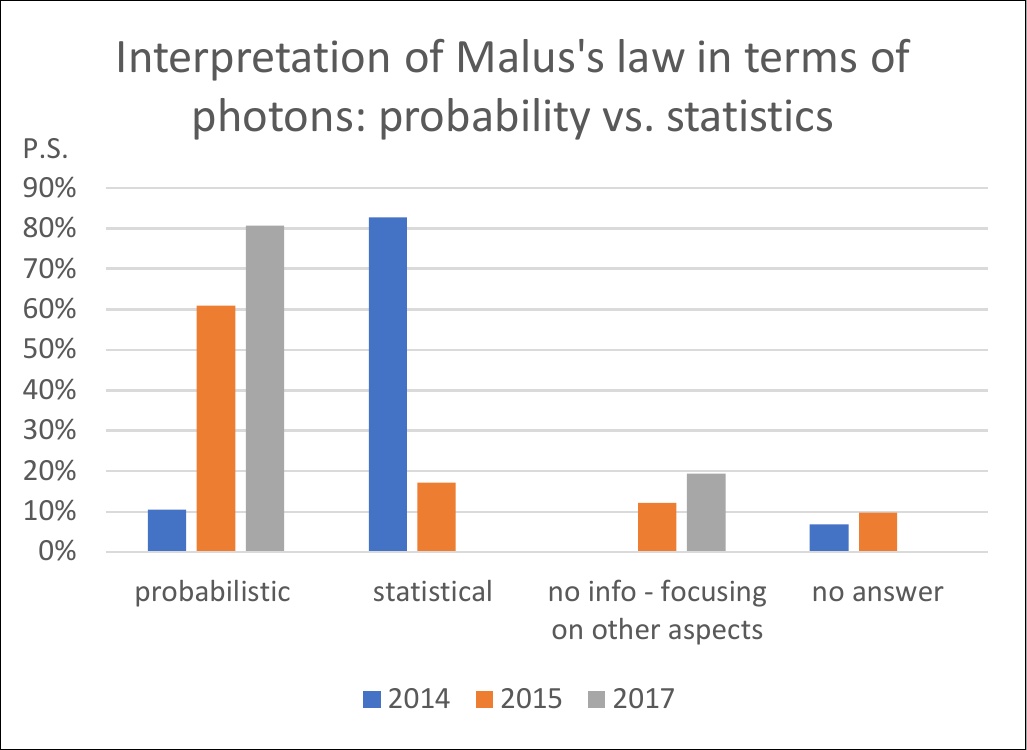}
    \caption{Comparison table: statistical vs. probabilistic.}
    \label{FIG:37}
\end{figure}

\begin{figure}[!htpb]
    \includegraphics[width=\columnwidth]{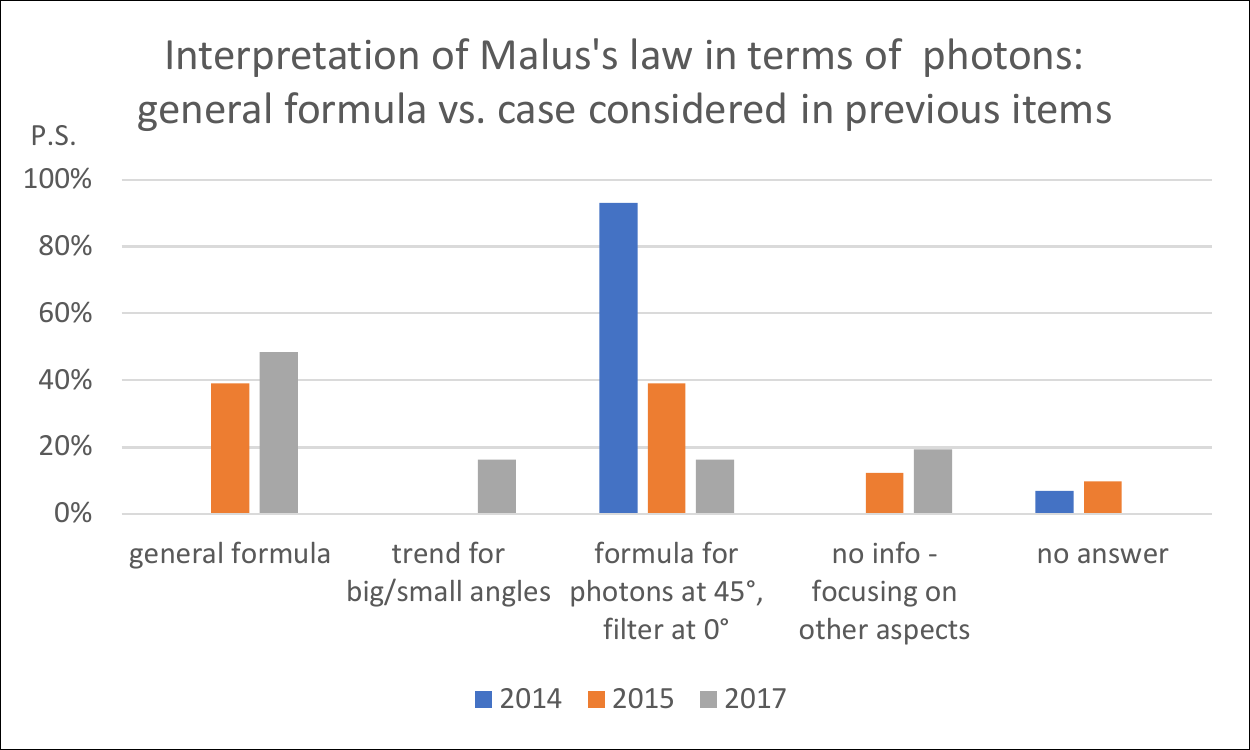}
    \caption{Comparison table: local vs. global reasoning.}
    \label{FIG:38}
\end{figure}

\emph{Final Version - Results}. Data of 2017 on the Horoscope of the photon (\textbf{B3}) are displayed in comparison with results of the equivalent item in the design experiments of 2014 and 2015.
Specifically, we compare two different dimensions of student reasoning on Malus's law: 1) the alternative between a probabilistic and a statistical interpretation (Fig. \ref{FIG:37}); 2) local reasoning vs. global reasoning, i.e., focusing on the transition probability for an angle of $45^{\circ}$ between the initial property and the outcome-property or on the general formula (Fig. \ref{FIG:38}).
Since we promote a probabilistic interpretation and a global form of reasoning, we see that in both respects there has been a steady improvement over the years. After administering the Horoscope, generalizing the formula of Malus's law to arbitrary angles (\textbf{B4}) was a trivial task: all but one student answered correctly. This means that, if we consider the two items as components of a single task on the interpretation of Malus's law, virtually all students developed a consistent understanding of the topic. By using this sequence of items and, in particular, by discussing transition in terms of the angle between the initial property and the outcome-property, question \textbf{C2} on the transition probability in the context of crystals became straightforward. All students correctly answered the item, even if there was a long teaching/learning session in between.

\subsubsection{Thought experiment: Superposition as statistical mixture of component states?} \label{Sec:5.4.3}

\emph{Goals}. Distinguishing between superposition states and mixed states. Developing an awareness of the implications of interpreting a superposition as a mixture.

\emph{Path of Refinement}. For the design of the activity, see Section \ref{Sec:3.4.4}. The first version was administered in Liceo Alessi, Perugia, 2019. The activity started with questions on the implications of the mixture hypothesis: measurement on a single system is deterministic, the observable to be measured is definite, the use of probability is due to lack of knowledge about the state of the system. For instance: ``What are the implications of the hypothesis on the nature of the interaction of each photon with the device that measures the $0,90$ observable: is it deterministic or stochastic?''. Issues were evidenced in the interpretation of the item, with some students focusing on beams instead of the single photon, and thus coming to the opposite conclusion: measurement is probabilistic, the $0,90$ observable is indefinite. Most issues however, concerned the last item, asking to design an effective thought experiment to assess the hypothesis.
For instance, some students either used a filter at $0^{\circ}$ or a crystal at $0^{\circ}$ and $90^{\circ}$, thus coming to the conclusion that the hypothesis is confirmed: the testing experiment does not work because it is identical to the hypothesis itself. On the positive side, others consistently completed the task by using a logical consequence of the hypothesis: the fact that if it is true, only photons polarized at $0^{\circ}$ and $90^{\circ}$ should exist. The idea was rejected by having the beam interact with arbitrarily oriented filters. We decided to take advantage of productive reasoning and to provide more content support both verbally and in the items text, also with the addition of visual clues.

\begin{figure*}[!htpb]
       \includegraphics[width=\textwidth]{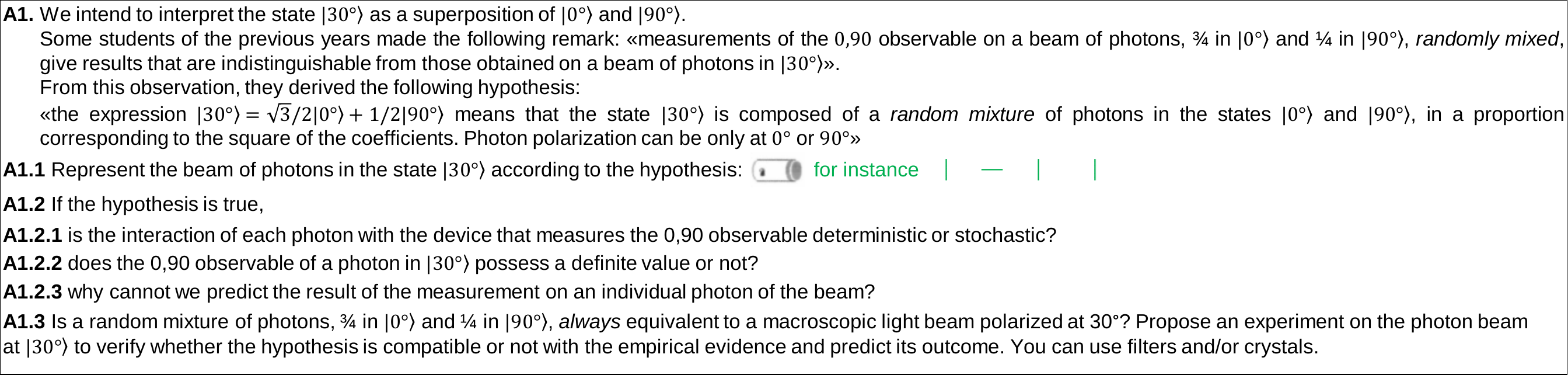}
    \caption{Worksheet block on the interpretation of quantum superposition: Liceo Galilei, Trieste, 2019.}
    \label{FIG:45}
\end{figure*}

\emph{Final Version - Description of the activity}. The task corresponds to activity 3.2 in Fig. \ref{FIG:9}, and is displays in Fig. \ref{FIG:45}. It was administered in Liceo Galilei, 2019 (16 students attending the lesson).
A logical consequence was added to the formulation of the mixture hypothesis (``photons are polarized only at $0^{\circ}$ or $90^{\circ}$''), which has been productive for some students. The new item \textbf{A1.1} provides a visual support for this statement. In \textbf{A1.1.1} (now \textbf{A1.2.1}), we added a verbal prompt for focusing on the single photon, asking students to think back to the initial task on measurement performed on mixtures of photons already prepared at 0° and 90°, which is isomorphic to the present case. Finally, we explained to students that the design and conduction of the thought experiment tests the validity of the hypothesis in the measurement of observables different from $0,90$.

\emph{Final Version - Results}. All the students consistently answered the first three questions, and 75\% of them successfully ran the thought experiment, most by using a filter with axis at $30^{\circ}$. The only student who designed an experiment with a crystal gave a very detailed answer: ``I rotate the crystal by $30^{\circ}$, so that the channels are at $30^{\circ}$, $120^{\circ}$. In this case, the property of a beam polarized at $30^{\circ}$ is one of the outcome-properties $\Rightarrow$ certain result. Instead, not all photons of the random mixture are transmitted, the process is stochastic. False.'' Of the 4 students who did not answer consistently, two designed reliable experiments that tested the hypothesis, but did not make a correct prediction; one proposed an experiment for measuring the $0^{\circ}$,$90^{\circ}$ observable, thus confirming the hypothesis. The last one left the answer blank.

\subsubsection{Identifying and interpreting mathematical constructs for describing physical situations and deriving new results: Entangled superposition} \label{Sec:5.4.4}

\emph{Goals}. Identifying and interpreting the mathematical expression of an entangled superposition of modes.

\begin{figure*}[!htpb]
       \includegraphics[width=\textwidth]{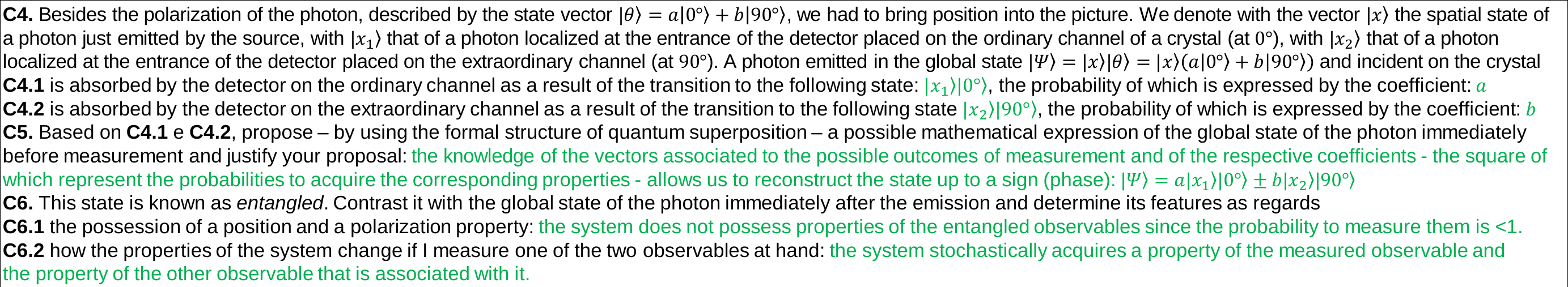}
    \caption{Worksheet block on the derivation and interpretation of entangled superposition: Liceo Galilei, Trieste, 2019.}
    \label{FIG:47}
\end{figure*}

\emph{Description of the activity}. The tasks correspond to activities 4.7, 4.8 in Fig. \ref{FIG:9}, and are displayed in in Fig. \ref{FIG:47}.
After two oral questions on the mathematization of two special cases (\textbf{C4.1}, \textbf{C4.2}), we ask students to model the general case in a written item (\textbf{C5}). Two interpretive items on the proposed mathematical model follow (\textbf{C6.1}, \textbf{C6.2}). They were experimented for the first time in Liceo Galilei, Trieste, 2019 (17 students attending the lesson).

\emph{Results of the activity}. 76\% of the students consistently answered \textbf{C5}), proposing a superposition state that is compatible with the situation at hand: $a|x_1\rangle|0^{\circ}\rangle+b|x_2\rangle|90^{\circ}\rangle$. Another student wrote a similar expression, but using the square of the coefficients: $a^2|x_1\rangle|0^{\circ}\rangle+b^2|x_2\rangle|90^{\circ}\rangle$. More than half of them added consistent explanations, either focusing on the interpretation of the components and coefficients of the superposition (``the state of the photon is a superposition of the state corresponding to $0^{\circ}$, that is $|x_1\rangle|0^{\circ}\rangle$ and the state corresponding to $90^{\circ}$, that is $|x_2\rangle|90^{\circ}\rangle$. The probability to find the photon in that states are $a^2$ and $b^2$''), or on the change in state from the initial situation to the final one (``we do not have $|x\rangle|\theta\rangle$ anymore because the photon is after the crystal, and since it is only probabilistic, both outcomes must be included [in the superposition].''). Inconsistent answers included students who wrote separable states: $(1/\sqrt{2}|0^{\circ}\rangle+1/\sqrt{2}|90^{\circ}\rangle)(1/\sqrt{2}|x_1\rangle+1/\sqrt{2}|x_2\rangle)$, or $|x\rangle (1/2|0^{\circ}\rangle+1/2|90^{\circ}\rangle)$.
Negative transfer from the context of the hydrogen-like atom affected both this question and \textbf{C6.2}. See, e.g., a very inconsistent expression in terms of quantum numbers: $a^2b^2|n_1+n_2, l_1+l_2, m_1+m_2, s_1+s_2\rangle$.

Moving on to examine the interpretive task on the possession of properties of position and polarization before measurement (\textbf{C6.1}), also in this case a large majority of students gave consistent answers (71\%). These students displayed different forms of productive reasoning: some used the definition of the possession of a property, others linked superposition to uncertainty, or focused on the change in state from the initial situation to the final one. Some students also added that position and polarization properties are compatible and correlated. As we talked about compatibility between position and spin properties only in Unit 1 (see Fig. \ref{FIG:9} and Section \ref{Sec:5.3.2} for data), this is a case of productive transfer from the context of the hydrogen-like atom. The remaining students either said that the system possesses the involved properties, or focused on the relation between the two observables.

The last item (\textbf{C6.2}) asked how system properties change if we measure one of the two observables at hand. This question was by far the least successful of the three:  only 47\% of the students gave consistent answers. In general, these answers were very concise: ``by measuring one observable, we acquire both properties.'' One students highlighted the correlation between the two observables: ``I acquire a property and the corresponding property of the other observable.'' Another one productively referred to the discussion of Experience 2: ``if I measure position, I also measure polarization and vice versa.'' Inappropriate transfer from the context of the hydrogen-like atom affects 75\% of the inconsistent answers: ``nothing changes, as the properties of position and spin are compatible.'' In a future version, we could add a reference to Experience 2, which concerns only the photon and not the atom.


\subsection{Epistemological debates} \label{Sec:5.5}
Since epistemological debates are addressed in a whole class discussion without the use of worksheets or other written assignments, their refinement could be based only on the instructor's diary.

Here we limit ourselves to report on the revision of the first debate on the problem of indefiniteness and uncertainty. As described in Section \ref{Sec:3.5.2}, in the initial versions of the course, we discussed the Heisenberg's microscope thought-experiment and Bohr's criticism of it in Unit 1, after the application of the relations between properties to the case of position and velocity. Students were generally at ease with an interpretation of uncertainty as caused by measurement disturbance, that they could reconcile with their classical intuition on point-like particles. However, when this view was questioned, raising the possibility that uncertainty is an intrinsic property of quantum systems, some students clearly showed their discomfort. In Thiene, one of the best performing students explicitly complained that, if that was the case, we should conclude that QM is an absurd theory and makes no sense. Up until then, she had taken active part to all the worksheet activities and to the whole class discussions that ensued. After that, and for the rest of the lesson, the level of her engagement significantly declined.

In the retrospective analysis, we considered the possibility to add more content support to this activity, including an anticipation of the discussion on the wave-particle duality. However, this would have subverted the structure of the course which, in agreement with the gradual construction of content in spin-first approaches \cite{Zuccarini2020} and textbooks written in collaboration with physics education researchers \cite{McIntyre2012}, scheduled the discussion of propagation only after a careful examination of the system at a point in time and of its behavior in measurement.

For this reason, we opted for providing students with an operative idea of indefinite quantity, that could give empirical meaning to this situation and be immediately connected with the now familiar context of polarization: ``a quantity of a system is called indefinite when the ideal measurement of this quantity on a large ensemble of identical systems gives different results according to a probabilistic distribution'' (see Fig. \ref{FIG:9}, activity 1.14). Of course, this begged the question of how to establish whether two systems are identical according to QM. Therefore, we told students that this would have been the driving question of the next unit since, in order to give a reasonable answer, we would have needed the concept and the formal representation of the quantum state.

The discussion of Heisenberg's microscope was moved to Unit 4, activity 4.5, where, based on the adoption of a field ontology, it was possible to contemplate the idea that quantum uncertainty is due to a measurement disturbance, and to reject it without regret.

With this revision, the discussion of the issue at hand did not cause any visible discomfort.

\section{Conclusions}

Learning quantum mechanics involves overcoming manyfold challenges, among which knowledge revision, fragmentation, and the construction of a personally plausible picture of the quantum model. Extensive investigations of cognitive and epistemological challenges, as well of the use of student resources and intuition in the discussion of introductory science topics have been conducted in the framework of CC \cite{Vosniadou2008, diSessa2014}, leading to the development of effective approaches to science teaching \cite{Amin2014}. Also in the case of QM, various researchers have considered the problem of teaching the subject as the design of strategies to effectively promote a CC in individual learners \cite[e.g.,][]{Thagard1992, Kalkanis2003, Tsaparlis2013, Malgieri2017}. In general, the progress from a classical to a quantum perspective requires changes in knowledge structures about a scientific theory, which have been developed as a result of instruction. Since educational models of CC have been introduced to account for a different kind of change (the transition from na\"{\i}ve to scientific knowledge), there is a need to revise the framework, examining the differences involved in learning a successive theory. In addition, while studies on CC are undergoing a systemic turn \cite{Amin2014}, there is a lack of educational proposals on QM that coordinate multiple interacting aspects at different levels of analysis. Last, in accordance with the shift in focus from difficulties to resources \cite{Goodhew2019}, there is a need to identify the links between student intuition and strategies for learning QM.

In this article, we describe the development of a course for secondary school that is based on an examination of CC in the transition from CM to QM, and integrates cognitive and epistemological aspects including the adoption of epistemic practices of theoretical physicists, and a structured approach to interpretive themes. The aim is to help fill the gaps in analysis and design, addressing all the challenges mentioned at the beginning of this section, and taking advantage of available knowledge elements and intuition in the task.

This systemic approach led to the derivation of the following design principles: \emph{Principle of Knowledge Revision}, \emph{Principle of Knowledge Organization}, \emph{Epistemic Principle}, and \emph{Epistemological Principle}.

The first one relies on the analysis of continuity and change in basic concepts and constructs to promote the understanding of their quantum counterparts and the ability to discriminate between aspects of the old and the new notions, identifying their correct context of application. The instruments used in this process suggest strategies to leverage prior knowledge according to specific patterns of change in the trajectory of each notion.

The second principle concerns the development of conceptual tools denoted as \emph{relations between properties}, that act as organizing elements in the construction of a unifying picture of quantum measurement across contexts, and help address student's need of comparability with CM by offering interpretive keys for the shift from the classical to the quantum task.

The third one proposes to design the course around a modelling process that includes epistemic practices of the theoretical physicist, with the goal to help students accept the quantum description of the world as a plausible and reliable product of their own inquiry.

The last principle proposes to work in the context of a clearly specified form of interpretation, so as to identify and discuss the facets of the foundational debate that are triggered by each choice, with the aim to help students develop an awareness of the cultural significance of the debate, of the limits the chosen stance, of the open issues.

The sequence has been informed by the interplay of the first three principles, and has been developed in the template of the model of modelling \cite{Gilbert2016}, a perspective devised to analyze the process of modelling in science education. The result is a model that starts from the description of a property of an object (photon polarization), and is extended and revised through a process conducted by means of theoretical epistemic practices (\emph{Epistemic Principle}), gradually incorporating quantum measurement, state, superposition, propagation and entanglement (\emph{Principle of Knowledge Revision}). Thanks to the \emph{Principle of Knowledge Organization}, each step allows students to advance in parallel in the development of an elementary model of the hydrogen-like atom.

To illustrate in full the implementation of the principles, we show how each of them guided us in the design of a set of individual activities. A special attention is devoted to the conversion of epistemic practices into active learning strategies: inquiry-based learning \cite{Llewellyn2012}, the model of modelling, and different research perspectives on the role of mathematics in physics \cite{Uhden2012, Redish2015} converge to structure the chain of activation used for mathematical modelling in a purely theoretical context. Then, these frameworks are blended together
with aspects of the ISLE learning system, such as the conduction of testing experiments and the rubrics of scientific abilities \cite{Etkina2006, Etkina2015}, to convert thought experiments into theoretical inquiry activities. Last, we show how the \emph{Epistemological Principle} guided us to strengthen the coherence of the course and to design the discussion of epistemological themes.

The course is presented in Section \ref{Sec:4}, which includes an outline of its structure, of the types of activities that are designed to implement the principles, of the instruments and methods. A bird's eye view of the sequence and the types of activities is provided in Fig. \ref{FIG:9}.

The second part of the article describes the cycles of refinement of the same set of activities previously discussed. In this work, we do not test the global effectiveness of the course, but show how a wide range of different inquiry and modelling activities has been made effective in addressing the challenges at a local level, thanks to the cycles of preparation, experiment, and analysis of the learning outcomes conducted in the framework of design-based research \cite{Bakker2015}. This concerns in particular the refinement of theoretical epistemic activities, that have become innovative and effective forms of inquiry for engaging students in the development of theoretical skills (e.g., generating and/or running thought experiments).

Future directions include the analysis of a pre-post-test administered in regular classrooms, in order to evaluate
the effectiveness of the course as a whole.


\begin{thebibliography}{99}
\bibitem{Zhu2012} G. Zhu and C. Singh, Improving students' understanding of quantum measurement. I. Investigation of difficulties, Phys. Rev. ST Phys. Educ. Res., \textbf{8}, 010117 (2012).
\bibitem{Ayene2011} M. Ayene, J. Kriek, and D. Baylie, Wave-particle duality and uncertainty principle: Phenomenographic categories of description of tertiary physics students' depictions, Phys. Rev. ST Phys. Educ. Res., \textbf{7}, 020113 (2011).
\bibitem{Pospiech2021} G. Pospiech, A. Merzel, G. Zuccarini, E. Weissman, G. Katz, I. Galili, L. Santi, and M. Michelini, The role of mathematics in teaching quantum physics at high school, in \textit{Teaching-Learning Contemporary Physics: From Research to Practice}, edited by B. Jarosievitz and C. S\"{u}k\"{o}sd (Springer Nature Switzerland AG, Cham, Switzerland, 2021), pp. 47-70.
\bibitem{Passante2015} G. Passante, J. Emigh, and P.S. Shaffer, Student ability to distinguish between superposition states and mixed states in quantum mechanics. Phys. Rev. ST Phys. Educ. Res., \textbf{11}, 020135 (2015).
\bibitem{Vosniadou2008} S. Vosniadou, The framework theory approach to the problem of conceptual change, in  \textit{International handbook of research on conceptual change, 1st edition}, edited by S. Vosniadou (Routledge, New York and London, 2008), pp. 3-34.
\bibitem{diSessa2014} A.A. diSessa, A history of conceptual change research: Threads and fault lines, in \textit{The Cambridge handbook of the learning sciences, Second edition}, edited by R. K. Sawyer (Cambridge University Press, 2014), pp. 265-281.
\bibitem{Johnston1998} I.D. Johnston, K. Crawford, and P.R. Fletcher, Student difficulties in learning quantum mechanics, Int. J. Sci. Educ., \textbf{20}, 427-446 (1998).
\bibitem{Marshman2015} E. Marshman and C. Singh, Framework for understanding the patterns of student difficulties in quantum mechanics, Phys. Rev. ST Phys. Educ. Res. \textbf{11}, 020119 (2015).
\bibitem{Malgieri2017} M. Malgieri, P. Onorato, and A. de Ambrosis, Test on the effectiveness of the sum over paths approach in favoring the construction of an integrated knowledge of quantum physics in high school, Phys. Rev. Phys. Educ. Res. \textbf{13}, 010101 (2017).
\bibitem{Griffiths2018} D.J. Griffiths, and D.F. Schroeter, \textit{Introduction to quantum mechanics}, Third Edition (Cambridge University Press, Cambridge, United Kingdon, 2018).
\bibitem{Ravaioli2017} G. Ravaioli and O. Levrini, Accepting quantum physics: Analysis of secondary school students' cognitive needs, in \textit{Electronic Proceedings of the ESERA 2017 Conference. Research, Practice and Collaboration in Science Education, Part 2}, edited by O. Finlayson, E. McLoughlin, S. Erduran, and P. Childs (2017). \url{https://www.dropbox.com/s/t1ri3ql7ufpihun/Part_2_eBook.pdf?dl=0}
\bibitem{Posner1982} G.J. Posner, K.A. Strike, P.W. Hewson, and W.A. Gertzog, Accomodation of a scientific conception: Toward a theory of conceptual change, Sci. Educ., \textbf{66}(2), 211-227 (1982).
\bibitem{Lewerissa2017} K. Krijtenburg-Lewerissa, H.J. Pol, A. Brinkman, and W.R. Van Joolingen, Insights into teaching quantum mechanics in secondary and lower undergraduate education, Phys. Rev. Phys. Educ. Res., \textbf{13}, 010109 (2017).
\bibitem{Wittmann2020} M.C. Wittmann and J.T. Morgan, Foregrounding epistemology and everyday intuitions in a quantum physics course for nonscience majors, Phys. Rev. Phys. Educ. Res. \textbf{16}, 020159 (2020).
\bibitem{Baily2015} C. Baily and N.D. Finkelstein, Teaching quantum interpretations: Revisiting the goals and practices of introductory quantum physics course, Phys. Rev. ST Phys. Educ. Res. \textbf{11}, 020124 (2015).
\bibitem{Coppola2013} P. Coppola and J. Krajcik, Discipline-centered post-secondary science education research: Understanding university level science learning, J. Res. Sci. Teach., \textbf{50}, 627-638 (2013).
\bibitem{Goodhew2019} L.M. Goodhew, A.D. Robertson, P.R.L. Heron, and R.E. Scherr, Student conceptual resources for understanding mechanical wave propagation, Phys. Rev. Phys. Educ. Res. \textbf{15}, 020127 (2019).
\bibitem{Dreyfus2017} B.W. Dreyfus, A. Elby, A. Gupta, and E.R. Sohr, Mathematical sense-making in quantum mechanics: An initial peek, Phys. Rev. Phys. Educ. Res. \textbf{13}, 020141 (2017).
\bibitem{Dini2017} V. Dini and D. Hammer, Case study of a successful learner's epistemological framings of quantum mechanics, Phys. Rev. Phys. Educ. Res. \textbf{13}, 010124 (2017).
\bibitem{Kuhn1962} T. Kuhn, \textit{The Structure of Scientific Revolutions} (University of Chicago Press, Chicago, 1962).
\bibitem{Thagard1992} P. Thagard, \textit{Conceptual revolutions}, (Princeton University Press, Princeton, New Jersey, 1992).
\bibitem{Tsaparlis2009} G. Tsaparlis and G. Papaphotis, High school Students' Conceptual Difficulties and Attempts at Conceptual Change: The case of basic quantum chemical concepts, Int. J. Sci. Educ. \textbf{31}, 895-930 (2009).
\bibitem{Singh2015} C. Singh and E. Marshman, Review of student difficulties in upper-level quantum mechanics, Phys. Rev. ST Phys. Educ. Res., \textbf{11}, 020117 (2015).
\bibitem{Potvin2020} P. Potvin, L. Nenciovici, G. Malenfant-Robichaud, F. Thibault, O. Sy, M.-A. Mahhou, A. Bernard, G. Allaire-Duquette, J. M. Blanchette Sarrasin, L. M. Brault Foisy, et al., Models of conceptual change in science learning: Establishing an exhaustive inventory based on support given by articles published in major journals. Stud. Sci. Educ., 56(2), 157–211 (2020).
\bibitem{Tsaparlis2013} G. Tsaparlis, Learning and teaching the basic quantum chemical concepts, in \textit{Concepts of matter in science education}, edited by G. Tsaparlis and H. Sevian (Springer, Dordrecht, 2013), pp. 437-460.
\bibitem{Kalkanis2003} G. Kalkanis, P. Hadzidaki, and D. Stavrou, An instructional model for a radical conceptual change towards quantum mechanics concepts, Sci. Educ. \textbf{87}, 257-280 (2003).
\bibitem{Amin2014} T.G. Amin, C. Smith, and M. Wiser, Student conceptions and conceptual change: Three overlapping phases of research, in \textit{Handbook of research on science education, Volume II}, edited by N. G. Lederman and S. K. Abell (Routledge, New York, 2014), pp. 71-95.
\bibitem{Carey1999} S. Carey, Sources of Conceptual Change, in \textit{Conceptual development: Piaget's legacy}, edited by E. K. Scholnick, K. Nelson, S. A. Gelman, P. H. Miller (Lawrence Erlbaum Associates Publishers, 1999), pp. 293-327.
\bibitem{Chi2013} M. Chi, Two Kinds and Four Sub-Types of Misconceived Knowledge, Ways to Change It, and the Learning Outcomes, in \textit{International Handbook of Research on Conceptual Change, 2nd edition}, edited by S. Vosniadou (Routledge, New York and London, 2013), pp. 49-70.
\bibitem{Zuccarini2022} G. Zuccarini and M. Malgieri, Modeling and Representing Conceptual Change in the Learning of Successive Theories, Sci. Educ. (Dordr.), (2022). \href{https://doi.org/10.1007/s11191-022-00397-1}{DOI:10.1007/s11191-022-00397-1}
\bibitem{Arabatzis2020} T. Arabatzis, What are scientific concepts?, in \textit{What is scientific knowledge? An introduction to Contemporary Philosophy of Science}, edited by K. McCain and K. Kampourakis (Routledge, New York and London, 2020), pp. 85-99.
\bibitem{Hoyningen1993} P. Hoyningen-Huene, \textit{Reconstructing scientific revolutions: Thomas S. Kuhn's philosophy of science} (University of Chicago Press, Chicago, 1993).
\bibitem{Andersen2006} H. Andersen, P. Barker, and X. Chen, \textit{The cognitive structure of scientific revolutions} (Cambridge University Press, Cambridge, 2006).
\bibitem{Potvin2015} P. Potvin, \'{E}. Sauriol, and M. Riopel, Experimental evidence of the superiority of the prevalence model of conceptual change over the classical models and repetition, J. Res. Sci. Teach., \textbf{52}, 8, 1082–1108 (2015).
\bibitem{Henderson2017} J. B. Henderson, E. Langbeheim, and M.T. Chi, Addressing robust misconceptions through the ontological distinction between sequential and emergent processes, in \textit{Converging perspectives on conceptual change: Mapping an emerging paradigm in the learning sciences}, edited by T. G. Amin and O. Levrini (Routledge, 2017), pp. 26–33.
\bibitem{Ballentine1994} L.E. Ballentine, Y. Yang, and J.P. Zibin, Inadequacy of Ehrenfest's theorem to characterize the classical regime, Phys. Rev. A, \textbf{50}, 4 (1994).
\bibitem{Klein2012} U. Klein, What is the limit $\hslash \to 0$ of quantum theory?, Am. J. Phys., \textbf{80}, 1009 (2012).
\bibitem{Stadermann2019} H. K. E. Stadermann, E. van den Berg, and M. J. Goedhart, Analysis of secondary school quantum physics curricula of 15 different countries: Different perspectives on a challenging topic, Phys. Rev. ST Phys. Educ. Res., \textbf{15}, 010130 (2019).
\bibitem{Zuccarini2020} G. Zuccarini, Analyzing the structure of basic quantum knowledge for instruction, Am. J. Phys., \textbf{88}, 385 (2020).
\bibitem{diSessa1998} A. A. diSessa and B. Sherin, What changes in conceptual change?, Int. J. Sci. Educ., \textbf{20}, 1155 (1998).
\bibitem{diSessa2016} A. A. diSessa, B. Sherin, and M. Levin, Knowledge analysis: An introduction, in {Knowledge and interaction: A synthetic agenda for the learning sciences}, edited by A. A. diSessa, M. Levin, and N. J. S. Brown  (Routledge, New York, 2016), pp. 30-61.
\bibitem{Levrini2008} O. Levrini and A. A. diSessa, How students learn from multiple contexts and definitions: Proper time as a coordination class, Phys. Rev. ST Phys. Educ. Res., \textbf{4}, 010107 (2008).
\bibitem{Debianchi2011} M. Sassoli De Bianchi, Ephemeral properties and the illusion of microscopic particles, Found. Sci.,  \textbf{16}, 393 (2011).
\bibitem{Elby2016} A. Elby, C. Macrander, and D. Hammer, Epistemic cognition in science, in  \textit{Handbook of Epistemic cognition}, edited by J.A. Greene, W.A. Sandoval, and I. Br{\aa}ten (Routledge New York, NY, 2016), pp. 113-127.
\bibitem{Sandoval2016} W.A. Sandoval, J.A. Greene, and I. Br{\aa}ten, Understanding and promoting thinking about knowledge: Origins, issues, and future directions of research on epistemic cognition, Rev. Res. Educ., \textbf{40}, 457-496 (2016).
\bibitem{Uhden2012} O. Uhden, R. Karam, M. Pietrocola, and G. Pospiech, Modelling Mathematical Reasoning in Physics Education, Sci. Educ. (Dordr.), \textbf{21}, 485-506 (2012).
\bibitem{Redish2015} E. F. Redish and E. Kuo, Language of Physics, Language of Math: Disciplinary Culture and Dynamic Epistemology, Sci. Educ. (Dordr.), \textbf{24}, 561-590 (2015).
\bibitem{Stephens2012} A. L. Stephens and J. J. Clement, The Role of Thought Experiments in Science and Science Learning, In \textit{Second international handbook of science education}, edited by B. J. Fraser, K. T. Campbell, and J. McRobbie (Springer, Dordrecht, 2015), pp. 157-175.
\bibitem{Gilbert2000} J. K. Gilbert and M. Reiner, Thought Experiments in Science Education: Potential and Current Realisation, Int. J. Sci. Educ., \textbf{22}, 265 (2000).
\bibitem{Etkina2015} E. Etkina, Millikan Award lecture: Students of Physics - Listeners, Observers, or Collaborative Participants in Physics Scientific Practices?, Am. J. Phys., \textbf{83}, 669 (2015).
\bibitem{Bub1997} J. Bub, \textit{Interpreting the quantum world} (Cambridge University Press, Cambridge, UK, 1997).
\bibitem{Schlosshauer2007} M. Schlosshauer, \textit{Decoherence and the quantum-to-classical-transition} (Springer, 2007).
\bibitem{Hobson2013} A. Hobson, There are no particles, there are only fields, Am. J. Phys., \textbf{81}, 211 (2013).
\bibitem{Ghirardi1996} G. Ghirardi, R. Grassi, and M. Michelini, A Fundamental Concept in Quantum Theory, in \textit{Thinking Physics for Teaching}, edited by C. Bernardini, C. Tarsitani, and M. Vicentini (Springer, New York, 1996) pp. 329-334.
\bibitem{Michelini2004} M. Michelini, R. Ragazzon, L. Santi, and A. Stefanel, Discussion of a Didactic Proposal on Quantum Mechanics with Secondary School Students, Il Nuovo Cimento C, \textbf{27}, 555 (2004).
\bibitem{Michelini2019} M. Michelini and A. Stefanel, A path to build basic Quantum Mechanics ideas in the context of light polarization and learning outcomes of secondary students, J. Phys.: Conf. Ser., 1929 012052 (2021).
\bibitem{Michelini2002} M. Michelini, L. Santi, A. Stefanel, and G. Meneghin, A resource environment to introduce quantum physics in secondary school, in \textit{Proceedings of MPTL-7 International Workshop} (2002), Retrieved from: \url{www.ud.infn.it/URDF/ffc/quanto/articoli/art_11.pdf}
\bibitem{Gilbert2002} R. Justi and J. K. Gilbert, Modelling, teachers' views on the nature of modelling, and implications for the education of modellers, Int. J. Sci. Educ., \textbf{24}(4), 369-387 (2002).
\bibitem{Gilbert2016} J. K. Gilbert and R. Justi, \textit{Modelling-Based Teaching in Science Education} (Springer, Switzerland, 2016).
\bibitem{Grangier1986} P. Grangier, G. Roger, and A. Aspect, Experimental evidence for a photon anticorrelation effect on a beam splitter: a new light on single-photon interferences, EPL (Europhysics Letters), \textbf{1}, 173 (1986).
\bibitem{Grangier2005} P. Grangier, Experiments with single photons, Seminaire Poincar\'{e} (2005), retrieved from: \url{http://www.bourbaphy.fr/grangier}
\bibitem{Heyde2020} K. Heyde and J. L. Wood, \textit{Quantum Mechanics for Nuclear Structure, Volume 1} (IOP Publishing, Bristol, UK, 2020).
\bibitem{Dirac1967} P. A. M. Dirac, \textit{The principles of quantum mechanics, 4th ed. (revised)} (Clarendon Press, UK, 1967)
\bibitem{Llewellyn2012} D. Llewellyn, \textit{Teaching high school science through inquiry and argumentation, 2nd edition} (Corwin Press, Rochester, NY, 2012).
\bibitem{Galili2007} I. Galili, Thought Experiments: Determining Their Meaning, Sci. Educ. (Dordr.), \textbf{18}, 1-23 (2007).
\bibitem{Brown1991} J. Brown, \textit{The Laboratory of the Mind: Thought Experiments in the Natural Sciences} (Routledge, London, UK, 1991).
\bibitem{Etkina2006} E. Etkina, A. Van Heuvelen, S. White-Brahmia, D. T. Brookes, M. Gentile, S. Murthy, D. Rosengrant, and A. Warren, Scientific abilities and their assessment, PRST-PER, \textbf{2}, 020103 (2006).
\bibitem{Michelini2014} M. Michelini and G. Zuccarini, University students’ reasoning on physical information encoded in quantum state at a point in time, in \textit{Proceedings of PERC 2014}, edited by P. V. Engelhardt, A. Churukian, and D. L. Jones. (Minneapolis, USA, 2014), pp. 187-190.
\bibitem{Passon2019} O. Passon, T. Z\"{u}gge and J. Grebe-Ellis, Pitfalls in the teaching of elementary particle physics, Phys. Educ., \textbf{54}, 015014 (2019).
\bibitem{Norsen2017} T. Norsen, \textit{Foundations of Quantum Mechanics: An Exploration of the Physical Meaning of Quantum Theory} (Springer International Publishing, Cham, Switzerland, 2017).
\bibitem{Debianchi2013} M. Sassoli De Bianchi, Quantum fields are not fields; comment on ``There are no particles, there are only fields,'' by Art Hobson [Am. J. Phys. \textbf{81}, 211 (2013)], Am. J. Phys., \textbf{81}, 707 (2013).
\bibitem{Isham1995} C. J. Isham, \textit{Lectures on quantum theory: Mathematical and structural foundations} (Imperial College Press, London, UK, 1995).
\bibitem{Schilpp1998} P. A. Schilpp, \textit{Albert Einstein, Philosopher-Scientist: The Library of Living Philosophers Volume VII, 3rd edition} (Open Court Publishing Co, US, 1998).
\bibitem{Bell2010} J. S. Bell and A. Aspect, \textit{Speakable and Unspeakable in Quantum Mechanics, 2nd edition, with a new introduction by Alain Aspect} (Cambridge University Press, UK, 2010).
\bibitem{Tanona2004} S. Tanona, Uncertainty in Bohr's response to the Heisenberg microscope, Stud. Hist. Philos. Sci. \textbf{35}, 483-507 (2004).
\bibitem{House2018} House of Representatives 6227-National Quantum Initiative Act (2018) 115th Congress, 164:132 STAT. 5092. \url{www.congress.gov/bill/115th-congress/house-bill/6227}.
\bibitem{Bub1998} J. Bub, Quantum measurement problem,  in \textit{Routledge Encyclopedia of Philosophy}, edited by E. Craig (Taylor and Francis, 1998), URL: \url{https://www.rep.routledge.com/articles/thematic/quantum-measurement-problem/v-1}
\bibitem{Kragh1992} H. Kragh, A Sense of History: History of Science and the Teaching of Introductory Quantum Theory, Sci. Educ. (Dordr.) \textbf{1}, 349-363 (1992).
\bibitem{McDermott2013} L. C. McDermott, Improving the teaching of science through discipline-based education research: An example from physics, EJMSE, \textbf{1}(1), 1-12 (2013).
\bibitem{Bakker2015} A. Bakker and D. van Eerde (2015), An introduction to design-based research with an example from statistics education, in \textit{Approaches to qualitative research in mathematics education}, edited by A. Bikner-Ahsbahs, C. Knipping, and N. Presmeg (Springer Netherlands, 2015), pp. 429-466.
\bibitem{Erickson2012} F. Erickson, Qualitative research methods for science education, in \textit{Second international
handbook of science education}, edited by  B.J. Fraser, K. Tobin, and C.J. McRobbie (Springer, Dordrecht, 2012), pp. 1451-1469.
\bibitem{White1992} R. White and R. Gunstone, \textit{Probing understanding} (The Falmer Press, London, UK, 1992).
\bibitem{McIntyre2012} D. H. McIntyre, C. A. Manogue, C. A., and J. Tate, \textit{Quantum Mechanics: A Paradigms Approach} (Pearson Addison-Wesley, San Francisco, CA, 2012).


\end{thebibliography}
\end{document}